\documentclass[10pt]{article}

\usepackage{adjustbox}
\usepackage{amsmath}
\usepackage{amssymb}
\usepackage{amsthm}
\usepackage{authblk}
\usepackage{graphicx}
\usepackage{fullpage}
\usepackage{textcomp}
\usepackage{url}
\usepackage{natbib}
\usepackage[colorlinks=true,linkcolor=blue,citecolor=blue,urlcolor=blue]{hyperref}
\usepackage[x11names]{xcolor}
\usepackage[utf8]{inputenc}
\usepackage[normalem]{ulem}
\usepackage{comment}
\usepackage{tcolorbox}
\usepackage{subcaption}
\usepackage{geometry}
\usepackage{caption}
\usepackage{threeparttable}
\usepackage{xcolor}
\usepackage{soul}
\usepackage{float}
\usepackage{mdframed}
\usepackage{lipsum}
\usepackage{tabularx} 
\usepackage{array}
\usepackage{tikz}
\usepackage{longtable}
\usepackage{tcolorbox}  
\soulregister{\citep}{1} 
\usepackage{comment}

\mdfdefinestyle{reviewer}{
    hidealllines=true,
    linecolor=black,
    linewidth=1pt,
    backgroundcolor=yellow!70,
    innertopmargin=10pt,
    innerbottommargin=10pt,
    innerrightmargin=10pt,
    innerleftmargin=10pt,
    skipabove=10pt,
    skipbelow=10pt,
}

\mdfdefinestyle{improve}{
    hidealllines=true,
    linecolor=black,
    linewidth=1pt,
    backgroundcolor=DarkSeaGreen1!70,
    innertopmargin=10pt,
    innerbottommargin=10pt,
    innerrightmargin=10pt,
    innerleftmargin=10pt,
    skipabove=10pt,
    skipbelow=10pt,
}

\DeclareCaptionFormat{myformat}{\textbf{#1}#2#3\par}
\setcitestyle{aysep={,}}

\theoremstyle{definition}

\title{\Large \bf A new framework to predict and visualize technology acceptance: A case study of shared autonomous vehicles}

\author[a]{Lirui Guo\footnote{Corresponding author. E-mail address:  {\tt\small lirui.guo@monash.edu (L.Guo)}. Present address: Department of Civil Engineering, Monash University. Wellington Rd, Clayton VIC 3800. © 2024. This manuscript version is made available under the CC-BY-NC-ND 4.0 license https://creativecommons.org/licenses/by-nc-nd/4.0/}} 
\author[b]{Michael G. Burke}
\author[a,b]{Wynita M. Griggs}
\affil[a]{\small Department of Civil Engineering, Monash University, Clayton, VIC 3800, Australia}
\affil[b]{\small Department of Electrical and Computer Systems Engineering, Monash University, Clayton, VIC 3800, Australia}
\date{}
\begin{document}

\maketitle
\pagestyle{plain}

\begin{abstract}
Public acceptance is critical to the adoption of Shared Autonomous Vehicles (SAVs) in the transport sector. Traditional acceptance models, primarily reliant on Structural Equation Modeling, may not adequately capture the complex, non-linear relationships among factors influencing technology acceptance and often have limited predictive capabilities. This paper introduces a framework that combines Machine Learning techniques with chord diagram visualizations to analyze and predict public acceptance of technologies. Using SAV acceptance as a case study, we applied a Random Forest machine learning approach to model the non-linear relationships among psychological factors influencing acceptance. Chord diagrams were then employed to provide an intuitive visualization of the relative importance and interplay of these factors at both factor and item levels in a single plot. Our findings identified Attitude as the primary predictor of SAV usage intention, followed by Perceived Risk, Perceived Usefulness, Trust, and Perceived Ease of Use. The framework also reveals divergent perceptions between SAV adopters and non-adopters, providing insights for tailored strategies to enhance SAV acceptance. This study contributes a data-driven perspective to the technology acceptance discourse, demonstrating the efficacy of integrating predictive modeling with visual analytics to understand the relative importance of factors in predicting public acceptance of emerging technologies.

\end{abstract}

\noindent\textbf{Keywords:} Shared autonomous vehicles; Technology acceptance; Predictive modeling; Machine learning; Chord diagram visualization

\section{Introduction}\label{sec: Introduction}

Over the past decades, consumer behavior has shifted from ownership-centric to access-centric models, notably with the rise of shared mobility services \citep{lirui-020}. This shift is evidenced by a decline in vehicle ownership and the rising relevance of shared mobility \citep{lirui-080, lirui-081, lirui-015}. In this context, Shared Autonomous Vehicles (SAVs) emerge as a transformative innovation \citep{lirui-020}. SAVs promise enhanced efficiency, reduced congestion and accidents, lower environmental impact, and improved accessibility compared to traditional mobility solutions \citep{lirui-055, lirui-025, lirui-024, lirui-021, lirui-043}.

However, like all other new technology and novel products, the success of SAVs critically hinges on public acceptance, a challenge often overlooked in the face of technological advancements \citep{lirui-019}. Existing models, like the Technology Acceptance Model (TAM) and the Unified Theory of Acceptance and Use of Technology (UTAUT), offer theoretical frameworks to understand this acceptance \citep{lirui-053, lirui-085}. These frameworks almost all rely on factor analysis tools like Structural Equation Modeling (SEM) for validation. Yet, traditional validation methods, notably explanatory modeling such as SEM, exhibit limitations in capturing the complex, non-linear relationships essential in understanding individual acceptance decisions for these advanced systems \citep{OMALLEY2014131}. 
For example, a population comprising adopters and non-adopters with polarised views may result in factors driving acceptance being masked. In addition, existing factor analysis visualizations produced using these tools often fail to intuitively present the impact of individual predictors, a vital aspect to better understanding SAV acceptance.

Research has traditionally tended towards explanatory modeling, but the field of predictive analytics holds significant potential as a data-driven technique. SEM primarily focuses on either confirming (i.e., covariance-based SEM) or explaining (i.e., partial least squares SEM) variable relationships based on theoretical constructs and achieving a good model fit \citep{PLS-SEM_R_book}, rather than on predictive accuracy. This focus can result in models that fit the data well but have limited ability to predict new observations or capture the nuanced factors influencing individual acceptance decisions. \cite{breimanStatisticalModelingTwo2001} argues that higher predictive accuracy provides more reliable insights into the underlying data mechanisms, while weak predictive accuracy can lead to unreliable conclusions. Predictive analytics, especially using machine learning methods, can offer better predictive accuracy than traditional data models and bridge the gap between theoretical constructs and practical application \citep{breimanStatisticalModelingTwo2001, Alwabel_data-driven_modeling_2021}. Therefore, adopting predictive analytics emerges as a compelling avenue for advancing technology acceptance research, offering insights that are both data-rich and pragmatically relevant \citep{Alwabel_data-driven_modeling_2021}.

Furthermore, while SEM is a widely used approach for exploring factors influencing Autonomous Vehicle (AV) acceptance, the visualization strategies commonly employed often only show the weights of the overall factors. This limitation makes it difficult to visualize the relative influence of each item on AV acceptance, particularly when attempting to discern how specific items within key factors---such as Trust, Perceived Risk, or Attitude---uniquely influence SAV acceptance. The inability of SEM to succinctly visualize these intricate relationships points to a significant methodological gap. This underscores the need for more intuitive and revealing visualization tools that can offer deeper insights into the complex dynamics underpinning SAV acceptance.

Addressing the identified gaps, this paper introduces an innovative framework for predicting public acceptance of technology. We adopt a Machine Learning approach complemented by chord diagram visualizations to analyze factor relationships and forecast public acceptance of emerging technologies like SAVs. Our framework thoroughly evaluates the influence of the identified predictors of technology acceptance derived from the literature and provides an intuitive, user-friendly visualization of their importance. By allowing for both factor-level and item-level analysis, it provides detailed insights into which specific aspects within each factor are most influential in predicting acceptance.
This approach provides policymakers, designers, and stakeholders with essential insights, fostering strategic initiatives to enhance public acceptance of new technological innovations. 

Specifically, this research delves into the complex and non-linear psychological factors affecting the acceptance of SAVs as a case study to validate our proposed framework. We focus on single-request Car-Sharing systems and SAE Level 5 SAVs \citep{lirui-038, SAEwebsite}, representing the forefront of shared mobility autonomy. 

The rest of this paper unfolds as follows: Section \ref{sec: Literature review} reviews relevant literature on key psychological factors influencing public acceptance of SAVs, contrasts explanatory and predictive modeling approaches, and explores data visualization in SAV acceptance studies. Section \ref{sec: Framework} presents our proposed framework. Section \ref{sec: Methodology} reveals the application of our proposed framework in the SAV case study, which also includes the questionnaire design and data collection details. Section \ref{sec: Results} presents the results and discussion, outlining factor importance and model performance validation. Section \ref{sec: Discussion} provides concluding remarks, discussing key implications of our findings, practical implications for stakeholders, and directions for future research.

\section{Related Literature} \label{sec: Literature review}

\subsection{Private AVs or SAVs}

Diverse perceptions exist among the public regarding the preference for private autonomous vehicles (AVs) over SAVs. While many studies have not explicitly focused on specific SAE levels of autonomy, descriptions typically imply higher levels of automation. In Australia, there appears to be a trend favoring ride-sharing AVs over the purchase of private AVs, especially among those already using car-sharing services \citep{lirui-025, lirui-021}. Conversely, a study in Austin, Texas, indicates a preference for owning AVs, with individuals interested in SAVs expressing a desire to try these before considering purchase \citep{lirui-026}. This finding aligns with surveys by \cite{lirui-014} and \cite{lirui-082}, where 63\% of New Zealand participants showed a preference for owning AVs despite higher costs, and approximately 60\% of American participants expressed disinterest in SAVs. A study by \cite{lirui-055} comparing preferences among 542 Chinese respondents further underscores these trends. This research evaluated preferences for owning a private human-driven vehicle, a private AV, using an SAV, or engaging in car-sharing. The findings suggest a general inclination towards retaining personal vehicles or purchasing AVs over utilizing shared options. However, SAVs emerged as more favorable compared to traditional car-sharing models, highlighting a nuanced preference landscape in the context of AV acceptance.

\subsection{Theoretical Models in AV Acceptance Research}

Survey studies and theoretical models play a pivotal role in understanding the factors influencing public acceptance of AVs. Research consistently identifies certain demographic factors, such as being younger, male, and residing in urban areas, as contributing to higher AV acceptance \citep{lirui-010, lirui-017, lirui-078}. Additional factors include higher education and income levels \citep{lirui-008}, frequency of driving \citep{lirui-017}, shorter driving histories \citep{lirui-021}, individuals with physical limitations impeding driving, and owners of vehicles with advanced automated features \citep{lirui-010}. 

Several theoretical models have been developed to analyze human behavior and the psychological factors influencing the acceptance of emerging technologies, including AVs. Prominent among these are the Technology Acceptance Model (TAM, \citet{lirui-053}), the Theory of Planned Behavior (TPB, \cite{lirui-084}), and the Unified Theory of Acceptance and Use of Technology (UTAUT, \cite{lirui-085}). While TPB offers a general framework for understanding human behavior, TAM and UTAUT were specifically formulated for technology acceptance in Information Systems research \citep{lirui-005}. 

In recent years, TAM has been extensively applied to study public acceptance of AVs due to its simplicity, efficacy in explaining various information systems' technology acceptance, and its proven effectiveness in interpreting driver acceptance of new in-vehicle technologies \citep{lirui-002, lirui-005, lirui-019, lirui-036, lirui-037, lirui-043, lirui-071, lirui-091, lirui-092}. 
TAM primarily employs perceived ease of use (PEOU), perceived usefulness (PU), and attitude towards using (A) as the drivers of behavioral intention to use (BI).
Other psychological factors that have been widely explored include social trust \citep{lirui-001}, initial trust \citep{lirui-019, lirui-002, lirui-004, lirui-005}, perceived risk (PR) \citep{lirui-030, lirui-001, lirui-005}, and psychological ownership \citep{lirui-036, lirui-037, lirui-092}.
These factors have also been validated in studies focusing on public acceptance of SAVs \citep{lirui-037, lirui-043, lirui-092}, highlighting their relevance in this evolving field.

\subsection{Psychological Factors}
In this work, we propose methods to explore several psychological factors crucial in predicting public acceptance of SAVs. These include elements from the TAM (i.e., PEOU, PU, A, BI), alongside trust, PR, and psychological ownership.

\subsubsection{Technology Acceptance Model}

The original TAM identifies three primary determinants influencing technology acceptance (i.e., BI): PEOU, PU, and A. PEOU refers to the degree to which a person believes that using a particular system would be effortless, while PU is the degree to which a person believes that using a particular system would enhance job performance \citep{lirui-053}. Attitude represents an individual’s positive or negative feelings towards using the technology \citep{lirui-019}. These determinants have been robustly validated in numerous studies, making them primary predictors of users’ intention to use SAVs.

\subsubsection{Trust}\label{subsubsection:trust}

Trust, defined by \cite{lirui-097} as “a psychological state comprising the intention to accept vulnerability based upon positive expectations of the intentions or behavior of another”, is a critical factor in understanding public acceptance of AVs. Various aspects of trust have been explored in previous research. \cite{lirui-004} identified key sources of trust in AVs as system transparency, technical competence, and situation management, though their survey primarily emphasized dependence and reliability. \cite{lirui-019} underscored initial trust as the most pivotal factor in fostering a positive attitude towards AVs, subsequently influencing users' intention to use AVs, aligning with findings from \cite{lirui-010}. 

\cite{lirui-001} explored trust in the context of limited access to new technologies, highlighting the importance of trust in manufacturers (automakers and IT companies involved in AV production) and government authorities responsible for regulation and oversight, known as social trust. Their research indicated that social trust directly and indirectly influences acceptance of fully autonomous vehicles, with its indirect effects notably shaping overall acceptance and its direct effects influencing willingness to pay and behavioral intention to use.

\cite{lirui-047} delved deeper, delineating three dimensions of trust: trust in AV performance, trust in AV manufacturers, and trust in regulatory institutions, alongside two risk dimensions (performance risk and privacy risk). Their comprehensive study revealed that performance trust directly impacts the intention to use AVs, while trust in manufacturers affects intention through privacy risk (defined below in Section \ref{subsubsection:PR}) mediation. Notably, trust in regulatory institutions was found to have neither direct nor indirect effects on behavioral intention to use. 

Our research aims to thoroughly investigate these various aspects of trust through a series of targeted survey questions, predictive analysis, and visualization, to better understand its nuanced role in shaping public acceptance and SAV usage intention.

\subsubsection{Perceived Risk (PR)}\label{subsubsection:PR}

Despite AVs potentially offering safer transportation than human drivers \citep{lirui-033, lirui-036}, public concerns persist about the risks associated with AVs \citep{lirui-019, lirui-074, lirui-034, lirui-017, lirui-030}. A primary issue is the perceived loss of vehicle control during AV operation, which leads to diverse perceptions regarding risk and safety \citep{lirui-033}. Studies suggest that the public often places more emphasis on the potential risks, such as accidents, associated with AVs than on their benefits \citep{lirui-030}. \cite{lirui-047} have identified two types of risk impacting the intention to use AVs: performance risk, encompassing safety and functionality concerns, and privacy risk, risks pertaining to security and privacy of user data. Interestingly, their findings indicate that while performance trust significantly influences user intentions, perceived performance risk does not have a notable effect. This suggests that trust in AV performance plays a more pivotal role than risk perception in determining user acceptance.
This insight underscores the importance of thoroughly understanding public perceptions of risks and their influence on the behavioral intention to use SAVs. Addressing these perceived risks, particularly by enhancing trust in AV performance, is crucial for fostering wider acceptance of this emerging technology.

\subsubsection{Psychological Ownership}

Psychological ownership, as defined by \cite{lirui-061}, refers to the state where individuals feel a sense of ownership over a target, asserting `It is mine!' even in the absence of legal ownership. This sensation can develop towards any tangible or intangible object that an individual finds visible, attractive, and engaging, whether it be legally owned, shared, or abstract \citep{lirui-061, lirui-040}. 

Research has demonstrated the broad implications of psychological ownership. \cite{lirui-066} found that heightened psychological ownership led to increased care for public goods, while \cite{lirui-064} linked it to a boost in self-esteem. \cite{lirui-062} observed that interventions fostering psychological ownership increased interest in claiming government benefits. In the realm of marketing, psychological ownership has been shown to influence consumer behavior, such as willingness to pay higher prices and purchase intentions \citep{POchapter5}. 
 
Understanding psychological ownership requires an exploration of its underlying human motives. According to \cite{POchapter1}, these include effectance motivation (effective interaction with one's environment leading to efficacy and pleasure), self-identity (using possessions for self-definition and expression), home (creating a sense of belonging in a specific space and time), and stimulation (seeking arousal and activation). The development of psychological ownership is facilitated through control over the target, intimate knowledge of it, and investment of the self into the target, covering a range of activities from customization to care and personalization \citep{lirui-040, lirui-039}.  

In the context of AVs, research has extensively investigated car ownership dynamics \citep{lirui-031, lirui-025, lirui-021, lirui-026, lirui-015}. However, the concept of psychological ownership in relation to SAVs presents a novel area of exploration with promising potential. Studies such as those by \cite{lirui-036, lirui-037} and \cite{lirui-092} have shed light on the role of psychological ownership in shaping behavioral intentions towards SAVs and suggested a positive relationship between psychological ownership and behavioral intention to use SAVs. Given the emerging trends towards shared mobility and the advancements in vehicle autonomy, the concept of psychological ownership offers a compelling perspective. It suggests the possibility of psychological ownership emerging as an influential factor, potentially supplanting the traditional notion of physical car ownership in the realm of SAVs. This shift highlights the evolving nature of consumer attitudes in a landscape increasingly dominated by high levels of automation and shared transport solutions. Our case study aims to investigate a potential triggering mechanism to explore the impact of psychological ownership in shaping public acceptance of SAVs, testing our framework's ability to delve into a relatively uncharted aspect of SAV acceptance.

\subsubsection{Relationship Among Constructs}\label{relationships_b/w_constructs}

Understanding the relationships among constructs is crucial in acceptance studies. From a methodological perspective, we face two key challenges: ensuring that our data collection instruments capture information relevant to these constructs and identifying the relationships between them.

In the field of AV acceptance research, there is often a tendency to assume fixed linear relationships between various constructs and follow a data modeling approach to propose theoretical models that explain the public acceptance of AVs. However, this approach can lead to theoretical conflicts, as multiple models may fit the data equally well, raising questions about which model truly captures the underlying mechanisms. When different models pass goodness-of-fit tests or statistical measures, it becomes difficult to select one over the others purely based on theoretical grounds \citep{breimanStatisticalModelingTwo2001}.

As a foundational framework, TAM traditionally connects constructs such as PU, PEOU, Attitude, and BI in a cohesive manner. However, the role of the Attitude construct in TAM has been a subject of debate. Some researchers, such as \cite{lirui-098} and \cite{Davis_A-critical-assessment_1996}, argued that excluding Attitude from TAM does not diminish the model's explanatory power. This claim has been supported by recent studies on AV acceptance \citep{lirui-002, lirui-004}. Conversely, other researchers advocate for the inclusion of Attitude as a direct antecedent of BI, emphasizing its potential impact on users' intention to adopt SAVs \citep{lirui-019, lirui-037, lirui-043}. These conflicting perspectives highlight the variability in theoretical approaches within the field.

Moreover, constructs related to attitudes have been treated as antecedents of PEOU and PU in previous literature. For example, \cite{venkateshTechnologyAcceptanceModel2008} included factors like computer anxiety, playfulness, and perceived enjoyment as antecedents of PEOU in the Technology Acceptance Model 3 (TAM3). These factors are closely related to individuals' feelings towards using technology, i.e. their attitudes. Similarly, \cite{yuenDeterminantsPublicAcceptance2020} included belief in benefits (e.g., ``I believe I can benefit from using AVs") in the Observability factor and showed that it serves as an antecedent of Perceived Value of AVs, similar to PU.

These examples illustrate that Attitude and its related items have been used as both antecedents and consequences in relation to PEOU and PU, highlighting the variability in theoretical approaches. The same variability is observed in the relationship between Trust and PR. \cite{lirui-047} identified models where PR is an independent variable with Trust acting as a mediator, and others where PR serves as a mediator between Trust and acceptance.

Recognizing the lack of consensus in the literature, our study adopts a data-driven approach to explore these relationships from a predictive modeling perspective, while ensuring that our analysis is informed by existing theoretical frameworks.

\subsection{Data analysis}

\subsubsection{Structural Equation Modeling (SEM)}\label{subsub:SEM}

SEM is a widely adopted statistical approach in researching the acceptance of AVs, valued for its ability to model complex relationships between observed and latent variables and to test theoretical models \citep{mda_2010, Principles_and_Practice_of_SEM_2022}. It enables researchers to explore and confirm hypothesized relationships among constructs, making it a powerful tool for theory testing and development.

However, there are limitations in applying SEM to AV acceptance research. A primary concern is its assumption of linear relationships between variables \citep{OMALLEY2014131}, which may not fully capture the often non-linear nature of behavioral intentions towards AVs or SAVs. This issue is particularly evident in the context of PR and Trust, where linear assumptions may oversimplify complex dynamics, as discussed in Section \ref{relationships_b/w_constructs}. The sample size is another crucial aspect of SEM, as estimates such as standard errors for the effects of latent variables may be inaccurate when working with small samples. While there is no absolute minimum sample size standard, small sample size, complex models, non-normal distributed samples, or missing data can pose challenges in AV acceptance studies \citep{Principles_and_Practice_of_SEM_2022}. Moreover, the default estimation method in SEM, maximum likelihood, relies on multivariate normality, a condition rarely fulfilled in practice \citep{Principles_and_Practice_of_SEM_2022}. Furthermore, inaccurate model specifications, such as incorrect causal directions or omitted variables, can lead to misleading conclusions about variable relationships \citep{breimanStatisticalModelingTwo2001, tomarken_structural_2005}.

Advancements like partial least squares SEM (PLS-SEM) address some of these limitations by offering flexibility with small samples, no strict distributional assumptions, and the capability to handle non-linear effects \citep{PLS-SEM_R_book}. However, PLS-SEM is limited in its ability to model non-recursive structures and lacks a comprehensive goodness-of-fit measure, constraining its utility for theory testing \citep{PLS-SEM_R_book}. Additionally, the increasing complexity of SEM and the availability of user-friendly software tools necessitate a strong conceptual understanding of the methodology to avoid misleading results \citep{Principles_and_Practice_of_SEM_2022}. 

Moreover, SEM is primarily suited for explanatory modeling based on theoretical constructs and focuses on achieving a good model fit rather than predictive accuracy. Different theoretical models may fit the data equally well yet lead to different conclusions about underlying relationships \citep{breimanStatisticalModelingTwo2001}. Such situations may result in potential ambiguity or conflicting interpretations of constructs like Perceived Risk and Trust, as discussed in Section \ref{relationships_b/w_constructs}.

These considerations highlight the challenges associated with applying and interpreting SEM in acceptance research. While SEM remains a valuable tool for exploring theoretical relationships, its limitations suggest the need for complementary approaches that can handle non-linear relationships and provide robust predictive capabilities.

\subsubsection{Machine Learning Modeling}

Machine Learning (ML) offers a data-driven alternative to traditional statistical methods like SEM, particularly in handling complex, non-linear relationships without relying on strict assumptions. ML algorithms can be classified as supervised, unsupervised, and reinforcement learning based on the nature of the learning process, and the type of data and feedback used to train the algorithms. Particularly relevant to understanding public perceptions of SAVs, supervised ML methods are favored because they focus on predicting observable response variables, such as user acceptance, which are often identified from collected survey results. Common supervised learning techniques include Multiple Linear Regression, Logistic Regression, Tree methods, K-nearest neighbors, Support Vector Machines, and Neural Networks.

ML modeling differs distinctly from the regression-based approach of SEM. Unlike SEM, which relies on theoretical regression coefficients to explain relationships between variables, ML prioritizes predictive accuracy and gauges the effectiveness of its models by how well predictions correspond to unseen, real-world data. ML algorithms are particularly adept at identifying intricate patterns and interactions within data sets, without the necessity for pre-established theoretical models of variable relationships \citep{breimanStatisticalModelingTwo2001}. This data-centric approach enables ML models to identify and adapt to complex data relationships that may not be clearly defined or fully captured by existing theoretical models. ML modeling offers a more flexible and potentially more effective option to SEM, particularly in situations marked by complex interactions that cannot be explained by simple theory, or where the patterns that emerge are beyond the scope of current theories \citep{Alwabel_data-driven_modeling_2021}.

\begin{figure}[!h]
\captionsetup{font=small}
  \centering
  \begin{subfigure}[b]{0.44\textwidth}
    \centering
    \includegraphics[width=\textwidth]{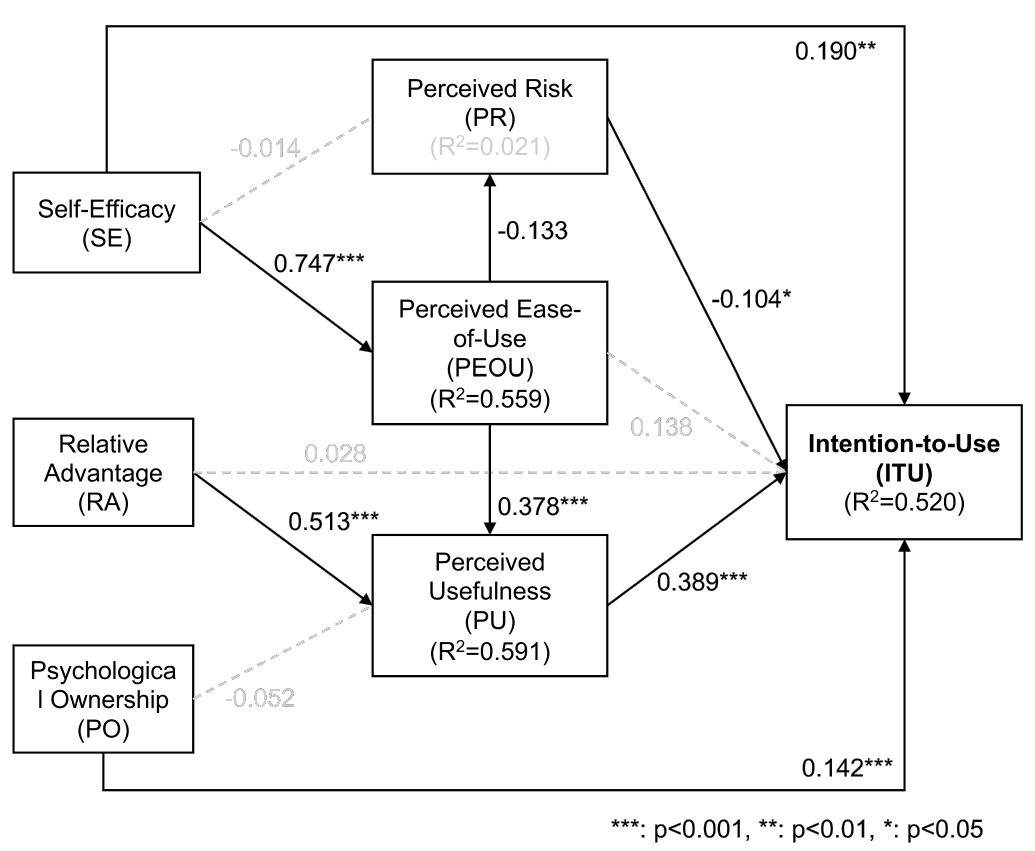}
    \caption{Estimated SEM by \citet{lirui-036} depicting the influential factors on the usage of AVs within a TAM framework.}
    \label{fig:Lee_model}
  \end{subfigure}
  \hfill
  \begin{subfigure}[b]{0.54\textwidth}
    \centering
    \includegraphics[width=\textwidth]{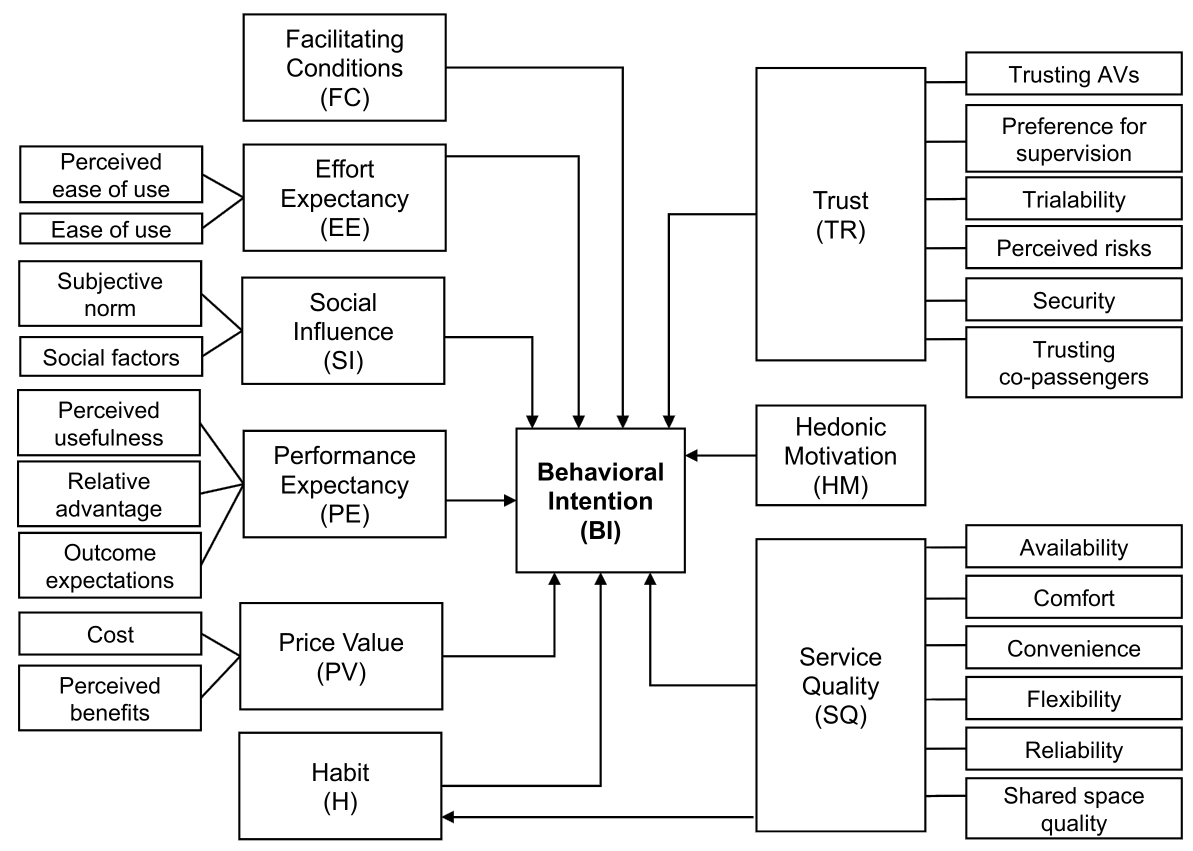}
    \caption{Conceptual acceptance model for SAVs based on UTAUT2, as proposed by \citet{Dichabeng_Factors_2021}.}
    \label{fig:Dichabeng_model}
  \end{subfigure}
  \hfill
  \begin{subfigure}[b]{0.8\textwidth}
    \centering
    \includegraphics[width=\textwidth]{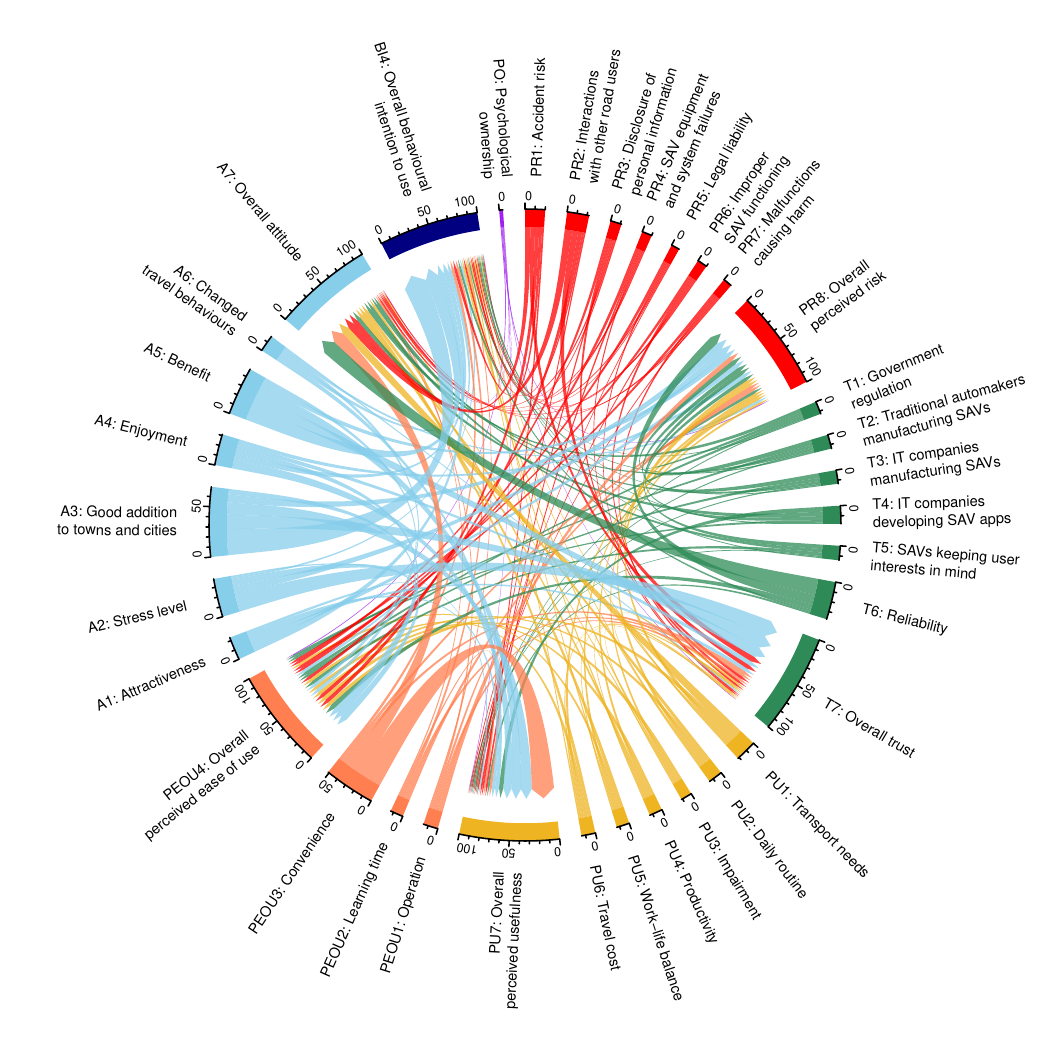}
    \caption{Chord diagram displaying the relative importance of various factors in predicting overall questions in the SAV case study, showcasing the application of the proposed framework in this study.}
    \label{fig:cir_plot_compare}
  \end{subfigure}
  \caption{Comparison of Factor Relationship Visualization: Subfigures \ref{fig:Lee_model} and \ref{fig:Dichabeng_model} depict previous models by \citet{lirui-036} and \citet{Dichabeng_Factors_2021}, respectively. Subfigure \ref{fig:cir_plot_compare} showcases the Chord Diagram derived from the current case study, emphasizing the methodological advancements and findings by applying the proposed framework in this study, as detailed in Section \ref{sec: Framework} and \ref{sec: Methodology}.}
  \label{fig:previous_models}
\end{figure}

While the integration of ML methods into SAV acceptance research is still in its early stages, existing applications in similar domains demonstrate significant promise. For instance, \citet{lirui-094} showcased the effectiveness of ML in predicting user acceptance of connected and autonomous vehicles (CAVs). Their research utilized a range of ML techniques, including Neural Networks, Random Forests, Naive Bayes, and Fuzzy Logic Models, and achieved high predictive accuracy across these methods. Similarly, \citet{Alwabel_data-driven_modeling_2021} developed a data-driven approach using ML and predictive analytics to predict end-users' acceptance of consumer-use technology in a non-organizational setting. Their findings highlight the ability of ML to construct sophisticated models that can advance traditional technology acceptance theories.

However, the application of ML models specifically in the context of SAVs remains underexplored, particularly regarding understanding the key drivers of acceptance and providing actionable insights at both the factor and item levels. Our research aims to bridge this gap by applying ML methodologies to delve deeper into the multifaceted nature of SAV adoption and acceptance.

While offering notable advantages, ML models are not without challenges. The `black box' nature of some ML algorithms can obscure the interpretability of results, posing a significant challenge \citep{rudin_stop_2019}. To mitigate this limitation, this work introduces the use of chord diagrams to enhance the interpretability and transparency of ML model results, providing a single-figure overview of acceptance and the hierarchies of constructs affecting it.

\subsection{Data Visualization}

Previous research in the field of SAVs has primarily utilized theoretical models, such as SEM, to examine the effects of selected factors on target variables like users' behavioral intention to use AVs (see Figures \ref{fig:Lee_model} and \ref{fig:Dichabeng_model}). These models offer valuable insights into the overall significance of broader constructs. However, challenges arise when attempting to analyze the contributions of numerous factor items or when dealing with complex, non-linear relationships. For example, while SEM can quantify the contributions of item---such as `trust towards government' within the broader `trust' category—if they are explicitly included in the model, incorporating a large number of item can make the model overly complex. This complexity can obscure the interpretability of the results and make it difficult to discern the relative importance of individual item. Moreover, SEM relies on several assumptions, including linearity, normality, and sufficient sample size, as discussed in Section \ref{subsub:SEM}. When data exhibit non-linear relationships or when dealing with bimodal distributions---common in studies contrasting adopters and non-adopters of new technology---SEM may not be the most suitable analytical tool.

To address these limitations, we employ algorithmic models such as the Random Forest, which prioritizes predictive accuracy and can efficiently handle large numbers of predictors and complex, non-linear relationships without relying on strict assumptions. However, as model complexity increases to achieve higher accuracy---such as with Random Forest, which builds an ensemble of decision trees by introducing randomness in their construction \citep{breimanStatisticalModelingTwo2001}---interpretability can decrease, making it challenging to visualize the intricate relationships between variables.

To enhance interpretability, we utilize chord diagrams, which have demonstrated innovative and powerful capabilities in various fields---from illustrating dynamics in contraceptive use \citep{Finnegan_using_2019} to mapping global bilateral migration flows \citep{lirui-093} and health monitoring applications \citep{Zhao_visual_2016}. Their versatility and effectiveness make them an excellent choice for visualizing complex systems with numerous interconnected factors.

In the context of SAV acceptance, chord diagrams offer a unique advantage. They can vividly illustrate how different factors, such as Trust, PR, PEOU, and PU, interact and collectively impact the acceptance of SAVs in a single, intuitive figure. This visualization approach allows for a more nuanced understanding of the interplay among these factors, surpassing the capabilities of traditional theoretical models in capturing the subtleties of user behavior and preferences in the realm of SAVs.

Figure \ref{fig:cir_plot_compare} presents the chord diagram obtained from our case study, in comparison with previous models based on SEM visualizations. The chord diagram compactly illustrates the importance of various factors and their items in predicting SAV acceptance, where the width of the chords indicates the relative importance of each predictor item. Detailed descriptions of how to interpret this diagram are provided in Section \ref{subsubsec: chord diagram visualisation}. By combining algorithmic models with chord diagrams, we aim to effectively predict and visualize technology acceptance, addressing both predictive accuracy and interpretability without proposing a new theoretical adoption model.

\section{Proposed Framework}\label{sec: Framework}

\begin{figure}[H]
\captionsetup{font=small}
    \centering
    \includegraphics[width=\linewidth]{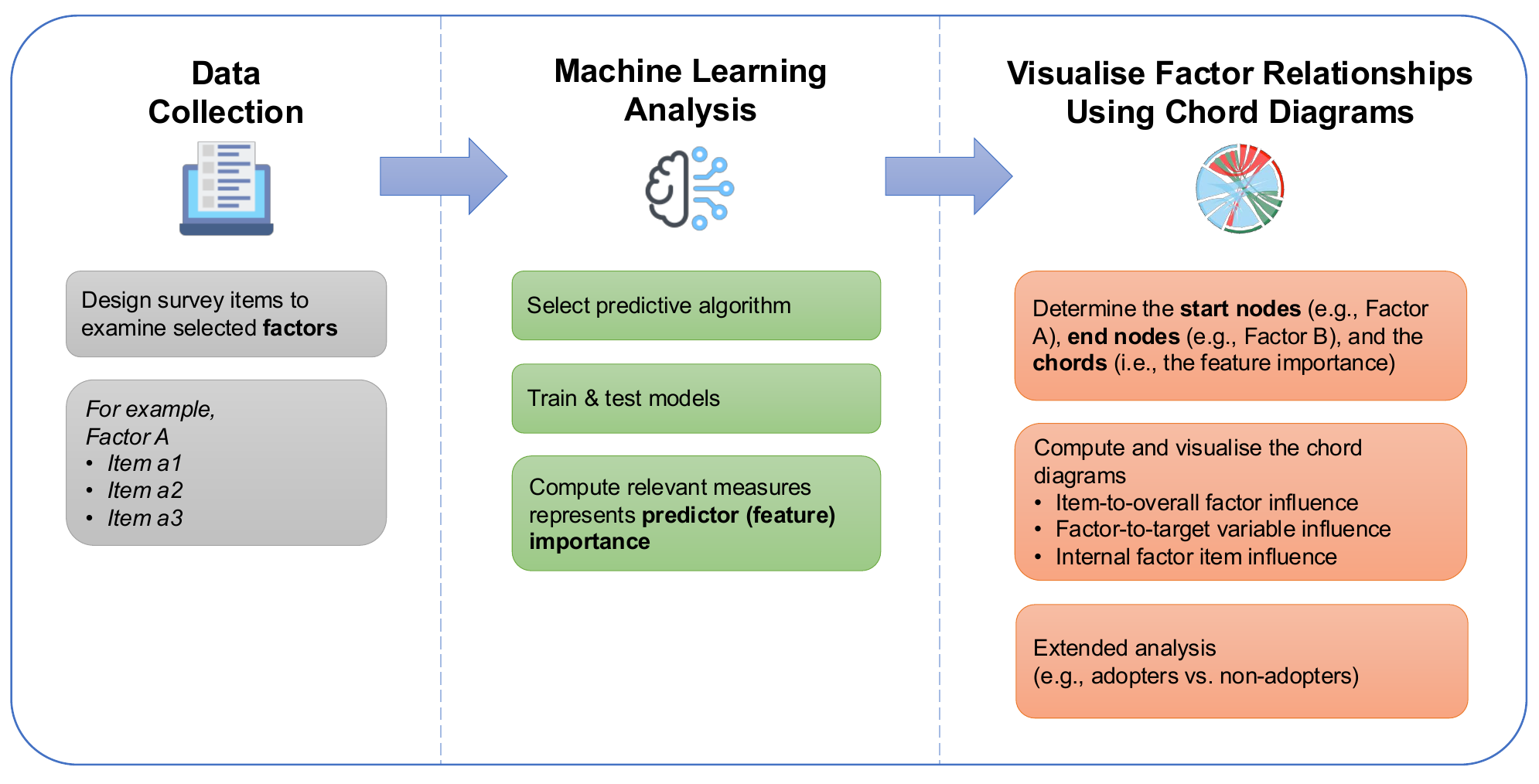} 
    \caption[Proposed Framework for Analyzing and Visualizing Technology Acceptance.]{Proposed Framework for Analyzing and Visualizing Technology Acceptance\footnotemark. This figure outlines a framework for predicting and visualizing the acceptance of new technologies. It encompasses the stages of data collection, Machine Learning analysis, and visualization through chord diagrams, offering a novel approach for acceptance studies across various technological domains.}
    \label{fig:framework}
\end{figure}

This research introduces a general framework designed to assist researchers across various fields in studying technology acceptance. This adaptable framework is graphically represented in Figure \ref{fig:framework}.

\footnotetext{The first two icons are from Icons8 (https://icons8.com).}

The framework encompasses three primary stages:

\begin{enumerate}
    \item \textbf{Data Collection:}
    \begin{itemize}
        \item \textbf{Factor Identification:} Researchers begin by identifying factors relevant to their study. This step may incorporate traditional methodologies to explore the factors that are critical in understanding technology acceptance.
        \item \textbf{Instrument Development:} Subsequently, researchers develop appropriate survey items corresponding to these selected factors. The data is then gathered through well-designed user studies.
    \end{itemize}

    \item \textbf{Machine Learning Analysis:}
    \begin{itemize}
        \item \textbf{Model Selection and Training:} Researchers employ suitable predictive models (e.g., Decision Trees, Support Vector Machines) to analyze the collected data. Each model is trained to predict outcomes based on the responses related to the identified factors.
        \item \textbf{Model Validation and Feature Importance:} The performance of these models is rigorously validated. Feature importance measures, such as Gini impurity in Decision Trees or factor weights in Support Vector Machines, are utilized to gauge the importance of each predictor.
    \end{itemize}

    \item \textbf{Visualize Factor Relationships Using Chord Diagrams:}
    \begin{itemize}
        \item \textbf{Chord Diagram Generation:} In the final stage, chord diagrams are created to visualize the relationships among and within individual factors. Researchers define the nodes (representing factors and items) and chords (representing feature importance and relationships) in these diagrams to analyze the interplay between different factors and items.
        \item \textbf{Extended Analysis:} This framework can also facilitate additional analysis, such as contrasting the perspectives of adopters and non-adopters of the technology under study.
    \end{itemize}
\end{enumerate}

While this framework is designed for broad applicability in acceptance studies, we contextualize its utility by demonstrating its effectiveness in forecasting the acceptance of SAVs. This case study illustrates the framework's capability to analyze and visualize complex acceptance dynamics in a specific technology domain.

\section{A Case Study of SAVs}\label{sec: Methodology}

To demonstrate the practical application of the proposed framework introduced above, this section delves into a case study focusing on Single-request Car-Sharing systems and SAE Level 5 SAVs, which represent the forefront of autonomy in shared mobility. The methodology adopted and application of the proposed framework in this case study, visually summarized in Figure \ref{fig:method}, involves several key steps:

\begin{figure}[!h]
\captionsetup{font=small}
    \centering
    \includegraphics[width=\linewidth]{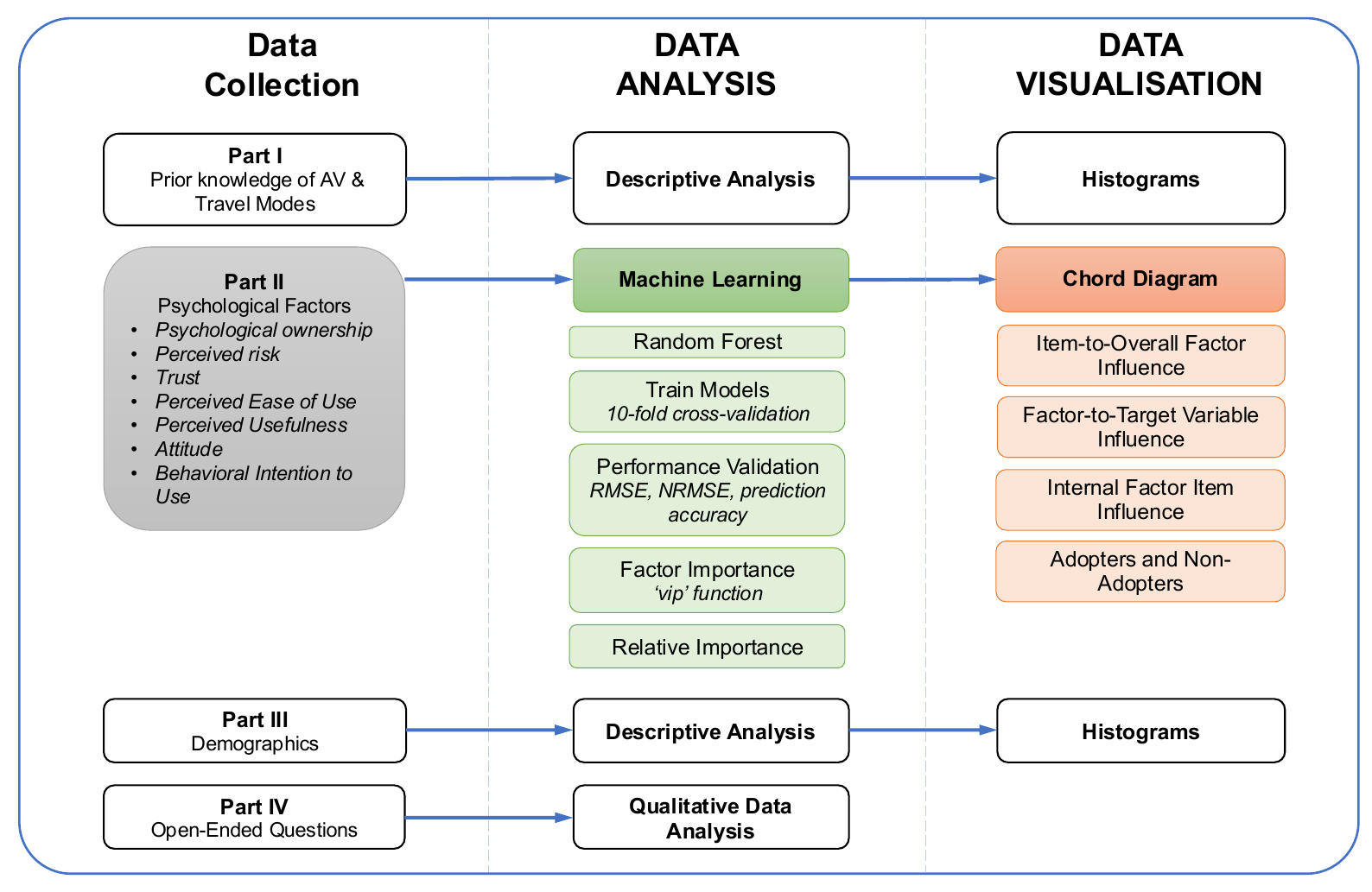} 
    \caption{Implementation of the Proposed Framework in the SAV Acceptance Case Study. This figure illustrates the practical application of the framework in analyzing and visualizing data specific to the acceptance of SAVs. It highlights the key steps undertaken in the case study, from questionnaire design to advanced data analysis and visualization techniques, to derive insights into SAV acceptance.}
    \label{fig:method}
\end{figure}

\begin{enumerate}
    \item \textbf{Literature Review and Questionnaire Design:} Drawing from the in-depth literature review in Section \ref{sec: Literature review}, we developed a detailed questionnaire aimed at assessing the impact of selected psychological factors on the public acceptance of SAVs.
    
    \item \textbf{Machine Learning Analysis:} The data collected was then analyzed using the Random Forest algorithm, a robust ML technique. This analysis predicted public acceptance of SAVs and computed the feature importance of each factor and its individual components (items) in the trained models.
    
    \item \textbf{Visualization Through Chord Diagrams:} The relationships among various factors, as well as their relative importance in influencing SAV acceptance, were visualized using chord diagrams. These diagrams offer an intuitive and clear representation of the interconnections and the relative importance of different predictors in the context of SAV acceptance. We also segmented the respondents into two groups---SAV adopters and non-adopters based on their overall acceptance (indicated by questionnaire item BI4, see Appendix \ref{Appendix_survey_item}). This categorization helped in understanding the distinct needs and perspectives of potential user groups.
\end{enumerate}

The detailed procedure is presented below.

\subsection{Questionnaire Design}

\subsubsection{Survey Structure}

To explore the relationships between the investigated factors and their importance in predicting SAV acceptance, a comprehensive, self-administered online questionnaire was developed using the Qualtrics Insight Platform. The survey (see Appendix \ref{Appendix_survey_item}), comprising 59 questions, was structured into four main parts to capture a broad spectrum of data:

\begin{itemize}
    \item \textbf{Part I: Travel Modes} - Focusing on respondents' existing knowledge about AVs, their current travel behaviors, car ownership, and satisfaction with shared mobility services (e.g., taxis, Uber, and Didi).

    \item \textbf{Part II: Be Driven Into The Future} - This section investigated the impact of the selected psychological factors on participants' intention to use SAVs. It commenced with a definition of fully autonomous vehicles (SAE Level 5 AVs) and employs stories (in both text and audio) to illustrate on-demand and reservation-based SAV services. Respondents then addressed questions (see Table \ref{tab:questionnaire}, Appendix \ref{Appendix_survey_item}) on PR (8 items including one overall item), trust (7), PU (7), PEOU (4), Attitude towards use (7), and behavioral intention to use SAVs (4). The questionnaire incorporated two distinct types of questions: specific questionnaire items (e.g., PR1, PR2) designed to explore various dimensions of each factor (e.g., PR), and overall items (e.g., PR8) aimed at gauging participants’ general perceptions of each investigated factor (e.g., PR).

    \item \textbf{Part III: Demographics} - Gathering socio-demographic data, this part inquired about gender, age, education, residence, disability status, and income. Positioned later in the survey, it aimed to reduce the fatigue of answering the main part of the survey (Part II).

    \item \textbf{Part IV: Open-Ended Questions and User Experience} - Comprising 4 open-ended and 1 multiple-choice questions, this section provided an opportunity for respondents to freely express their opinions, motivations, and satisfaction with this survey study. All open-ended questions were optional, respecting the respondents' preferences and time constraints. 
\end{itemize}

\subsubsection{Addressing Bias in Survey Design}

Recognizing the potential for bias in survey research, our study carefully considered past challenges faced by scholars in assessing public acceptance of AVs. Previous research typically employed five-point Likert scales (e.g., from “strongly disagree (=1)” to “strongly agree (=5)”) in their questionnaires \citep{lirui-019, lirui-036, lirui-037, lirui-043}. However, such scales can lead to interpretative discrepancies among different respondents, impairing the comparability and accuracy of responses \citep{lirui-042}. To enhance the precision of our data, we employed a nine-point Likert scale, ranging from -100 to 100 with intervals of 25, in our survey. This scale was designed to mitigate the issues of weight perception variability among respondents. Only the scale's extremes and midpoint were labeled, assigning an estimated magnitude to each response point. Additionally, a `N/A' option was included to distinguish between neutral and `no opinion' responses, addressing a common limitation in prior surveys where the absence of such options conflated indifference with a lack of opinion \citep{lirui-042}.

To avoid leading questions and suggestive information, neutral phrasing was used to capture respondents' genuine opinions without biasing them toward particular responses. This approach was particularly important for avoiding systematic errors such as leading questions, missing questions, and the presentation of partial information, issues identified by \cite{lirui-083} in their synthesis of 91 peer-reviewed survey studies on AV acceptance. Our survey included questions about respondents’ prior knowledge of AVs, their current travel behaviors, and car ownership to ensure a thorough understanding of the respondent's background and perspectives.

Lastly, to avoid confusion and suggestion bias, respondents were provided with clear definitions of fully autonomous vehicles (SAE Level 5) and car-sharing SAVs, as the case study focused solely on the public acceptance of fully autonomous car-sharing vehicles.

\subsubsection{Examining the Psychological Ownership Triggering Mechanism}

Previous studies, such as \cite{lirui-036} and \cite{lirui-037}, have examined psychological ownership by directly asking participants to rate statements like ``I would feel a very high degree of personal ownership for the autonomous vehicle." In our study, we employed a different approach by administering two distinct narratives (presented in Appendix \ref{Appendix_survey_story}) to participants after introducing the definition of SAVs. These narratives were designed to examine the influence of psychological ownership on participants' perceptions of SAVs. Participants were randomly assigned to one of two groups, each with a 50\% chance of receiving either Version 1 or Version 2 of the story. Version 1 served as the control condition, while Version 2 incorporated elements intended to trigger psychological ownership based on the three routes to psychological ownership (i.e., control, self-investment, and intimate knowledge \citep{POchapter1}). Therefore, participants who received Version 1 formed the Control Group, and those who received Version 2 formed the Psychological Ownership Group.

\subsection{Data collection}

The data collection for this research project was conducted from February 21 to June 26, 2023. In preparation, we secured the necessary ethical approval from the Monash University Human Research Ethics Committee (MUHREC) \footnote{Project ID: 36433}, ensuring that all procedures adhered to established ethical guidelines.

With a primary targeted group an Australian university community, to gather a diverse range of responses, we utilized a mixed approach in our data collection strategy. This included convenience sampling---reaching out to participants within immediate networks such as family, friends, and the university community (students and staff)---and random sampling methods. The latter involved disseminating the survey through various online channels. We leveraged social media platforms like Facebook Groups, specifically targeting student groups with active engagement such as `International Students of Melbourne and Victoria' and `Monash StalkerSpace'. LinkedIn also served as a valuable tool in attracting a wider audience. Additionally, we physically distributed survey flyers across multiple campuses in a large Australian university, which helped increase the survey’s visibility and participation.

We received a total of 507 questionnaire responses. Of these, 288 were fully completed, a completion rate of 56.8\%. Screening was applied to ensure data quality, with responses with more than 20\% `N/A' answers excluded as this is indicative of patterned responding. This resulted in 284 valid samples for analysis. This sample was composed of 137 responses from the Control Group (which were given Version 1 of the story; see Appendix \ref{Appendix_survey_story}) and 147 from the Psychological Ownership Group (which were given Version 2 of the story), with effective response rate of 56\%. An interesting aspect of our data collection was the engagement with the optional open-ended questions. Despite their optional nature, a significant 35.2\% of our valid respondents took the initiative to provide their inputs on these questions. These rich qualitative responses offer additional insights and are discussed in further detail in Section \ref{sec: Results}.

\subsection{Data Analysis}

\subsubsection{Descriptive Analysis and Examination of Psychological Ownership}

In our study, the initial step in data analysis involved presenting descriptive summaries of participants' prior knowledge about AVs, their travel habits, and demographic characteristics (Parts I and III of the survey). We utilized histograms for this purpose, as they offer a visual representation of data distribution, making it easier to identify patterns and outliers in the responses.

To evaluate the effectiveness of the storytelling method in triggering psychological ownership, we used the Mann-Whitney U test. This non-parametric test is well suited to our ordinal, non-normally distributed data (Figure \ref{fig:data_distribution}) and compared the responses of the Control and Psychological Ownership groups.  The null hypothesis (H0) posits that there is no significant difference between the two groups, while the alternative hypothesis (H1) suggests that the groups differ significantly. The statistical significance threshold was set at a 95\% confidence level for this non-parametric test. Accordingly, if the p-value is less than 0.05, there is sufficient evidence to reject the null hypothesis. This suggests a significant difference between the groups and affirms the impact of psychological ownership.

\subsubsection{Machine Learning Analysis with Random Forest}\label{subsubsec:ML with RF}

Several ML methods have been validated to be effective approaches in predicting user acceptance of CAVs in previous research \citep{lirui-094}, including Random Forests. The Random Forest, proposed by \citet{breiman_random_2001}, combines bootstrapping and random feature selection to predict outputs based on multiple decision trees \citep{lirui-094}. Recognized for its adaptability in both classification and regression tasks, Random Forest is particularly noted for its ease of implementation, capability to handle large datasets and numerous input variables without overfitting, and its proficiency in modeling complex, non-linear relationships \citep{breiman_random_2001, biau_2012_analysis}. This study utilizes Random Forest regression trees to examine numerical perception scales which range from -100 to 100 and were employed in the survey.

To ensure data reliability, responses containing `N/A' were excluded, resulting in a final sample size of 233. The Random Forest model was trained on 80\% of this dataset (186 responses), with the remaining 20\% (47 responses) utilized for validation. R programming language was used for analysis, with the `randomForest' package \citep{Package_randomForest} used to fit the Random Forest models and
`caret' package \citep{R_package_caret} for model training and validation procedures. All predictors identified from the literature---Perceived Risk (PR), Trust, Perceived Usefulness (PU), Perceived Ease of Use (PEOU), Attitude, and Psychological Ownership---were included to predict the intention to use SAVs, and the interactions among the predictors were observed as well. To enhance the model's reliability and generalizability, 10-fold cross-validation was performed during model training. This approach involves partitioning the training data into ten subsets, training the model on nine subsets, and validating it on the remaining subset, repeating this process ten times \citep{R_package_caret}.

The performance of the fitted models was evaluated using the Root Mean Square Error (RMSE) and the Normalized Root Mean Square Error (NRMSE), as defined in equations \eqref{eq.RMSE} and \eqref{eq.NRMSE}, respectively. In these equations, $y_i$ represents the actual value of the $i$-th observation (i.e., the target variable); $\hat{y}_i$ denotes the predicted value for the $i$-th observation obtained from our model; $n$ is the total number of observations, and $y_{\text{max}}$ and $y_{\text{min}}$ are the maximum and minimum values of the actual target variable, respectively. The RMSE measures the average magnitude of the prediction errors, providing insight into the model's overall performance. The NRMSE offers a normalized measure of error relative to the range of the observed data, allowing for comparisons between models or datasets with different scales.

\begin{equation}
    \text{RMSE} = \sqrt{\frac{1}{N} \sum_{i=1}^{N} (y_i - \hat{y}_i)^2}\label{eq.RMSE}
\end{equation}

\begin{equation}
    \text{NRMSE} = \frac{\sqrt{\frac{1}{N} \sum_{i=1}^{N} (y_i - \hat{y}_i)^2}}{\max(y) - \min(y)}\label{eq.NRMSE}
\end{equation}

Furthermore, a classification approach was applied to compute the prediction accuracy of the validation sample across various threshold values ranging from 0 to 200. To determine the accuracy, the constructs measured using nine scales ranging from -100 to 100 were considered correctly predicted if the predicted value fell within the range of [actual value - threshold, actual value + threshold]. Notably, when the threshold value is 200, the accuracy is always 1 regardless of the prediction value.
To ascertain the significance of the trees, these accuracy measures were compared against those obtained from a set of random sample data. 47 uniformly random data points were generated using \textit{`runif(47, min = -100, max = 100)'} command in R, and the accuracy of these randomly guessed data was calculated. To maintain objectivity, this process was repeated with 100 random samples, and their mean accuracy was used for comparison against the Random Forest predictions.


Based on the best Random Forest regression tree obtained from 10-fold cross-validation, we extracted the predictor importance when predicting the target variables and overall intention to use SAVs, using the \textit{`vip()'} function \citep{greenwell_package_2023}. To enhance interpretability, we further transformed the importance values into relative importance by computing the importance weight of each question, ensuring that the combined weight equaled 100\%. This normalization allows for an intuitive comparison of the predictors' contributions to the model.

\subsubsection{Chord Diagram Visualization}\label{subsubsec: chord diagram visualisation}

A key contribution of this work is the application of chord diagrams as a visualization tool to represent the complex interrelationships and relative importance of factors influencing technology acceptance. Chord diagrams are circular visualizations that display interrelated data, where different variables are arranged along the circumference of a circle, and relationships between variables are depicted as connecting chords \citep{Finnegan_using_2019}. This design is highly effective in illustrating the dependencies and interactions among multiple variables in a single comprehensive and intuitive layout.

In this case study, chord diagrams were constructed to visually represent the complex interrelationships and relative importance of factors influencing SAV acceptance as identified by the Random Forest models. Utilizing the \textit{`migest'} package in R \citep{website-chord}, these diagrams were constructed to effectively demonstrate the strength and directionality of connections between various predictors and SAV acceptance intentions.

In the chord diagrams, key components include nodes (start nodes and end nodes, representing factors or survey items) and chords (depicting the relationships between nodes). Nodes are positioned along the circumference of the circle and are color-coded to differentiate between variables. The predictors (factors or individual items) are represented as start nodes, while the target variables (factors or overall items, such as the overall behavioral intention to use SAVs) are represented as end nodes. Chords connect the start nodes to the end nodes, illustrating the relationships between predictors and the target variable. The thickness of each chord reflects the relative importance of the predictor in predicting the target variable, as determined by the Random Forest model. A thicker chord indicates a greater contribution of that predictor to the model's prediction. Some variables, such as psychological ownership, are represented by a single node due to having only one survey item, while others encompass multiple nodes corresponding to different survey items within that factor. Items from the same variable family (e.g., `PR1: Accident risk' from PR variable) are grouped together along with the overall item (e.g., `PR8: Overall perceived risk'), which serves as the target variable, and share the same color for visual consistency.

\begin{figure}[!htpb]
    \centering
    
    \begin{subfigure}{0.48\linewidth}
        \centering
        \includegraphics[width=\linewidth]{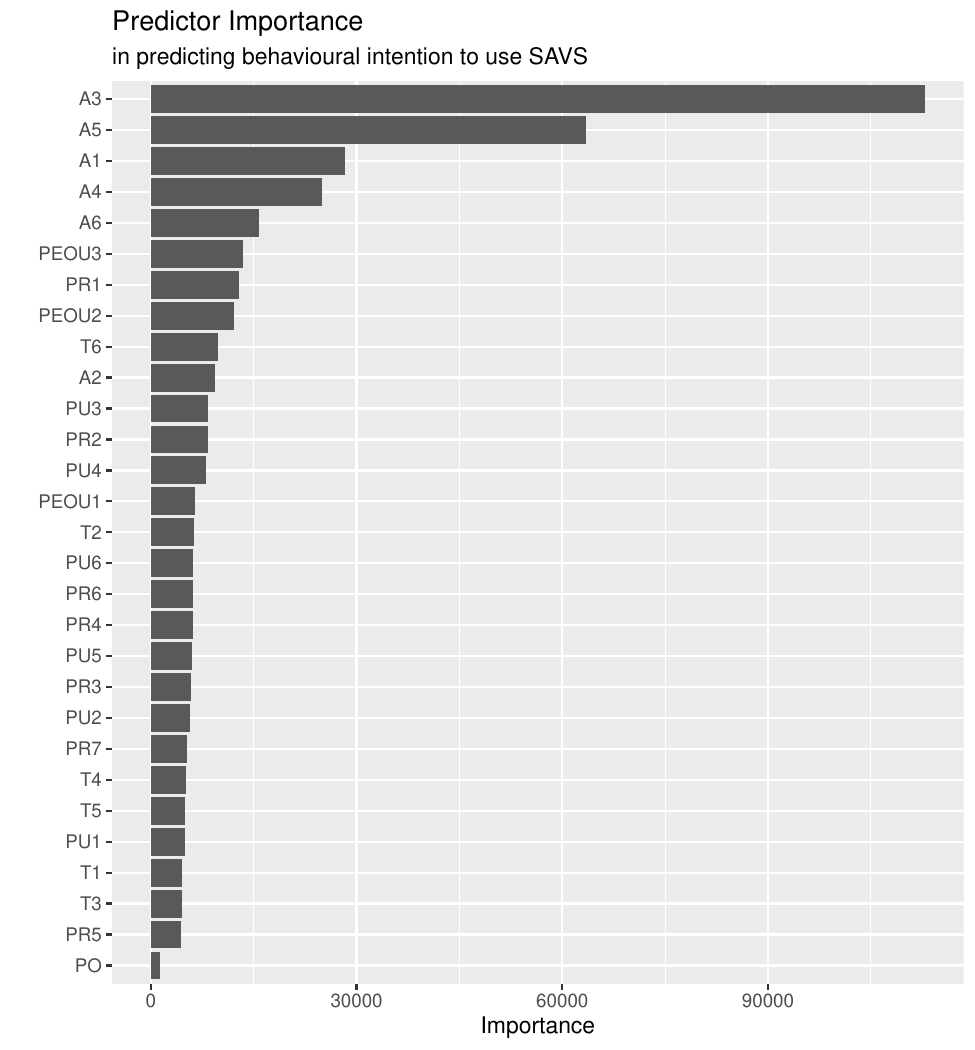}
        \caption{}
        \label{fig:imp_vip}
    \end{subfigure}
    \hfill
    \begin{subfigure}{0.48\linewidth}
        \centering
        \includegraphics[width=\linewidth]{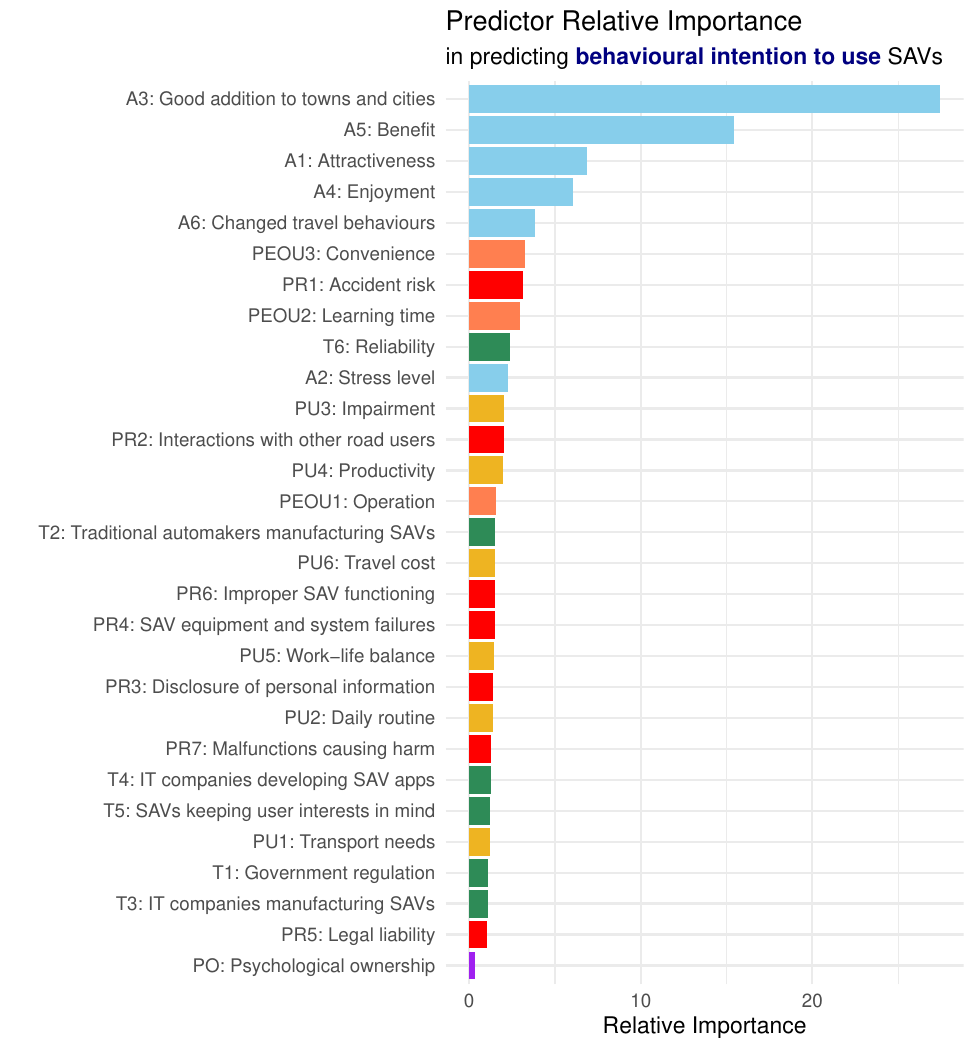}
        \caption{}
        \label{fig:imp_relative}
    \end{subfigure}

    \begin{subfigure}{0.48\linewidth}
        \centering
        \includegraphics[width=\linewidth]{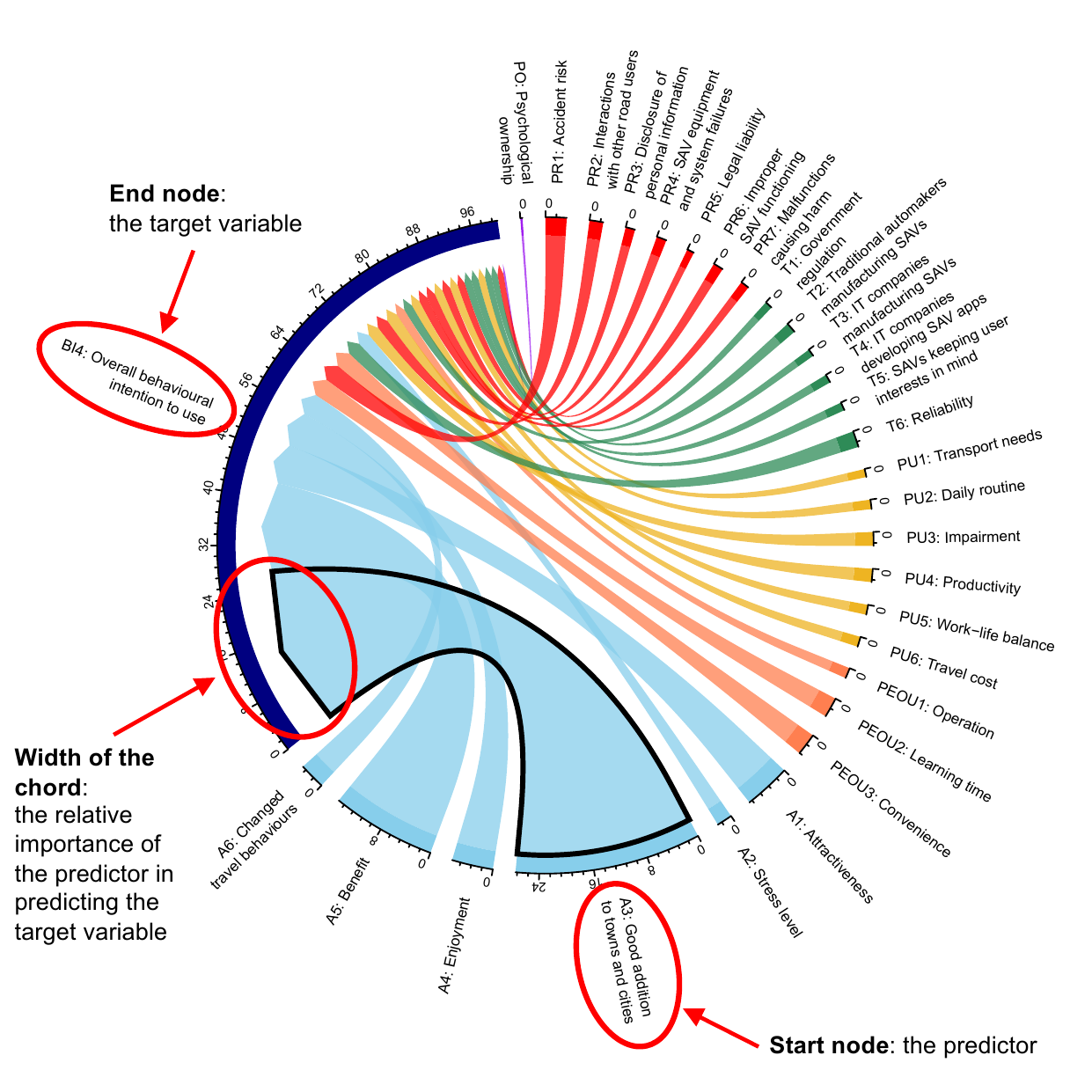}
        \caption{}
        \label{fig:circular_BI}
    \end{subfigure}
    \hfill
    \begin{subfigure}{0.48\linewidth}
        \centering
        \includegraphics[width=\linewidth]{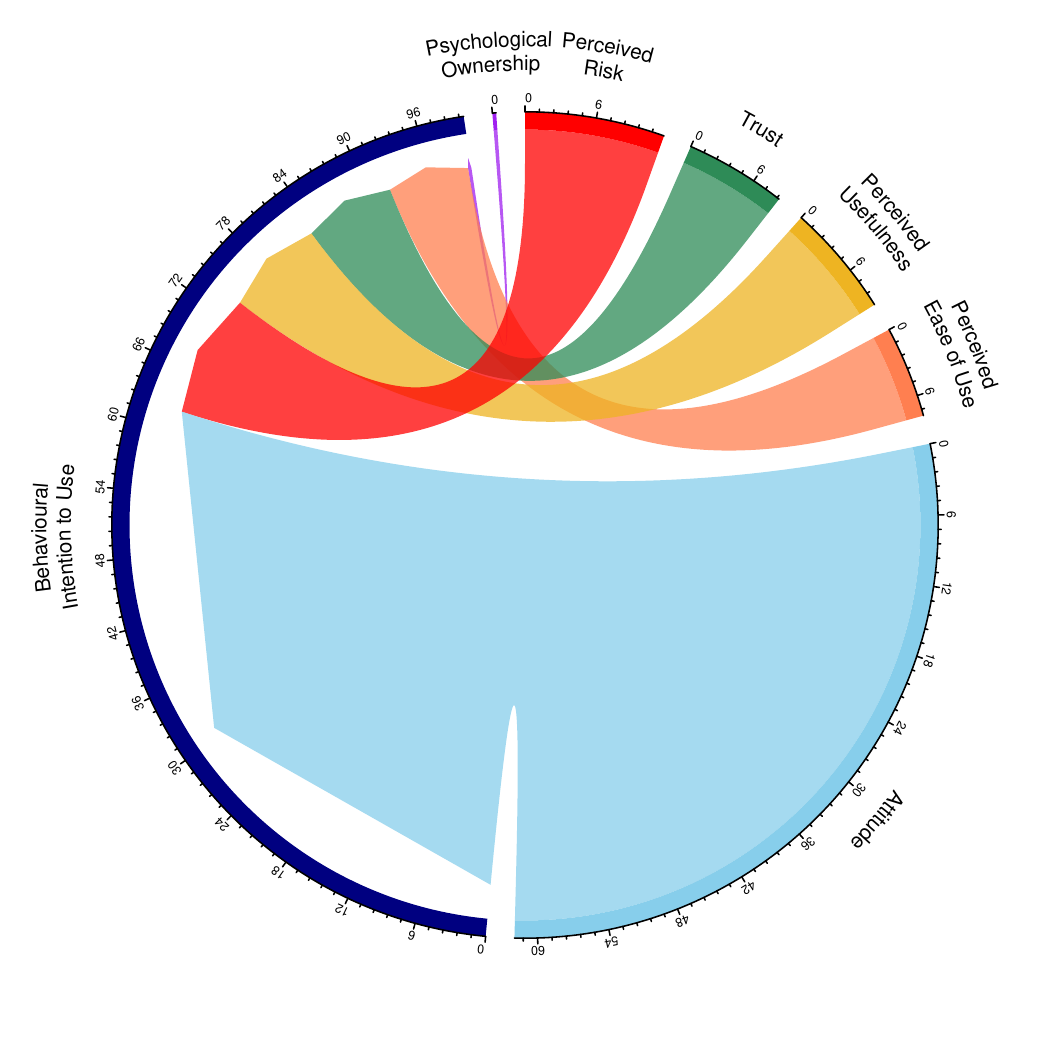}
        \caption{}
        \label{fig:circular_BI_overall}
    \end{subfigure}

    \caption{Example of Computing Relative Importance of the Questions and Factors When Predicting BI4. (a) The predictor importance was first computed using the `vip' function from the best-trained Random Forest regression trees. (b) Then, the relative importance for each predictor item was calculated by computing the weight of each item, ensuring the sum of all predictor items' relative importance is 100\%. (c) Next, the relative importance was transformed into the chord diagram, which shows the predictors (start nodes), the target variable (end nodes), and the relative importance of each item (chords). (d) The sum of all items' relative importance was computed to demonstrate the overall contribution of the predictor.}
    \label{fig:relative_imp}
\end{figure}

The relative importance of each factor item in predicting the target variables (as demonstrated in Section \ref{subsubsec:ML with RF}) was then transformed into the chord diagram, showcasing the predictors, the target variables, and each item's relative importance. An example of this process, detailed in Figure \ref{fig:relative_imp}, demonstrates how the relative importance values are converted into a chord diagram format when predicting the overall behavioral intention to use SAVs (BI4).

The process involves the following steps:
\begin{enumerate}
    \item \textbf{Computing Predictor Importance}: Using the `vip' package, we extracted the importance of each predictor item from the best-trained Random Forest regression model (Figure \ref{fig:imp_vip}).

    \item \textbf{Calculating Relative Importance}: The importance values were normalized to obtain the relative importance of each predictor item, ensuring that the sum of all predictor items' importance equals 100\% for a target variable (e.g., `BI4: Overall behavioral intention to use') (Figure \ref{fig:imp_relative}).

    \item \textbf{Creating the Chord Diagram}: The relative importance values were used to construct the chord diagram, where the predictors (start nodes) are connected to the target variable (end node) with chords representing their relative importance (Figure \ref{fig:circular_BI}). This step produces Figure \ref{fig:cir_plot}, which shows the relative importance of predicting the target variable (i.e., Overall items) at the item level.

    \item \textbf{Aggregating Factor Contributions}: The sum of the relative importance values of the items within each factor provides the overall contribution of that factor to the prediction of the target variable (Figure \ref{fig:circular_BI_overall}). This step produces Figure \ref{fig:cir_overall}, which shows the relative importance of predicting the target variable (i.e., factors) at the factor level.
\end{enumerate}

This method was applied to all target variables, resulting in comprehensive chord diagrams that visualize the contributions of individual predictor items and factors.

The chord diagrams reveal the contributions of each variable at both factor and item levels in predicting SAV acceptance. By visualizing the relative importance of individual survey items and their relationships with the target variable, the diagrams provide a holistic view of how different factors and their specific components influence behavioral intention. In contrast to the use of individual `vip' importance plots for each predictor or target variable, the chord diagram offers a compact representation that consolidates all this information into a single, comprehensive visualization. This allows for the summarization of variable importance and the interplay between predictors and target variables in one unified figure, eliminating the need to examine numerous separate plots.

Moreover, the chord diagrams can distinguish between adopters and non-adopters by generating separate diagrams for each group, highlighting differences in the factors influencing their acceptance of SAVs. This visualization enhances the understanding of each variable's impact and elucidates the complex interplay among various psychological factors affecting SAV acceptance.

By complementing the detailed numerical importance values obtained from the Random Forest analysis, the chord diagrams offer an intuitive and integrated representation of the data, facilitating better communication of findings to stakeholders and informing targeted strategies to enhance public acceptance.

\section{Results and Discussion}\label{sec: Results}

\subsection{Descriptive Summary}

\subsubsection{Demographics}

\begin{figure}[h!]
\captionsetup{font=small}
  \centering
  \begin{subfigure}[b]{\linewidth}
    \centering
    \includegraphics[width=\linewidth]{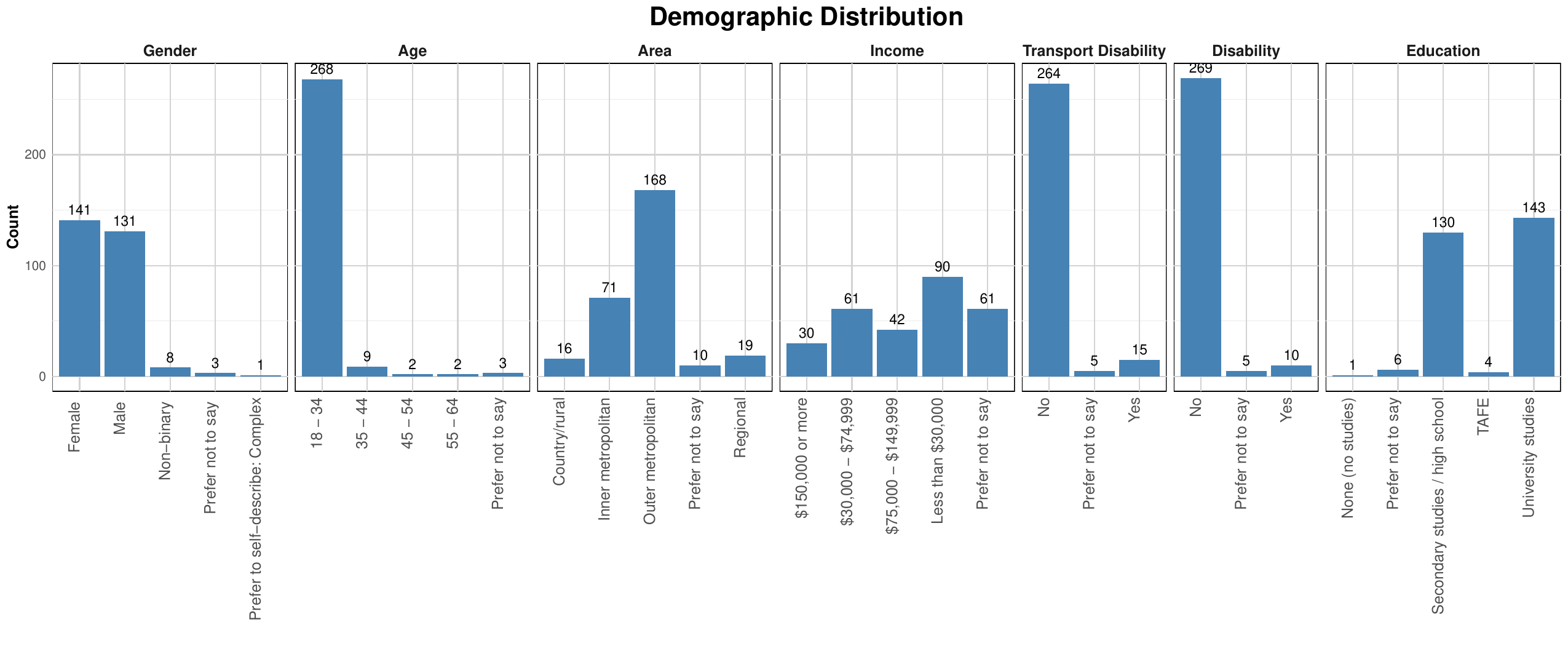} 
    \caption{Distribution of demography (N = 284).}
    \label{fig:demographic}
  \end{subfigure}
  
  \vspace{10pt}

  \begin{subfigure}[b]{\linewidth}
    \centering
    \includegraphics[width=\linewidth]{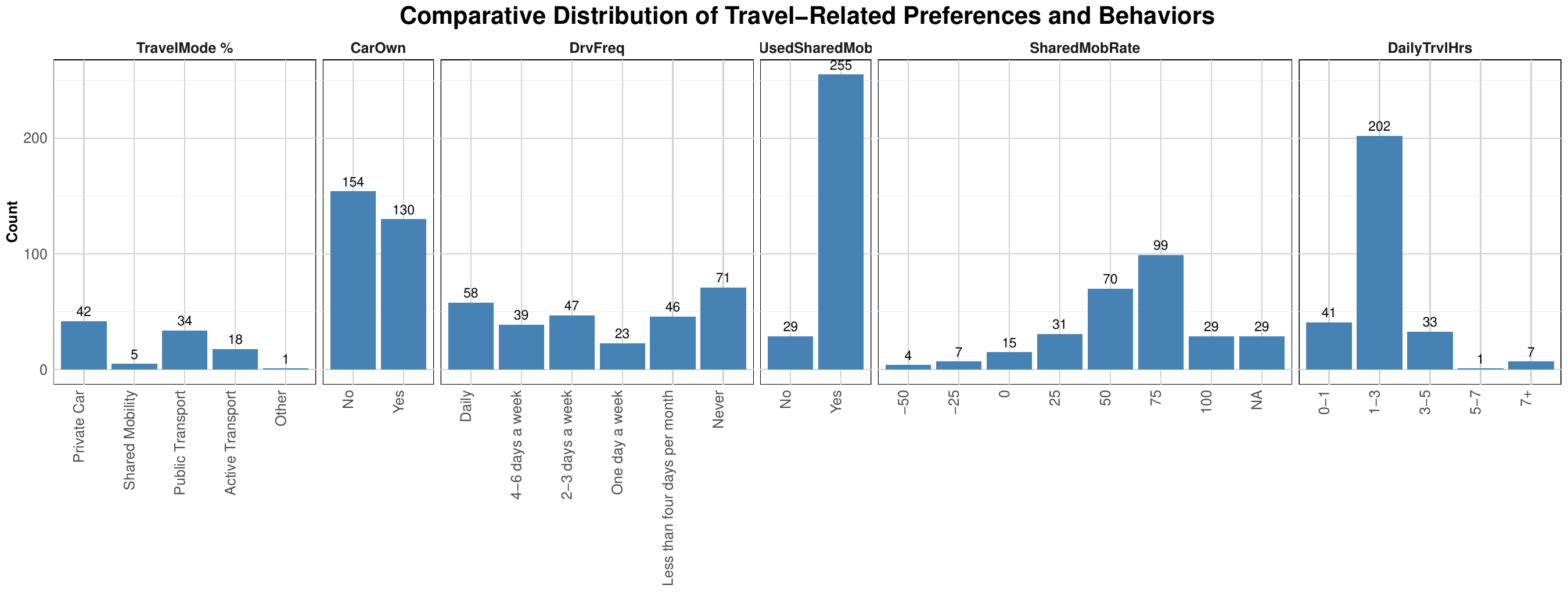} 
    \caption{Distribution of Travel Modes (N = 284).}
    \label{fig:travel_modes}
  \end{subfigure}
  \caption{Demographic and Travel Modes Distribution of the Collected Data.}
  \label{fig:descriptive_summary}
\end{figure}

The demographic profile of the respondents (Figure \ref{fig:demographic}) reveals a balanced gender distribution, predominantly younger participants, and a significant number residing in metropolitan areas. Most respondents reported moderate annual household incomes. Notably, a subset indicated responsibilities for transporting individuals with special care needs or faced personal driving limitations.

Our focus on an Australian university community naturally resulted in a participant pool that is younger and more educated than the general population. While factors such as age, education, and income are known to significantly influence technology acceptance \citep{lirui-010, lirui-017, lirui-078, lirui-008}, studying this demographic provides valuable insights. This group may represent potential early adopters of technologies like SAVs due to their openness to innovation and higher levels of education. Although this demographic skew may limit the generalizability of our findings to the broader population, it highlights important trends within a segment likely to influence initial market acceptance. Importantly, even within this relatively homogeneous and potentially highly accepting group, participants can still be classified as SAV adopters and non-adopters based on their survey responses, as detailed in Section \ref{subsubsec:adopters_non-adopters}. This intra-group variability emphasizes the need to understand acceptance dynamics even among potential early adopters.

\subsubsection{Awareness and Exposure to Autonomous Vehicle Technology}

The majority of the participants (97.54\%) reported having heard of AVs, indicating a high level of general awareness about these vehicles in the surveyed population. The main source of this knowledge was the news media (83.10\%), emphasizing the significance of mainstream media in introducing AVs to the public. The other sources of information were movies (56.69\%), science fiction (47.89\%), advertisements (33.80\%) and other 24.30\%. While most respondents reported having only a little (55.60\%) or a moderate amount (31.77\%) of knowledge about AVs, a smaller fraction believed they knew a lot or a great deal were relatively fewer, constituting 6.86\% and 1.81\% of the sample, respectively. Only a small segment (3.97\%) admitted to having no knowledge of AVs. Regarding practical exposure to automated driving technologies, 44.72\% had not used any driver assistance systems, while adaptive cruise control (41.90\%) and lane-keeping assistance (32.75\%) were the most commonly experienced features. A smaller segment experienced automated parking and Autopilot systems, constituting 17.96\% and 14.08\% respectively. This data reflects a general awareness of AVs among participants, coupled with varied levels of knowledge and practical exposure to related technologies.

\subsubsection{Current Travel Modes and Mobility Patterns}

Our survey also provides an insightful overview of the current travel modes and mobility preferences among the participants, as detailed in Figure \ref{fig:travel_modes}---a crucial step in understanding the potential acceptance of SAVs. To capture the complexity of modern travel behaviors, we asked participants to estimate the percentage of their total travel time spent using each transportation mode. This method accounts for multi-modal travel patterns and the varying reliance on different modes throughout daily activities. The full survey questionnaire can be found in the Appendix \ref{Appendix_survey_item}.

The data reveal that the majority of participants rely on private cars (41.78\%) and public transport (33.54\%), reflecting the principal transportation modes within the university community. These percentages represent the average proportion of travel time allocated to each mode across all respondents. For example, a participant might report spending 70\% of their travel time using a private car, 20\% using public transport, and 10\% using active transport modes like walking or cycling. 

\begin{figure}
    \centering
    \includegraphics[width=0.8\linewidth]{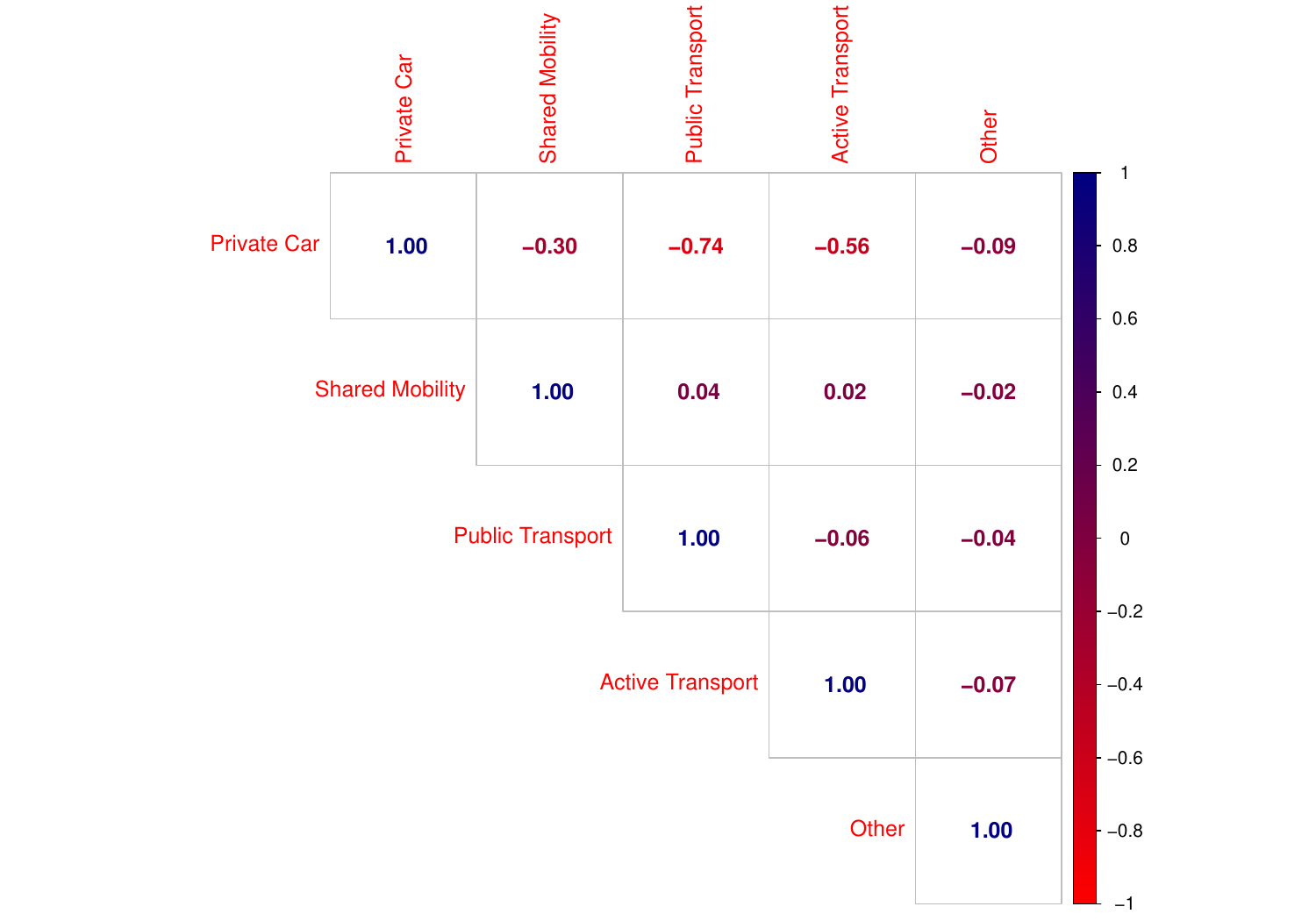}
    \caption{Correlation Matrix of Travel Modes Among Participants. he matrix displays correlation coefficients between the proportions of total travel time allocated to different travel modes. Positive values (0 to 1) indicate that participants who use one mode are more likely to use another, while negative values (0 to -1) suggest they are less likely to do so. Values near zero indicate little or no linear relationship between modes.}
    \label{fig:travelModes_corre}
\end{figure}

Further analysis of the travel modes reveals relationships between the usage of different modes, as illustrated in Figure \ref{fig:travelModes_corre}. The correlation matrix shows strong negative correlations between private car usage and other modes, particularly public transport (-0.74) and active transport (-0.56). This suggests that participants who spend a higher proportion of their travel time using private cars tend to use public transport and active transport modes less frequently. In contrast, the correlations between shared mobility, public transport, and active transport are close to zero, indicating minimal linear relationships between these modes. This suggests that participants who use shared mobility services do not significantly differ in their use of public transport or active transport compared to others. Understanding these relationships is essential for designing effective strategies to encourage the adoption of SAVs, as it highlights potential challenges in shifting users away from private car usage toward alternative modes of transportation.

A significant majority (89.79\%) of respondents have experienced shared mobility services, with 5.39\% regularly using these services. Among those who have used shared mobility options, 38.82\% rated their experience as 75 out of 100, indicating a generally positive perception of these services. The near-even division in car ownership among participants highlights diverse mobility needs, particularly among students who may have varying access to private vehicles. Additionally, the varied daily transportation times—most commonly around one hour---underscore the necessity for efficient and flexible mobility solutions. These insights are crucial in shaping strategies for SAV implementation, aligning with public needs and preferences to foster their acceptance.

\begin{figure}[!ht]
    \centering
    
    \begin{subfigure}{0.32\linewidth}
        \centering
        \includegraphics[width=\linewidth]{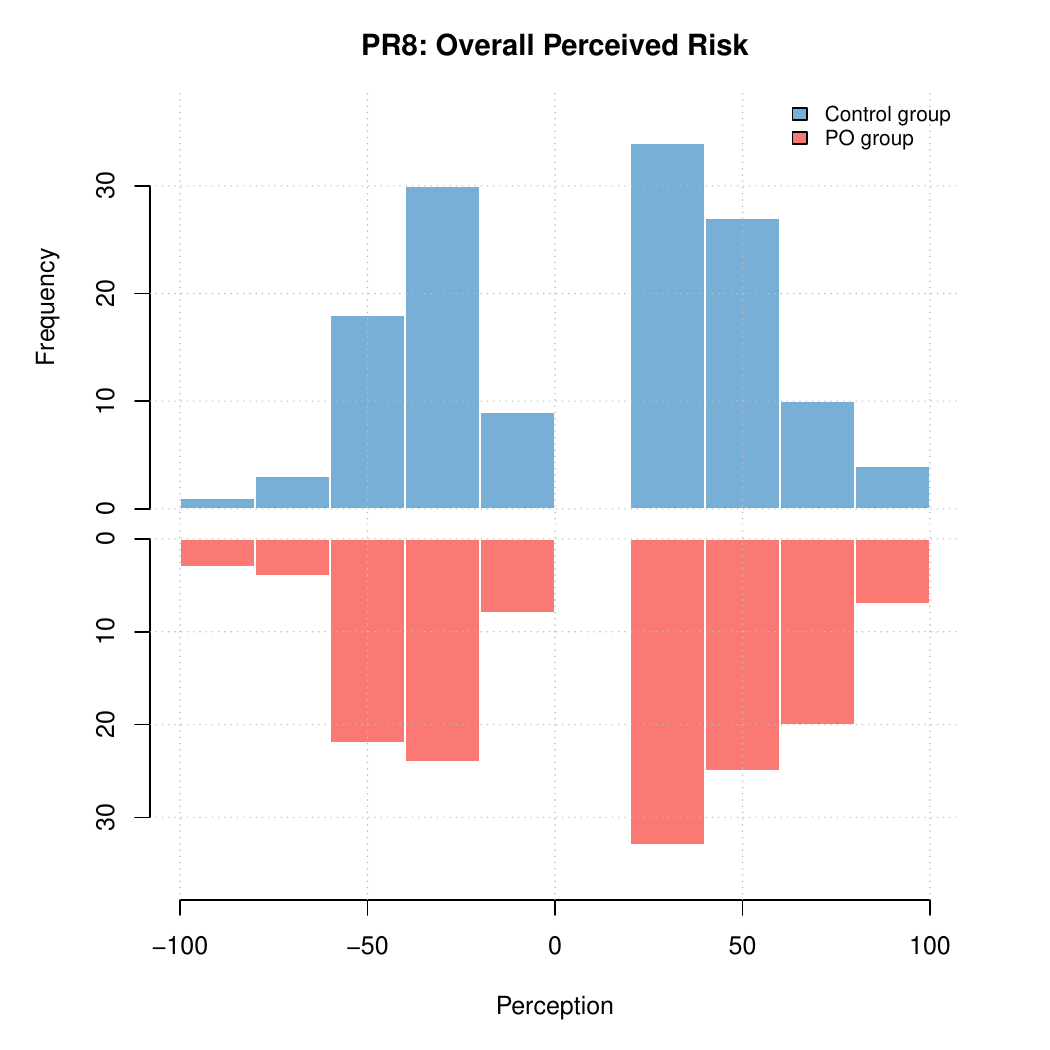}
        \label{fig:hist_PR}
    \end{subfigure}
    \hfill
    \begin{subfigure}{0.32\linewidth}
        \centering
        \includegraphics[width=\linewidth]{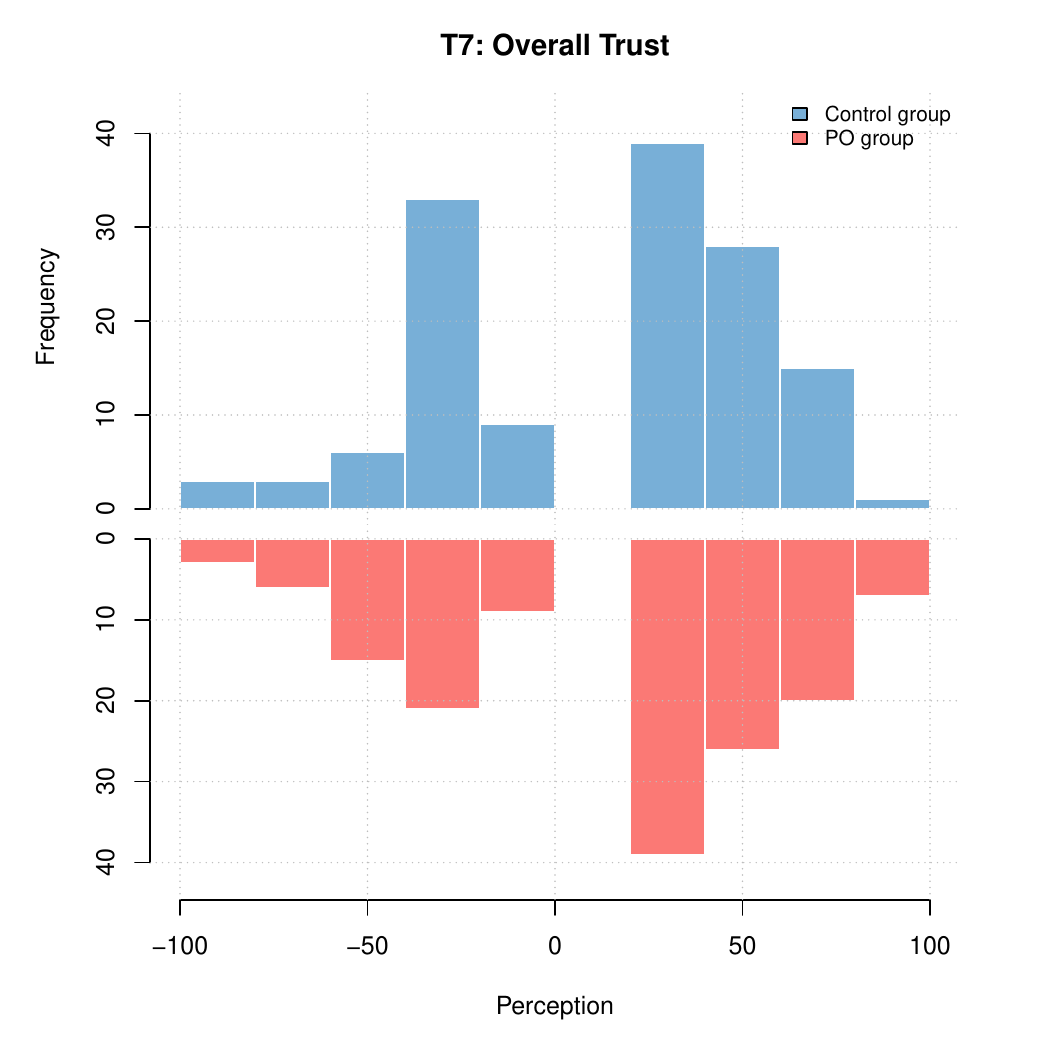}
        \label{fig:hist_T}
    \end{subfigure}
    \hfill
    \begin{subfigure}{0.32\linewidth}
        \centering
        \includegraphics[width=\linewidth]{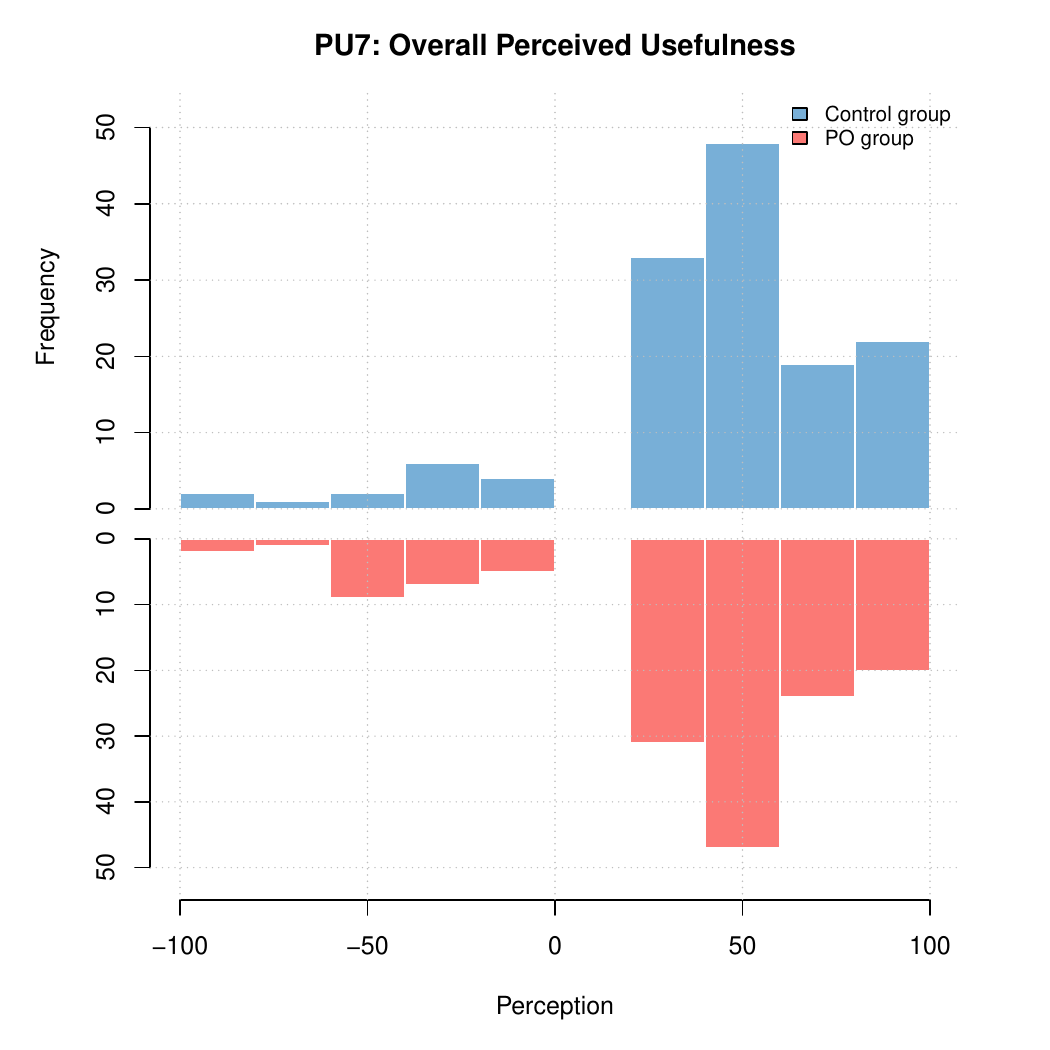}
        \label{fig:hist_PU}
    \end{subfigure}

    \begin{subfigure}{0.32\linewidth}
        \centering
        \includegraphics[width=\linewidth]{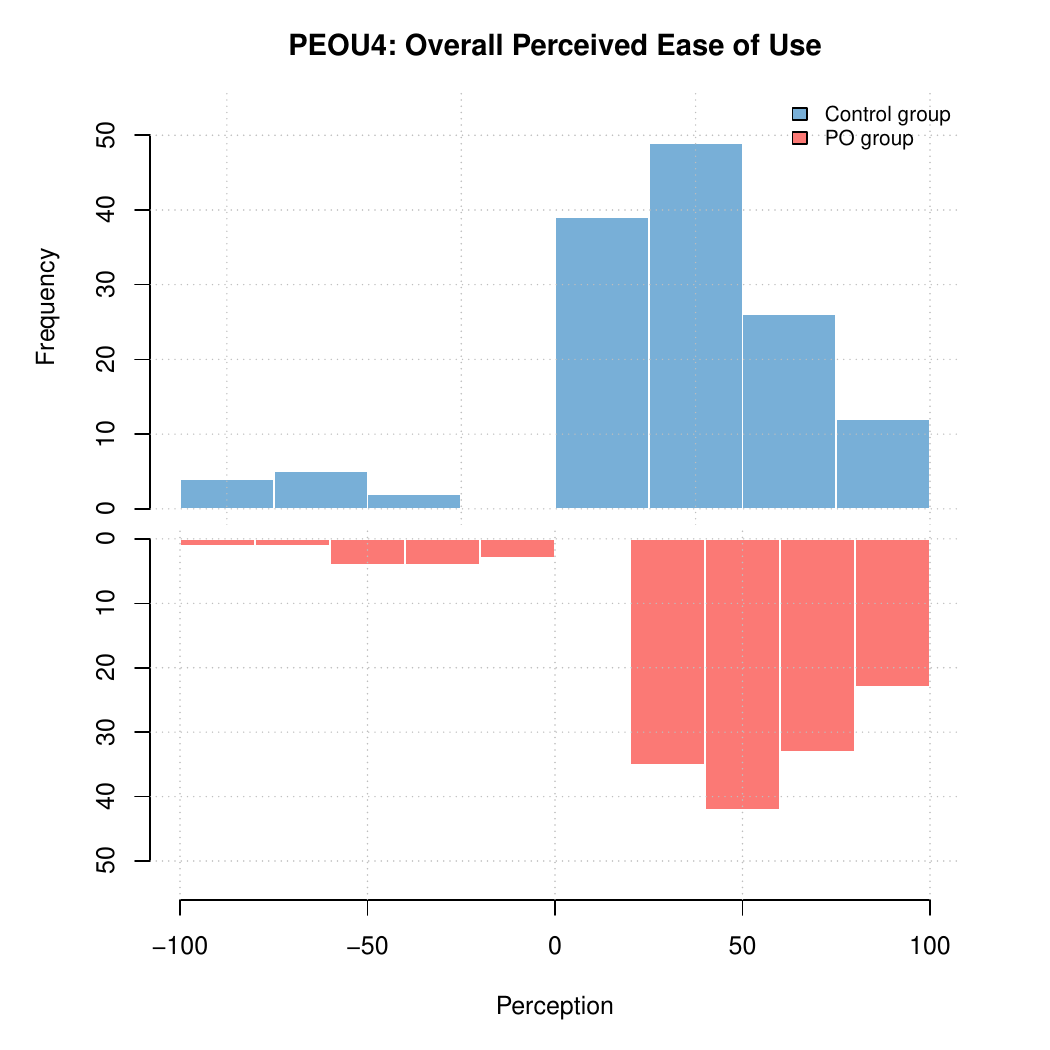}
        \label{fig:hist_PEOU}
    \end{subfigure}
    \hfill
    \begin{subfigure}{0.32\linewidth}
        \centering
        \includegraphics[width=\linewidth]{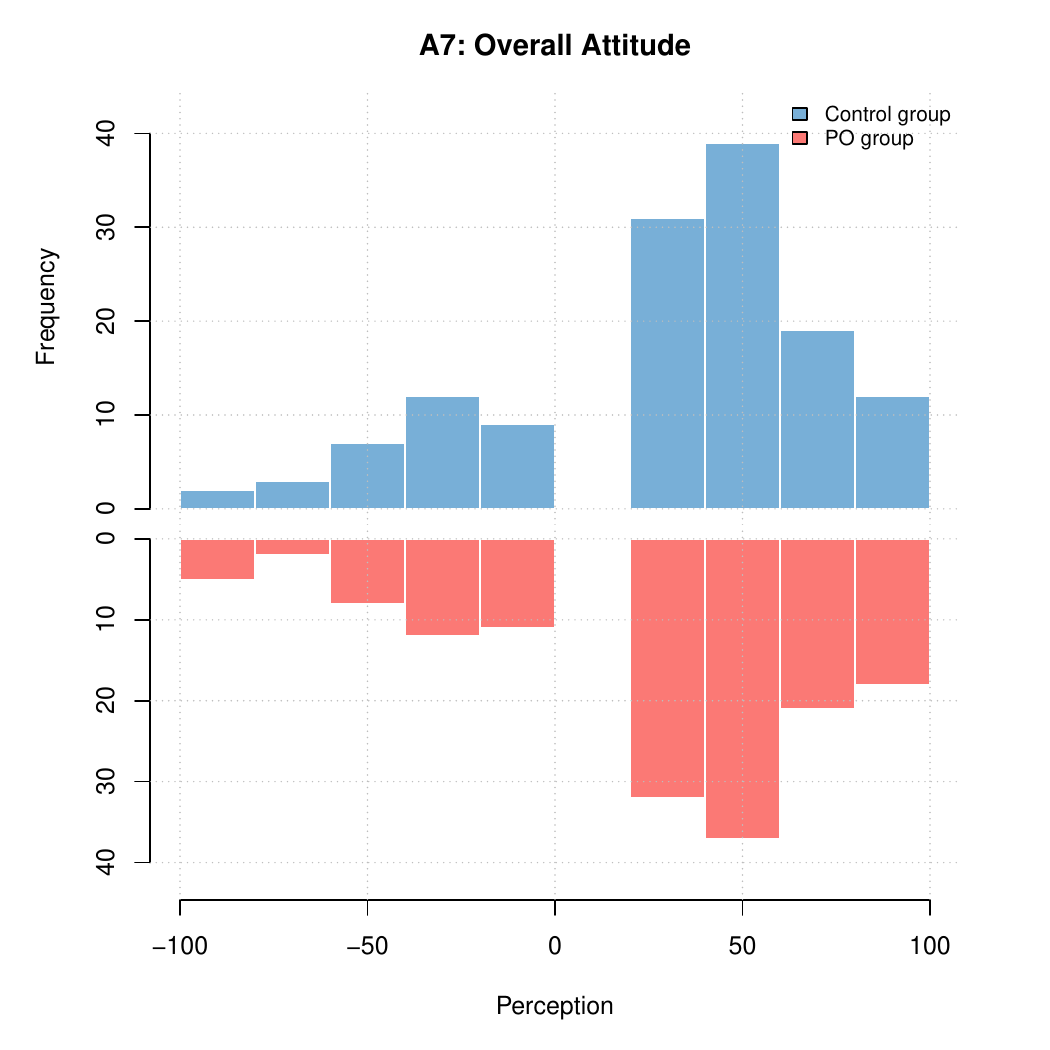}
        \label{fig:hist_A}
    \end{subfigure}
    \hfill
    \begin{subfigure}{0.32\linewidth}
        \centering
        \includegraphics[width=\linewidth]{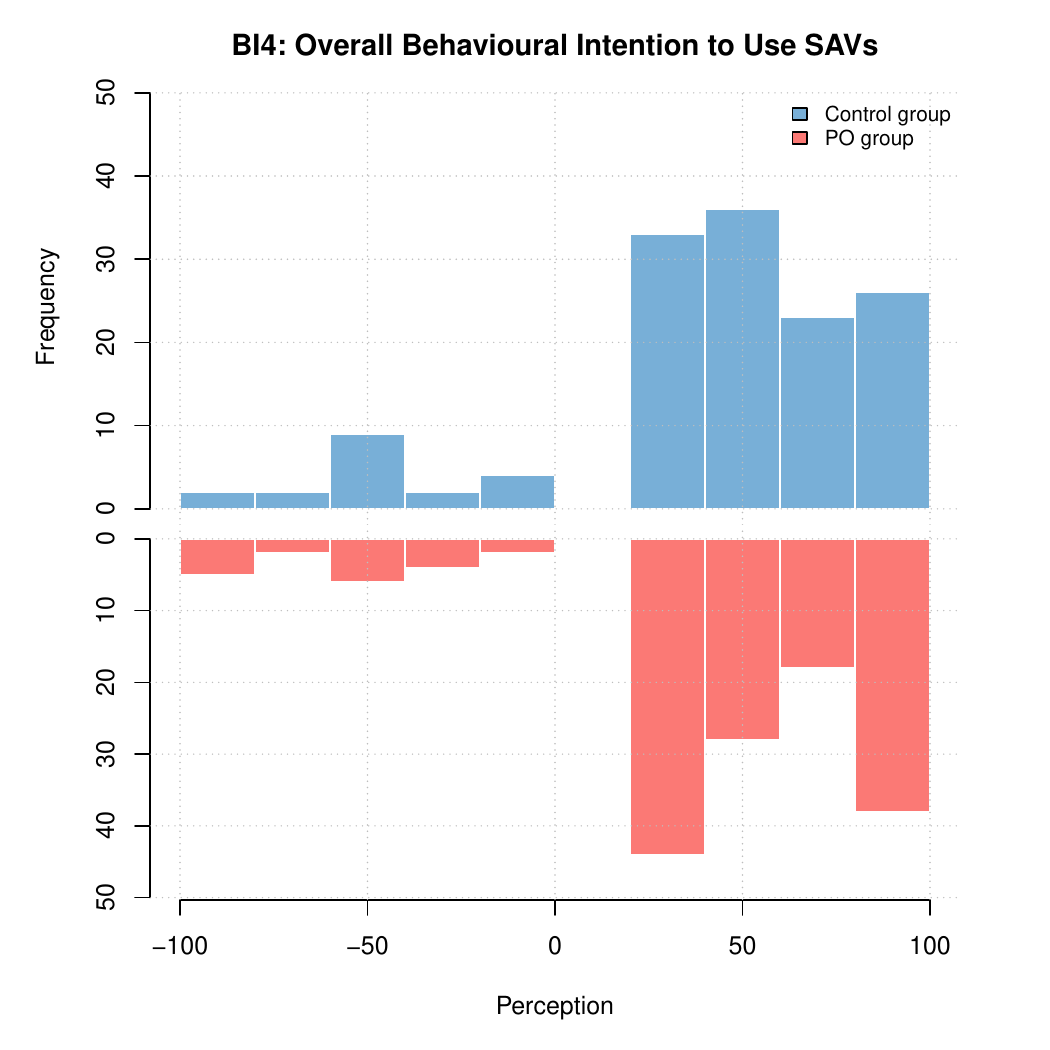}
        \label{fig:hist_BI}
    \end{subfigure}

    \caption{Comparative Response Distribution to the Overall Items for Psychological Factors - This figure presents a side-by-side comparison of response distributions for key psychological factors from the Control Group (top, in blue) and the Psychological Ownership Group (bottom, in pink).}
    \label{fig:data_distribution}
\end{figure}

\subsubsection{Distribution of Psychological Factors Data}

The data analysis on psychological factors indicated a notable deviation from the normal distribution for most of the Part II survey questions for both the Control Group and the Psychological Ownership Group. This trend is evident in Figure \ref{fig:data_distribution}, which provides an overview of the response distributions for the overall items of Part II of the survey. 

The data indicates that respondents tend to categorize themselves as either adopters or non-adopters of SAVs, resulting in a bimodal distribution. This raises a significant methodological issue as traditional analysis methods such as SEM often rely on assumptions of normality and may not perform optimally with non-normal or bimodal data distributions. Recognizing this pattern is crucial for interpreting the results and underscores the need for adaptable analytical approaches to accurately capture the nuances of public perception towards SAVs, particularly in the presence of complex, non-linear relationships.

\subsection{Effects of Psychological Ownership}

The Mann-Whitney U test analysis revealed significant differences only in specific aspects of perceived risk (PR): PR2, which relates to the perceived risk of SAVs interacting with other road users, and PR3, concerning the perceived risk of SAVs disclosing users' personal information, as detailed in Table \ref{tab:U-test}. These limited significant differences suggest that the emotional narratives employed may not have effectively triggered psychological ownership as intended, or that psychological ownership may not have a substantial influence on intentions to adopt SAVs among the targeted Australian university community. This underscores the need for further research to explore alternative methods for triggering psychological ownership and to examine its potential influence on the acceptance of SAVs.

\begin{table}[ht]
\captionsetup{font=small}
\centering
\caption{Mann-Whitney U test results. The threshold for statistical significance was set at a 95\% confidence level.}
\label{tab:U-test}
\resizebox{\textwidth}{!}{%
\begin{adjustbox}{max width=\textwidth}
\begin{tabular}{llrlrr}
  \hline
  Questionnaire items & p-value & Result \\ 
  \hline
PR2: Do you think SAVs interacting with other road users would be safe or dangerous? & 0.0184 & Passed \\
PR3: How concerned are you about the SAV disclosing my personal information? & 0.0125 & Passed \\
Others & - & Failed \\
  \hline
\end{tabular}
\end{adjustbox}
}
\end{table}

\subsection{Non-Linear Relationships Between Psychological Factors}\label{subsec: non-linear}

\begin{figure}[H]
\captionsetup{font=small}
    \centering
    \includegraphics[width=\linewidth]{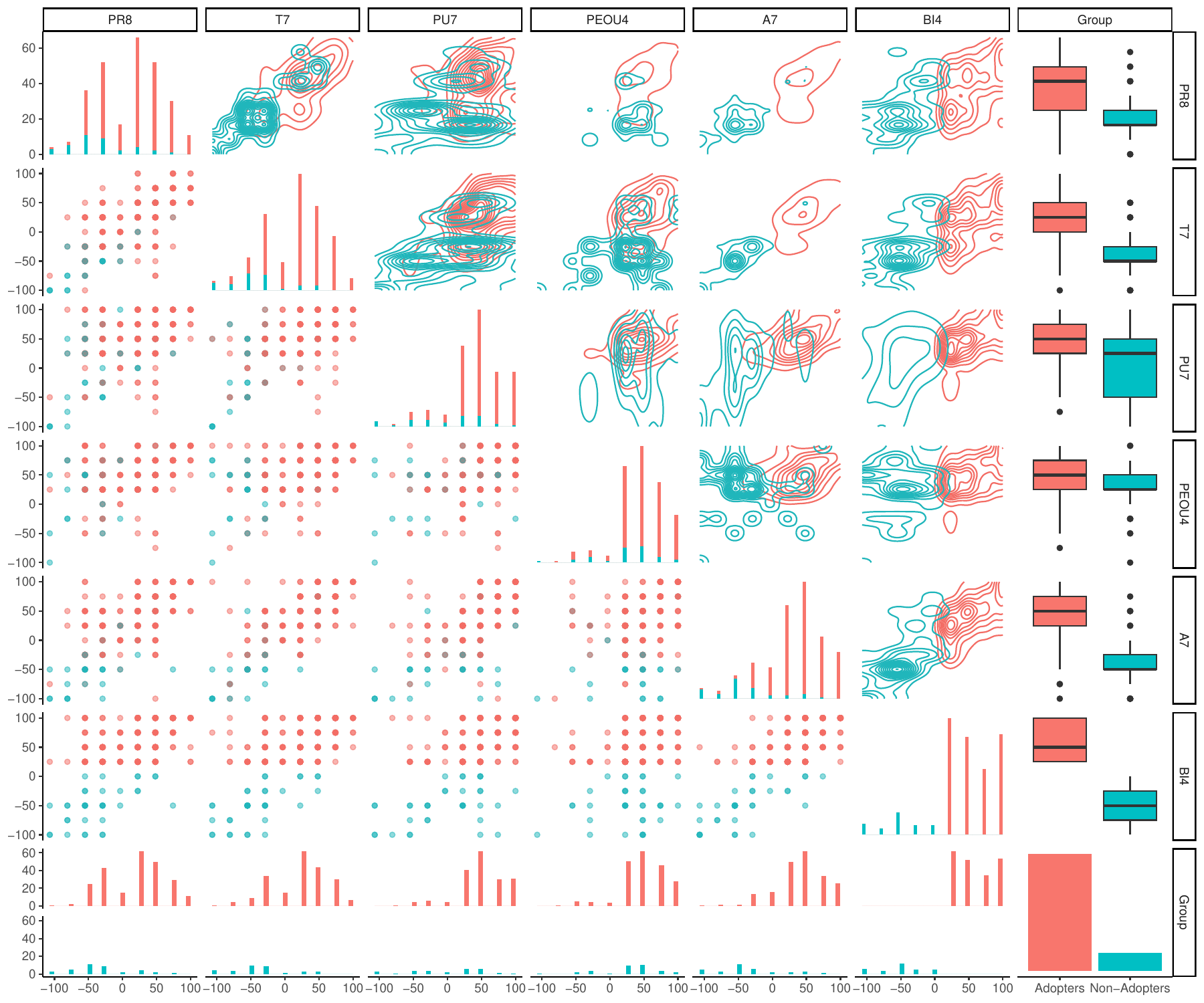} 
    \caption{Pairwise Relationship Plot of Psychological Factors by Adoption Group. The variables along the axes are the overall items of each factor. Adopters are presented in orange, and non-adopters are presented in blue.}
    \label{fig:ggpairs_plot_psy_factors}
\end{figure}

To further explore the relationships among key psychological factors, we plot pairwise relationships of the overall scores for perceived risk (PR8), trust (T7), perceived ease of use (PEOU4), perceived usefulness (PU7), attitude (A7), and behavioral intention (BI4) to use SAVs for both adopters (in orange) and non-adopters (in blue) (see Figure \ref{fig:ggpairs_plot_psy_factors}). The plot has four main components: (1) pairwise contour plots (upper right) which show the density and distribution of data points between each pair of variables; (2) pairwise scatter plots (lower left): which display individual data points, highlighting patterns and potential relationships between variables; (3) distribution plots (diagonal) which illustrate the distribution of each variable individually; and (4) boxplots of each factor (far right side) which summarize the central tendency and variability of each factor, comparing adopters and non-adopters.

The plots reveal that many relationships between these variables are non-linear. For instance, the relationship between perceived risk (PR8) and behavioral intention (BI4) shows variations that cannot be adequately captured with linear models. Similarly, a complex, curved pattern is observed between overall trust (T7) and perceived usefulness (PU7). This indicates that changes in trust do not directly correlate with proportional changes in perceived usefulness. In addition, the scatterplots between perceived ease of use (PEOU4) and other variables such as attitude (A7) exhibit clusters and non-uniform distributions. This suggests that participants' attitudes toward SAVs do not increase uniformly with perceived ease of use, further indicating non-linear dynamics.

These observations indicate that the relationships among these psychological factors are complex and cannot be adequately modeled using linear approaches. The presence of non-linear patterns, clusters, and varying rates of change supports our use of the Random Forest model, which is capable of capturing non-linear interactions and complex relationships among variables. By employing a model that accommodates non-linearity, we can better understand and predict the factors influencing users' acceptance of SAVs.

\subsection{Model performance}\label{subsec: RF_performance}

Model accuracy was gauged using RMSE and NRMSE (Table \ref{tab:RMSE}). The NRMSE values ranged from 0.1 to 0.2,  indicating good to medium performance levels \citep{lirui-095}. This suggests that the predictors used in the Random Forest regression trees are informative and effective in predicting the target variables. 

In the analysis, the `External' class refers to using all items except overall items to predict the target variable, while the `Internal' class refers to using all items of a factor to predict its overall item. The `mtry' parameter controls the amount of randomness added to the decision tree creation process, specifically how many of the input features a decision tree has available to consider at any given point in time \citep{website-mtry}. The `mtry' parameter was computed from the best model selected by minimizing RMSE after the 10-fold cross-validation. 

\begin{figure}[h!]
\captionsetup{font=small}
    \centering
    \includegraphics[width=\linewidth]{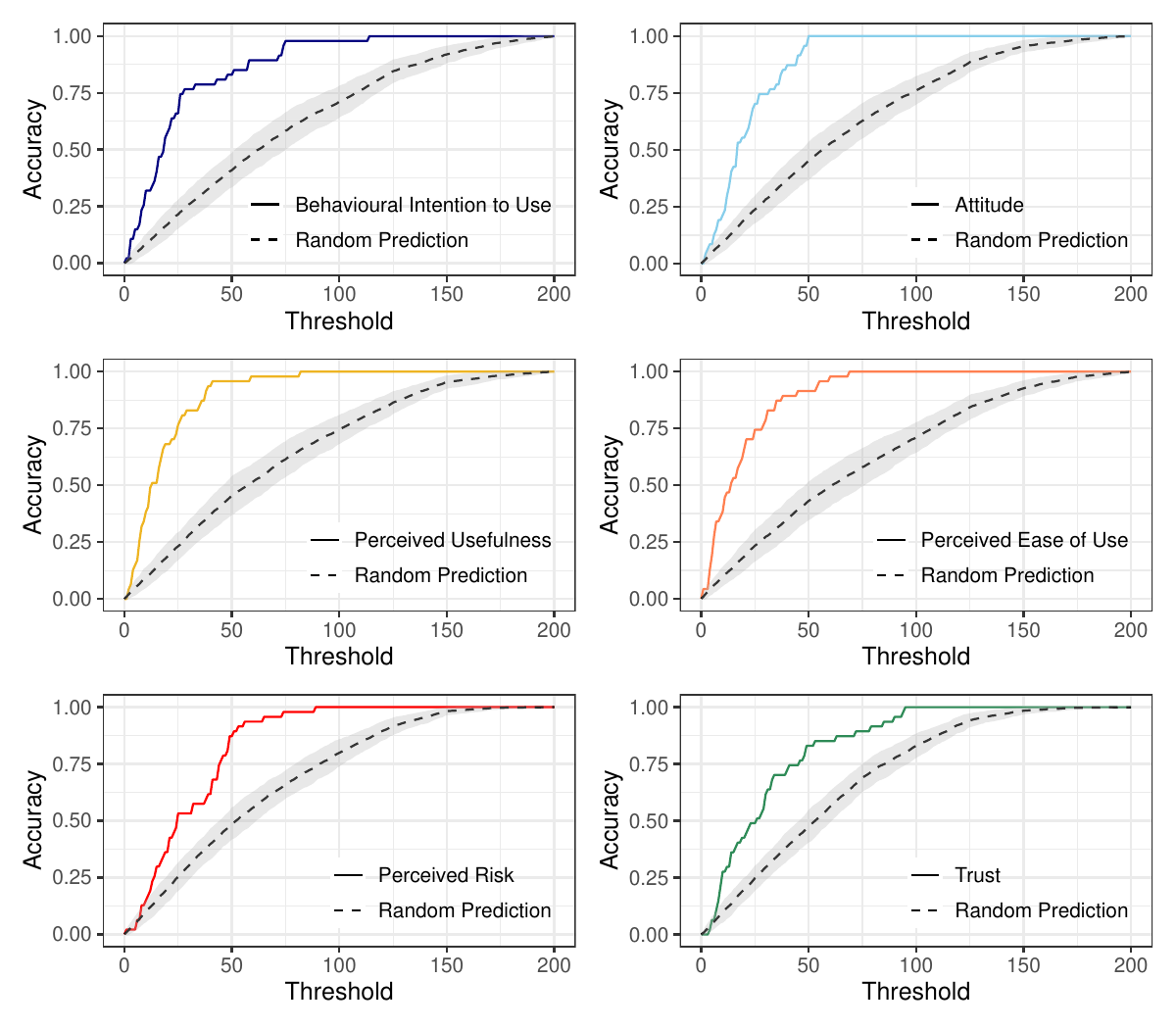} 
    \caption{Random Forest Regression Prediction Using Validation Sample (n = 47): Accuracy vs. Threshold. The colored lines indicate the performance of trained models in predicting target variables. The black dashed line represents the average accuracy of 100 randomly guessed predictions in predicting the target variables. The shadowed intervals indicate one standard deviation above and below the mean of the randomly predicted data.}
    \label{fig:accuracy_threshold}
\end{figure}

To further assess model performance, we evaluated the accuracy of predictions relative to varying threshold values and compared the results against random predictions (see Figure \ref{fig:accuracy_threshold}). For example, at a threshold value of 25, a prediction is considered accurate if it falls within ±25 units of the actual value on the scale from -100 to 100, corresponding to one bin to the left or right of the selection on the questionnaire. The accuracy at this threshold is summarized in Table \ref{tab:RMSE}. All trained Random Forest regression models outperformed random predictions, indicating that the models are informative and provide meaningful predictions. Notably, the model trained to predict Trust using PU, PEOU, Attitude, and PR exhibited relatively lower accuracy compared to the other models. At a threshold value of 25, the accuracy was 0.49. This suggests that the relationship between Trust and these predictors may be more complex or weaker than anticipated, and Trust may be influenced by additional factors not captured in the model. These findings highlight the importance of selecting appropriate predictors and suggest that further research may be needed to explore other variables that influence Trust in the context of SAV acceptance.

\begin{table}[h!]
\captionsetup{font=small}
\centering
\caption{Root Mean Squared Error (RMSE), Normalized Root Mean Squared Error (NRMSE), and Accuracy Results from Random Forest Regression. The accuracy was calculated at a threshold value of 25. The `External' class refers to using all items except overall items to predict the target variable, while the `Internal' class refers to using all items to predict the overall item. The `mtry' parameter was computed from the best model selected by minimizing RMSE after the 10-fold cross-validation.
}
\label{tab:RMSE}
\begin{tabular}{llrrrrr}
  \hline
  Target Variable & Predictors & mtry & sample & RMSE & NRMSE & Accuracy\\ 
  \hline
  Behavioural Intention to Use (BI) & External & 15 & train & 33.60 & 0.17 & 0.63 \\ 
   &   &   & test & 35.08 & 0.18 & 0.66 \\ 
   & Internal & 2 & train & 31.75 & 0.16 & 0.56 \\ 
   &   &   & test & 28.60 & 0.14 & 0.64 \\ 
  Attitude (A) & External & 23 & train & 32.42 & 0.16 & 0.63 \\ 
   &  &  & test & 25.17 & 0.13 & 0.70 \\ 
   & Internal & 2 & train & 25.78 & 0.13 & 0.67 \\ 
   &  &  & test & 31.99 & 0.16 & 0.79 \\ 
  Perceived Risk (PR) & External & 2 & train & 34.47 & 0.17 & 0.46 \\ 
   &  &  & test & 36.07 & 0.18 & 0.53 \\ 
   & Internal & 2 & train & 32.59 & 0.16 & 0.52 \\ 
   &  &  & test & 32.84 & 0.16 & 0.49 \\ 
  Perceived Usefulness (PU) & External & 12 & train & 26.37 & 0.13 & 0.7 \\ 
   &  &  & test & 23.74 & 0.12 & 0.77 \\ 
   & Internal & 2 & train & 32.46 & 0.16 & 0.66 \\ 
   &  &  & test & 28.27 & 0.14 & 0.66 \\ 
  Perceived Ease of Use (PEOU) & External & 2 & train & 29.14 & 0.14 & 0.69 \\ 
   &  &  & test & 24.82 & 0.12 & 0.74 \\ 
   & Internal & 2 & train & 25.84 & 0.13 & 0.70 \\ 
   &  &  & test & 22.88 & 0.11 & 0.70 \\ 
  Trust (T) & External & 12 & train & 32.42 & 0.16 & 0.60 \\ 
   &  &  & test & 40.11 & 0.20 & 0.49 \\ 
   & Internal & 2 & train & 32.56 & 0.16 & 0.59 \\ 
   &  &  & test & 30.58 & 0.15 & 0.64 \\ 
  \hline
\end{tabular}
\end{table}

\subsection{Chord diagram}

\begin{figure}[!h]
\captionsetup{font=small}
    \centering
    \includegraphics[width=\linewidth]{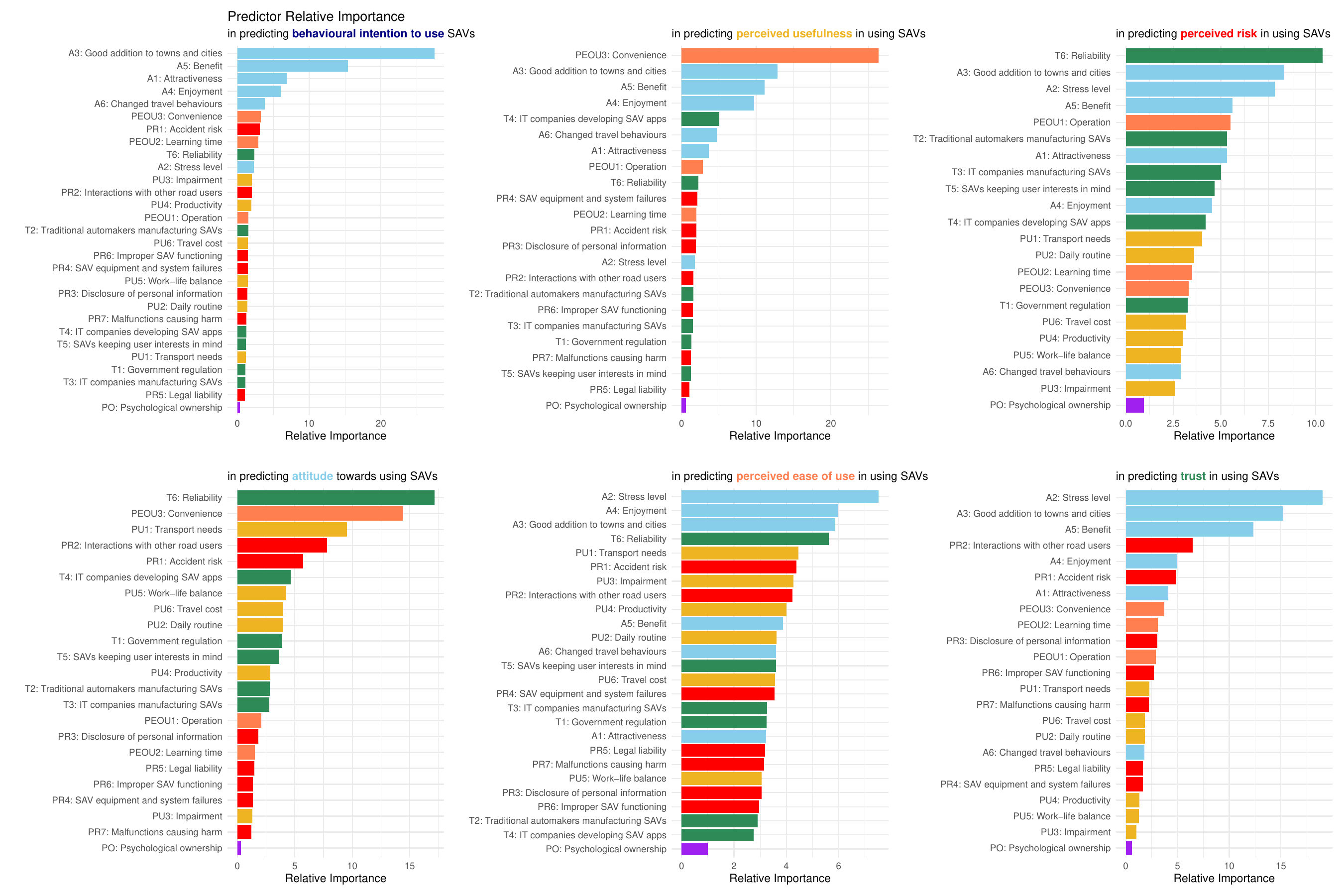} 
    \caption{Predictor Importance Obtained from the Random Forest Models.}
    \label{fig:imp_vip_all}
\end{figure}

\begin{figure}[h!]
\captionsetup{font=small}
  \centering
  \begin{subfigure}[b]{0.8\textwidth}
    \centering
    \includegraphics[width=\textwidth]{circular_plot_v2.pdf}
    \caption{Chord Diagram Showing Relative Importance of Predicting Overall Questions.}
    \label{fig:cir_plot}
  \end{subfigure}
  \begin{subfigure}[b]{0.48\textwidth}
    \centering
    \includegraphics[width=\textwidth]{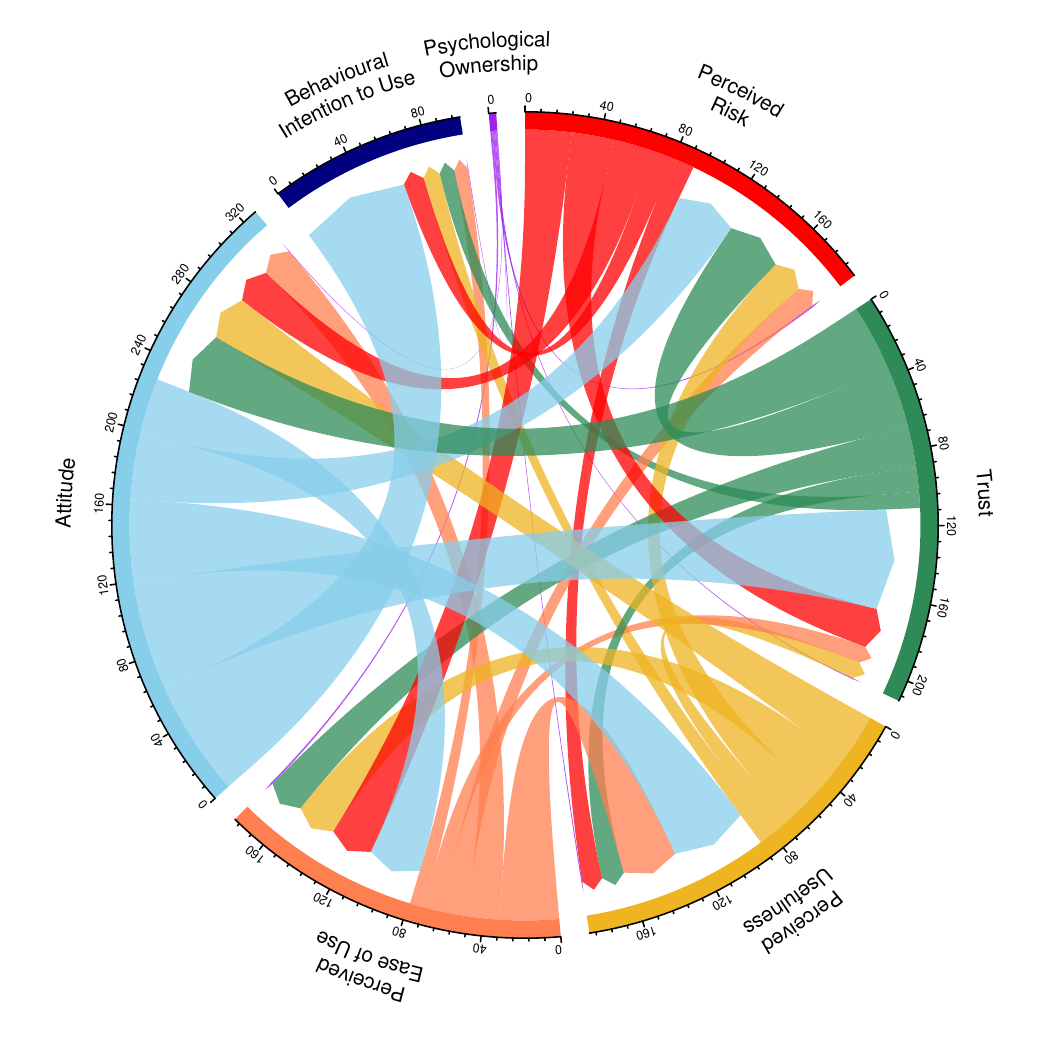}
    \caption{Chord Diagram Showing Overall Relative Importance of Predicting Overall Questions.}
    \label{fig:cir_overall}
  \end{subfigure}
  \hfill
  \begin{subfigure}[b]{0.48\textwidth}
    \centering
    \includegraphics[width=\textwidth]{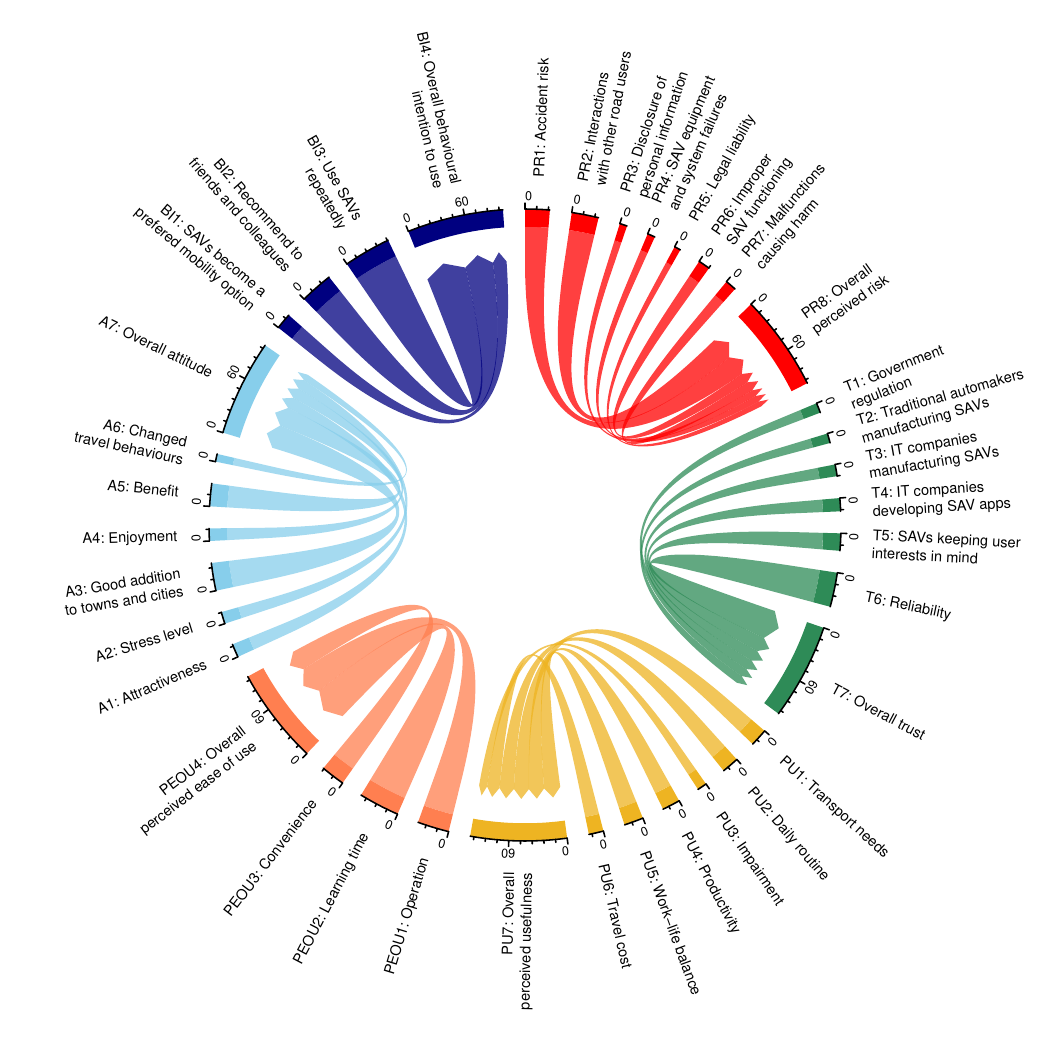}
    \caption{Chord Diagram Showing Relative Importance of the Individual Questionnaire Item on the Overall Question.}
    \label{fig:cir_question}
  \end{subfigure}
  \caption{Chord Diagram of Factor and Question Relative Importance. The arrows ( chords) connect the predictor items (start nodes) and the target variables (end nodes). The width of the chords indicates the relative importance of each predictor item. The sum of the relative importance of each target variable (i.e., the overall questions) is 100\%.}
  \label{fig:overall+question}
\end{figure}

The visualization of factor importance flows enables us to quantify the relationships between different factors and identify their individual contributions to the acceptance of SAVs. After extracting the importance results from all the Random Forest models, we summarized the data in a table with three columns---predictors, target variables, and relative importance---comprising 146 rows (Table \ref{table:item_impor}). Presenting this extensive table or generating numerous separate figures (such as six predictor importance plots using the `vip' package in Figure \ref{fig:imp_vip_all} and 29 predictor contribution plots in Figures \ref{fig:attitude_importance}--\ref{fig:trust_importance}) would be impractical and overwhelming for readers.  To address this, the chord diagram (Figure \ref{fig:cir_plot}) is employed as a concise and intuitive summary of this complexity, following the approach introduced in Figure \ref{fig:relative_imp}. 

This visual representation allows us to see at a glance which specific items have the most significant impact on the target variable. For example, a thick chord connecting the predictor item A3 (``SAVs would be a good addition to towns and cities") to the target variable BI4 (``Overall behavioral intention to use SAVs") indicates that A3 is a highly influential factor in predicting BI4. By examining the chord diagram, we can discern the contributions from both the target variable perspective (shown in Figure \ref{fig:imp_vip_all}) and the predictor perspective (detailed in Figures \ref{fig:attitude_importance}--\ref{fig:trust_importance}). This dual perspective provides a comprehensive understanding of the factors influencing SAV acceptance without overwhelming the reader with excessive data or figures.

To better visualize the overall relationship between the factors, we computed the sum of the relative importance of each construct for every predictor and depicted it using a chord diagram, as shown in Figure \ref{fig:cir_overall}. Additionally, we calculated the importance of each item in predicting the overall measurement of each factor, enabling us to visualize the contribution of different aspects of each factor. The results are summarized in Figure \ref{fig:cir_question}.

We categorized the respondents into SAV adopters and non-adopters based on their overall intention to use SAVs, as measured by their responses to item BI4 in Table \ref{tab:questionnaire}. Participants indicated their intention on a scale from -100 to 100, where:
\begin{itemize}
    \item Negative values indicated a negative intention to use SAVs (``No"),
    \item 0 indicated a neutral stance (``Neutral"),
    \item Positive values indicated a positive intention to use SAVs (``Yes").
\end{itemize}
For the purpose of our analysis, we classified respondents as follows:
\begin{itemize}
    \item \textbf{Non-adopters}: Participants who selected a value between -100 and 0, inclusive of 0.
    \item \textbf{Adopters}: Participants who selected a value between 1 and 100.
\end{itemize}
By including those who selected 0 (neutral stance) in the non-adopter group, we acknowledge that individuals without a positive intention to use SAVs may currently share similar characteristics or concerns as those who are negatively inclined. This grouping allows us to analyze factors that might influence these individuals to move toward adoption in the future.
To visualize the relative importance of predictors for each group, we computed chord diagrams, as shown in Figure \ref{fig:adopter+nonadopter}. These diagrams illustrate the connections between key factors and the intention to adopt SAVs for both adopters and non-adopters, providing insights into how different variables influence each group.

\bigskip

\subsection{Analysis of Factor Importance through Chord Diagrams}

This section delves into the influence of each factor in the context of SAV acceptance at both factor (e.g., Attitude) and item (e.g., A1) levels, as determined through the analysis of chord diagrams (Figure \ref{fig:overall+question}).

\subsubsection{Psychological Ownership Impact}

Our analysis revealed nuanced findings regarding psychological ownership's role in public acceptance of SAVs. While previous studies \citep{lirui-036, lirui-037, lirui-092} identified a positive influence of psychological ownership on the intention to use AVs, our study presents a contrast. We implemented the three routes of psychological ownership as defined by \cite{POchapter1}, integrating them into emotional stories to facilitate a sense of ownership among respondents. Despite this, our findings suggest that psychological ownership did not substantially sway PU, aligning with \citep{lirui-038}, who observed reluctance among users to incur additional costs for personalized services in ride-sharing or car-sharing scenarios.

More notably, our study did not find evidence to support the triggering of psychological ownership in influencing the intention to use SAVs through emotional stories. This observation highlights the challenge of triggering psychological ownership among users in the context of SAVs. Intriguingly, we did observe an impact of psychological ownership on specific aspects of perceived risk. When considering SAVs' interactions with other road users and the potential disclosure of personal information, psychological ownership showed contrasting effects---a positive influence on the former (Control Group mean = 4.93, Psychological Ownership Group mean = 18.15) and a negative impact on the latter (Control Group mean = 8.39, Psychological Ownership Group mean = -8.56).

Given the overall minimal triggering of psychological ownership in our study, this factor has been excluded from further discussions. However, the observed influences on specific risk perceptions point towards the nuanced and complex role psychological ownership may play in the broader context of SAV acceptance.

\subsubsection{Attitude (A)}

\textbf{\textit{Attitude as a Predictor}}

Contrary to suggestions by \cite{Davis_A-critical-assessment_1996} and \cite{lirui-002} to eliminate Attitude from the TAM due to its limited mediating effect of PU on BI, our study underscores the pivotal role of Attitude in determining users' intentions to use SAVs. We discovered that Attitude significantly contributes 61.94\% to the relative importance of predicting SAV usage intention. This dominant role of Attitude is further reinforced by its substantial contributions in shaping trust (57.45\%), PU (43.81\%), PR (34.59\%), and PEOU (30.08\%). This might suggest that, particularly before users gain actual experience with higher-level SAVs, their attitude towards this technology is a decisive factor. 

Notably, respondents predominantly viewed SAVs as valuable additions to urban landscapes (item A3, with an overall relative importance of 62.82\%), emphasizing the societal impact of SAVs. This finding suggests a potential paradigm shift, where SAVs are seen as complements rather than replacements to existing travel options. Integrating SAVs with public transport systems might pave the way for sustainable urban mobility, echoing insights from \cite{lirui-038}.

\bigskip

\textbf{\textit{Attitude as a Target Variable - A7: Overall Attitude}}

In examining Attitude as a target variable, all predictors, except psychological ownership, demonstrated informative effects. Trust, especially in the reliability of SAVs (item T6, 17.14\%), emerged as the most influential, indicating that reliability is more significant in shaping attitudes than factors like trust in IT companies (T4, 4.65\%) or government regulations (T1, 3.92\%). Interestingly, despite its lower overall importance, the convenience aspect of SAVs (PEOU3, 14.43\%) was a key factor in shaping positive attitudes. This suggests a nuanced interplay between ease of use and user's attitude towards using SAVs, highlighting the importance of user-centric designs in SAV development. Additionally, PR factors like safety in interaction with other road users (PR2, 7.78\%) and accident risk (PR1, 5.72\%) also notably influenced attitudes, reflecting public safety concerns.

\subsubsection{Perceived Risk (PR)}

\textbf{\textit{Perceived Risk as a Predictor}}

Our findings align with prior studies \citep{lirui-036, lirui-002, lirui-047}, identifying PR as a key predictor in shaping users’ intentions to adopt SAVs. Ranking as the second most influential factor, PR accounted for 11.85\% of the relative importance in determining SAV acceptance. Furthermore, it significantly influenced PEOU (24.51\%), Trust (22.61\%), Attitude (20.74\%), and PU (11.44\%). Notably, concerns regarding the safety of SAVs interacting with other road users (PR2) and risks associated with accidents (PR1) were paramount, underscoring the public's emphasis on safety. In contrast, data privacy risks (PR3) also surfaced as a notable concern, differing from findings of a 2017 Australian survey by \cite{lirui-017}. This suggests that younger and more educated populations may place a higher priority on privacy compared to the general population. Other perceived risks like SAV equipment failures (PR4) and system malfunctions (PR6, PR7) had lesser but still relevant impacts, with legal liability (PR5) emerging as the least influential factor.

\bigskip

\textbf{\textit{Perceived Risk as a Target Variable - PR8: Overall Perceived Risk}}

The analysis revealed that all predictors, barring psychological ownership, had substantial impacts on PR. Attitude stood out as the most influential factor (with an overall relative importance of 34.59\%), closely followed by Trust (32.87\%), PU (19.27\%), and PEOU (12.30\%). The minimal impact of psychological ownership (0.96\%) suggests its limited role in influencing PR. Intriguingly, respondents' trust in SAV reliability (T6) emerged as a primary factor in shaping PR. Additionally, perceptions of SAVs enhancing urban environments (A3), reducing stress levels (A2), and offering benefits (A5) also influenced PR concerns.

\subsubsection{Perceived Usefulness (PU)}

\textbf{\textit{PU as a Predictor}}

While \cite{lirui-043} highlighted PU as a key determinant in the public acceptance of SAVs, our study presents a nuanced view. While PU emerged as an informative predictor of SAV acceptance intention, it played a less dominant role compared to Attitude and PR, contributing 9.54\% to the relative importance. PU notably influenced other variables, ranking second in predicting Attitude (25.94\%) and third in predicting PEOU (22.98\%), PR (19.27\%), and Trust (9.60\%). Our findings challenge previous studies, showing that PU is not just influenced by PEOU but also acts as a determinant itself. A detailed analysis of individual questionnaire items showed a fairly even distribution of importance across various aspects of PU, with the most significant being the alignment of SAVs with users' transport needs (PU1, with an overall relative importance of 21.47\%), underscoring its critical role in shaping overall Attitude towards SAVs.

\bigskip

\textbf{\textit{PU as a Target Variable - PU7: Overall Perceived Usefulness}}

Contrasting with the TAM, our study found that Attitude was the most influential factor in shaping PU, accounting for 43.81\% of its overall importance. This was followed by PEOU (31.24\%). This variation may be attributed to the broader scope of Attitude, encompassing six items, compared to the three items of PEOU. Trust and PR also played significant roles as predictors, with the relative importance of 12.92\% and 11.44\% respectively. Notably, psychological ownership showed minimal influence on PU (0.59\%). Among the individual predictors, the convenience of SAVs compared to traditional travel modes (PEOU3) emerged as a key factor, significantly influencing the overall PU. Within the Attitude dimension, several items stood out, including the belief that SAVs would enhance urban environments (A3), their perceived benefits (A5), and the enjoyment of using SAVs (A4). Interestingly, trust in IT companies developing SAV applications (T4) was identified as a relatively more influential aspect, suggesting that confidence in the technological interface plays a crucial role in shaping the PU of SAVs.

\subsubsection{Trust (T)}

\textbf{\textit{Trust as a Predictor}}

In our study, the role of Trust in determining behavioral intention to use SAVs was relatively modest, contributing only 8.60\% to the overall prediction. However, its significant influence on PR (32.87\%) aligns with prior research suggesting risk as a mediator between Trust and SAV acceptance \citep{lirui-002, lirui-047, lirui-004, lirui-001}. Trust also notably impacted Attitude towards SAVs (34.96\%) and, to a lesser degree, PEOU (21.42\%) and PU (12.92\%). This aligns with \cite{lirui-019}'s findings on the primacy of initial trust in shaping attitudes towards AVs. A detailed examination of Trust dimensions revealed that reliability (T6) was paramount, followed by trust in IT companies developing SAV apps (T4), traditional automakers (T2), SAVs maintaining user interests (T5), and government regulation (T1). The emphasis on reliability underscores the criticality of performance trust over institutional trust in SAV acceptance among the younger and more educated populations.

\bigskip

\textbf{\textit{Trust as a Target Variable - T7: Overall Trust}}

When evaluating the overall Trust towards using SAVs, Attitude emerged as the most influential factor, with a relative importance of 57.45\%. This suggests that the public's overall sentiment towards SAVs significantly shapes their trust. Key aspects underpinning this Trust include perceived relaxation and stress levels while using SAVs (A2), societal benefits of SAVs (A3), and the perceived personal benefits of SAVs (A5). PR (22.61\%), particularly concerning safety in interactions with other road users (PR2) and accident risks (PR1), also significantly influenced Trust in SAVs, highlighting the importance of addressing safety concerns. Less impactful but still informative were PU and PEOU, reflecting a nuanced interplay of various factors in shaping Trust toward SAV usage. However, the lower predictive accuracy of the Random Forest model for trust, compared to other predictors, suggests the presence of other influencing factors or the possibility that trust functions independently in this context. This calls for further exploration to unravel the nuanced dynamics between various predictors and Trust in the SAV context.

\subsubsection{Perceived Ease of Use (PEOU)}

\textbf{\textit{PEOU as a Predictor}}

Contrary to some earlier studies which downplayed the significance of PEOU in the context of SAVs \citep{lirui-036}, our research identifies PEOU as a noteworthy, albeit the least dominant, predictor for the intention to use SAVs, holding an overall relative importance of 7.75\%. Interestingly, PEOU was found to be most influential in predicting PU, indicating a significant interrelation between ease of use and perceived benefits of SAVs. In descending order, PEOU also impacts Attitude, trust, and PR. The item analysis revealed that the convenience of SAVs, especially compared to traditional travel modes (PEOU3, with an overall relative importance of 51.11\%), was deemed more crucial than factors like the learning time required (PEOU2) and the operational ease of SAVs (PEOU1). This indicates a shift in user focus towards the practical benefits of SAVs over their operational complexity.

\bigskip

\textbf{\textit{PEOU as a Target Variable}---PEOU4: Overall Perceived Ease of Use}

When examining PEOU as a target variable, Attitude emerges as the predominant influence, reflecting the significance of user sentiment in shaping PEOU. Specifically, relaxation and stress levels when using SAVs (A2, 7.52\%), the enjoyment associated with SAV use (A4, 5.99\%), and the perceived societal benefits of SAVs (A3, 5.84\%) were the most impactful Attitude items. PR and PU followed as significant predictors, although none of their respective items exceeded a 5\% relative importance. Trust, particularly in the reliability of SAVs (T6, 5.63\%), also contributed notably, reinforcing the link between Trust and PEOU. The relative importance of psychological ownership was minimal (1.01\%), yet its impact was slightly more pronounced in the context of PEOU compared to other variables, suggesting nuanced influences in user perceptions.

\subsubsection{SAV Adopters vs Non-Adopters: Contrasting Perspectives}\label{subsubsec:adopters_non-adopters}

\begin{figure}[h!]
\captionsetup{font=small}
  \centering
  \begin{subfigure}[b]{0.48\textwidth}
    \centering
    \includegraphics[width=\textwidth]{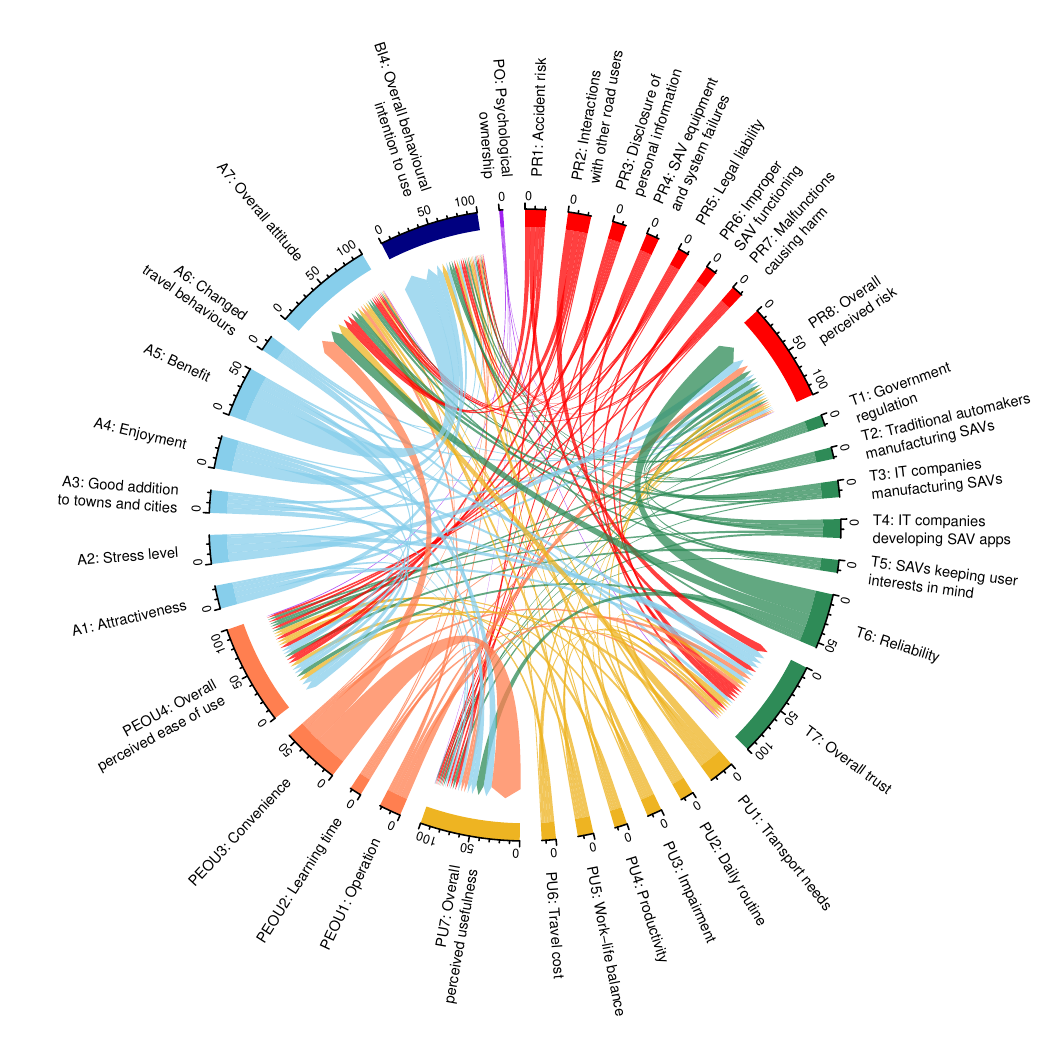}
    \caption{Adopters (n = 246): These respondents selected a value of 0 to 100 on the overall BI questionnaire item (i.e., BI4 in Table \ref{tab:questionnaire}).}
    \label{fig:circular_adopters}
  \end{subfigure}
  \hfill
  \begin{subfigure}[b]{0.48\textwidth}
    \centering
    \includegraphics[width=\textwidth]{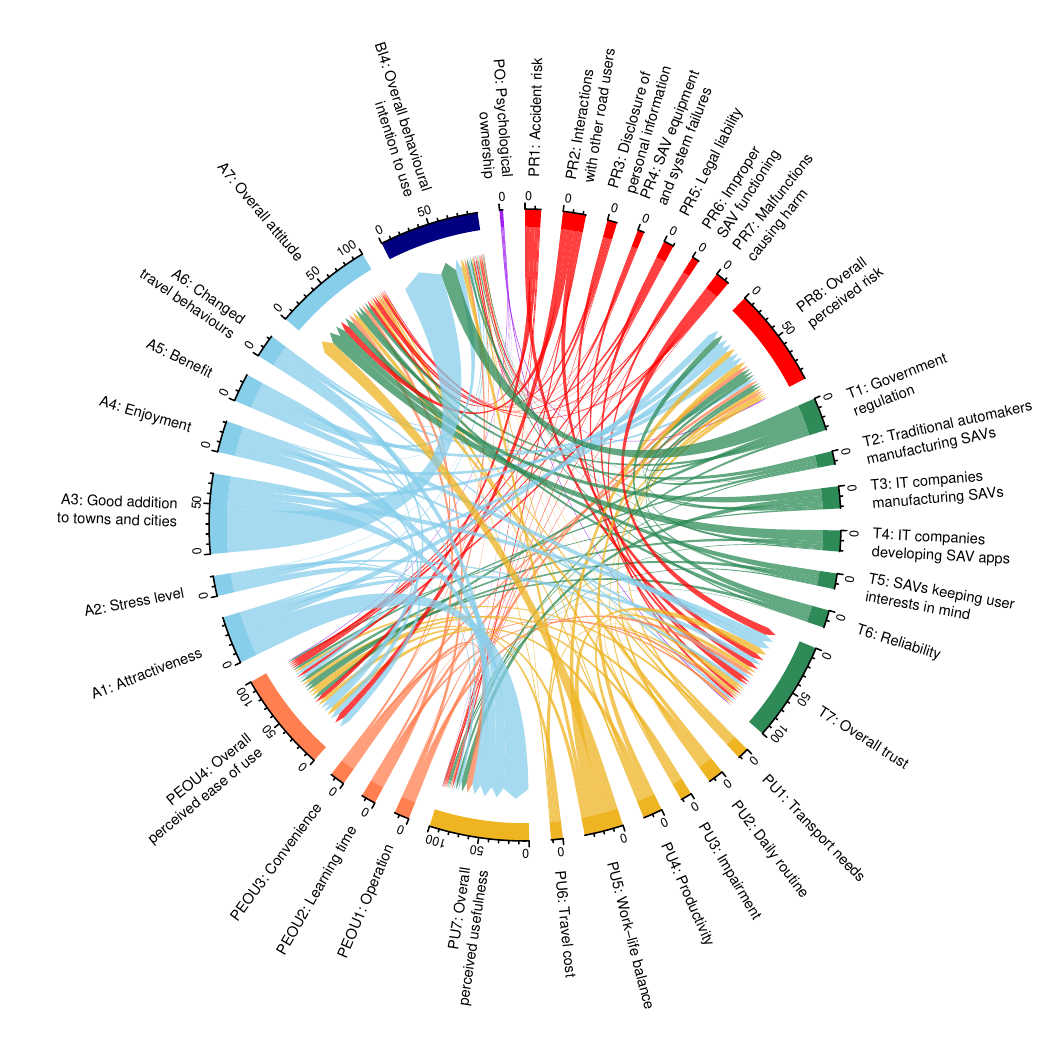}
    \caption{Non-adopters (n = 38): These are the respondents who selected a value between 0 to -100 (including 0) on the overall BI questionnaire item (i.e., BI4).}
    \label{fig:circular_nonadopters}
  \end{subfigure}
  \caption{Chord Diagram Showing Relative Importance of Predicting Overall Questions for SAV Adopters and Non-adopters.}
  \label{fig:adopter+nonadopter}
\end{figure}

In this analysis, respondents were categorized into SAV adopters and non-adopters based on their expressed willingness to use SAVs. This categorization, visualized through chord diagrams (Figure \ref{fig:adopter+nonadopter}), revealed distinct patterns in the factors influencing these two groups.

\textbf{\textit{SAV Adopters}}

For those inclined to use SAVs, the critical items shaping their intention were A5 (perception of SAVs being beneficial), A4 (enjoyable experience), and A1 (attractiveness of SAVs). This suggests that for adopters, the perceived direct benefits and enjoyment are primary motivators. Additionally, in predicting PU, the convenience of using SAVs (PEOU3) emerged as the most significant factor, indicating that practicality is a key consideration for this group. Additionally, the trust in the reliability of SAVs (T6) was crucial for shaping perceptions of risk and attitudes toward SAV usage, indicating a reliance on the reliability of this technology.

\bigskip

\textbf{\textit{SAV Non-Adopters}}

In contrast, the non-adopters prioritized different factors. The top predictors for their intention to use SAVs shifted to A3 (SAVs as a good addition to towns and cities), T1 (trust in government regulation), and A5 (benefits of SAVs). This shift highlights that non-adopters place greater emphasis on the societal impact and regulatory trust concerning SAVs. Similarly, when considering PU, the overall attractiveness of using SAVs (A1) was more influential for non-adopters, suggesting a focus on broader societal benefits. Additionally, PR for non-adopters was significantly influenced by A1 instead of Trust items, highlighting different risk perceptions compared to adopters.

\bigskip

However, it is important to note that our comparative analysis is based on an unequal distribution of participants between the two groups, with a larger number of adopters than non-adopters. This imbalance may affect the robustness of our comparative analyses and could obscure nuanced differences in attitudes between these groups. Nevertheless, the significant contrasts observed provide valuable insights into the factors influencing SAV acceptance and highlight the need for different strategies for promoting SAV adoption among different segments of the population.

\subsection{Open-Ended Question Responses: Insights into SAV Acceptance Factors}

Our survey included four optional open-ended questions, offering participants the opportunity to express their personal views on various aspects related to SAVs. These questions delved into factors influencing their intent to use SAVs, expectations from AV technology, envisioned features of an ideal AV/SAV, and any additional comments on the survey topic. These qualitative insights complement our quantitative findings and provide a richer understanding of public attitudes toward SAVs. They underline the importance of addressing safety, inclusivity, cost-effectiveness, and environmental considerations in the development and implementation of SAVs, while also acknowledging the diverse expectations and concerns of potential users.

\bigskip

\textbf{\textit{Influential Factors for SAV Acceptance (94 responses):}}

The majority of participants highlighted safety and cost as the primary influential factors impacting their intention to use SAVs. The concerns about safety were diverse and included general safety, reliability, data handling, interaction with other road users, and performance in unusual scenarios. This aligns with our quantitative findings where PR emerged as a significant determinant of SAV usage intention. Notably, a respondent highlighted the importance of testing safety features across diverse demographics, reflecting varied safety perceptions. On the other hand, while the cost factor showed less prominence in the chord diagram analysis, it was frequently mentioned in comparison to public transport in the open-ended responses, indicating its potential as a critical research area for future studies. The government's role in integrating SAVs into public transport systems and communicating cost benefits also emerged as vital. Additionally, factors such as waiting time and punctuality were noted as essential considerations for users.

\bigskip

\textbf{\textit{Expectations from AV Technology (86 responses):}}

Respondents expect AVs to independently handle driving and parking tasks under all conditions, enabling users to engage in personal activities while in transit. Noteworthy is the expectation of AVs catering to people with disabilities, underscoring the need for inclusive design considerations. One respondent's vision of envisioned AVs as extensions of personal living or working spaces, illustrating the desire for multifunctional, connected vehicle environments. However, some concerns were expressed about potential negative impacts, such as increased greenhouse gas emissions and socioeconomic disparities in service access. This indicates a complex perspective on AV technology, including skepticism regarding safety, evidenced by references to incidents involving automated driving systems such as Tesla Autopilot. These insights highlight the importance of addressing environmental sustainability and equitable access in the development of AV technologies.

\bigskip

\textbf{\textit{Ideal AV/SAV Conceptualization (74 responses):}}

Participants envisioned AVs and SAVs prioritizing clean energy, safety, efficiency, affordability, ease of use, and unique design. Several responses integrated SAVs within public transport systems. A few participants even foresaw applications in air travel, indicating interest in broader autonomous transportation modes. A personal perspective from a diabetic respondent with diabetes underscored the life-enhancing potential of SAVs for individuals with specific health conditions, highlighting the broader societal impact of this technology. This emphasizes the importance of considering the needs of various user groups in the design and implementation of SAVs.

\section{Concluding Remarks and Future Work}\label{sec: Discussion}

\subsection{Applications to Broader Research}

This study introduces an innovative framework that combines Machine Learning techniques with chord diagram visualizations to analyze factor importance in technology acceptance research. Rather than proposing a new theoretical model and causal relationships, this framework complements existing theories by examining the relative importance of various factors in predicting SAV acceptance through a predictive modeling perspective.  It effectively bridges the gap between the complex data analysis capabilities of predictive modeling and the user-friendly, insightful visual representations provided by chord diagrams.

As a case study of the proposed framework, Random Forest was selected to predict the behavioral intention to use SAVs using relevant psychological factors. Random Forest excels in processing medium-sized datasets and identifying complex, non-linear relationships, making it ideal for both regression and classification problems. Its robustness as a predictive model is enhanced by its ability to easily compute prediction accuracy and determine predictor importance effectively.

When integrated with chord diagrams, this approach allows for the simultaneous visualization of multiple model outputs, providing an intuitive representation of both inter-factor relationships and the importance of individual predictors. By combining the predictive power of Random Forest with the clarity of chord diagrams, our framework enables a thorough examination of the impact of each predictor and facilitates comparisons between different user groups, specifically SAV adopters and non-adopters in our study. This integration simplifies data analysis and enhances the presentation of complex insights, making them more accessible to a diverse audience, including those without extensive statistical knowledge.

Beyond the context of SAV acceptance, the framework has broader applicability to other fields that deal with complex datasets and multifaceted variables. In marketing, for instance, it could provide deeper insights into consumer behavior through a more intuitive understanding of factor importance. Education researchers could leverage this approach for a comprehensive visualization of factors affecting student performance. Similarly, consulting firms could adopt this user-friendly method to present complex data analyses to stakeholders in a more digestible format.

\subsection{Application to SAV Acceptance Problem}

In the case study, the proposed methodological approach has significantly enhanced our understanding of the factors influencing SAV acceptance. The integration of Random Forest and chord diagrams facilitated a detailed analysis of psychological factors, highlighting their complex interplay and influence on the behavioral intention to use SAVs. This analysis has provided in-depth insights into the drivers and barriers to SAV acceptance, which are essential for the further development of this emerging technology.

The ability of the Random Forest model to rank predictor variables and the chord diagrams' capacity to visually present these rankings have revealed the complex dynamics of SAV acceptance. The Random Forest model's effective predictive accuracy, demonstrated by RMSE and NRMSE values, along with validation at various thresholds, underscores the importance of Attitude, Trust, PR, PU and PEOU in determining SAV acceptance. 

While there has been debate in the literature regarding the role of Attitude in technology acceptance models---some studies have suggested eliminating attitude due to its limited mediating effect \citep{Davis_A-critical-assessment_1996, lirui-098, lirui-002, lirui-004}, our findings suggest that Attitude plays an important role in predicting behavioral intention. Attitude, specifically the public's perception of SAVs as valuable additions to urban settings and their potential individual benefits, are the key predictors in predicting PU, PEOU, Trust, PR, and  BI to use SAVs. This underscores the need for strategies that highlight both the shared and individual benefits of using SAVs, which could potentially influence perceptions across multiple psychological factors. Additionally, attitudes toward the stress level and enjoyment associated with using SAVs proved important in predicting PEOU, Trust, and PR. This suggests that SAV developers and manufacturers should consider user experience elements that promote relaxation and enjoyment, potentially improving overall acceptance. 

Our findings support the inclusion of attitude as a key factor in studying technology acceptance from a predictive modeling perspective, highlighting its critical role in predicting behavioral intention. This contrasts with prior studies that suggested eliminating attitude from models like TAM due to its limited mediating effect \citep{Davis_A-critical-assessment_1996, lirui-098, lirui-002, lirui-004}.

Our study also demonstrates that PU has both direct and indirect effects on behavioral intention to use SAVs, with attitude playing a crucial role in this relationship. At the item level, meeting users' transportation needs emerged as a significant predictor in shaping attitudes toward using SAVs. This finding emphasizes the importance of ensuring that SAVs effectively address the practical requirements of users. Furthermore, the perceived usefulness of SAVs in assisting individuals when they are impaired and in improving productivity were the most influential PU items in predicting BI. These insights suggest that highlighting these specific benefits could enhance users' perceived usefulness of SAVs and their willingness to adopt them.
Similar to previous studies that adopted theoretical models to study AV acceptance, PEOU showed a strong influence on Attitude and PU, particularly through the item concerning the additional convenience that SAVs could offer compared to traditional travel modes. Interestingly, our results also indicate that users' perceptions of the ease of operating SAVs significantly impact PR. This suggests a potential need for user-friendly interfaces and intuitive designs, which could influence perceived risks associated with operating SAVs.

Furthermore, our analysis reveals the complex interplay between Trust and PR in SAV acceptance. While Attitude items had the greatest overall effect on both PR and Trust, the influence between PR and Trust themselves was significant. At the item level, users' perceived risk of SAVs interacting with other road users and being involved in accidents were important predictors of Trust. Conversely, users' trust in the reliability of SAVs was the most influential predictor in predicting PR. These findings highlight the importance of addressing specific safety concerns and enhancing the reliability of SAVs to build trust and reduce perceived risks.

In addition, the distinct contrasts between SAV adopters and non-adopters highlight critical differences essential for crafting effective promotion strategies. This emphasizes the importance of delving into the nuances of each factor at the item level when examining public acceptance of SAVs, rather than considering these factors collectively.

\subsection{Practical Implications for Stakeholders}

The findings of this study have direct implications for SAV developers, policymakers, and marketers, particularly in regard to the younger and more educated population. Understanding that certain psychological factors and specific item-level perceptions are crucial in predicting public acceptance can guide the development of more user-centric SAV designs and informed policy-making.

For instance, emphasizing the shared benefits (e.g., improving urban environments) and individual benefits (e.g., enhancing personal transportation needs) that SAVs can offer could be an effective strategy to increase public acceptance. Marketing campaigns and informational materials should highlight how SAVs can contribute positively to urban settings and meet individual user needs.

Prioritizing enhancements in the convenience of using SAVs, such as demonstrating how SAVs offer greater convenience compared to traditional travel modes, can influence both PU and PEOU. SAV developers should focus on designing user interfaces and systems that are intuitive and easy to operate, thereby reducing perceived complexity and risks.

Building trust in the reliability of SAVs is paramount. Efforts to improve and communicate the reliability and safety features of SAV technology, such as advanced collision avoidance systems and rigorous testing protocols, can address users' concerns about accidents and interactions with other road users. Transparent communication about these features can enhance trust and reduce perceived risks. In addition, the role of mobile applications is noteworthy; fostering trust in IT companies developing SAV apps can influence users' attitudes and the perceived usefulness of SAVs. Ensuring that these applications are secure, user-friendly, and reliable can contribute to a positive user experience.

Additionally, the differential analysis between SAV adopters and non-adopters underscores the need for tailored strategies:
\begin{itemize}
    \item For Adopters: Emphasize the direct benefits and positive attributes of SAVs, such as enjoyment, stress reduction, and personal productivity enhancements. Marketing efforts could showcase testimonials and scenarios where users benefit directly from using SAVs.
    \item For Non-Adopters: Focus on illustrating how SAVs can contribute positively to urban landscapes and building trust in government regulations could be effective.
\end{itemize}

This nuanced understanding is crucial for devising targeted marketing and educational initiatives aimed at increasing the overall acceptance of SAVs.

\subsection{Future Research Directions}\label{sec: Limitations}

While our study contributes new insights and methodological approaches that could be valuable in exploring technology acceptance in various research disciplines, it is important to acknowledge its limitations and outline avenues for future research.

In the context of SAV acceptance, our case study's focus on an Australian university community provided valuable insights from younger and more educated participants compared to the general population. However, we acknowledge that this demographic bias may limit the generalizability of our findings, as different socio-demographic groups may have different perceptions of SAVs. Therefore, the practical implications discussed are most relevant to this specific group. Nevertheless, our proposed framework is applicable to any population group. Future studies could apply this framework to datasets with a wider and more diverse demographic cross-section to enhance the generalizability of the findings. Additionally, ensuring a more equitable distribution between SAV adopters and non-adopters would strengthen the robustness of comparative analyses, allowing researchers to uncover more nuanced differences in attitudes and acceptance levels and to develop more targeted strategies.

To further investigate the significance and triggering mechanisms of psychological ownership in the context of SAVs, future studies could augment survey-based methodologies with experiential approaches that provide participants with tangible interactions with SAV technology. This could potentially elicit a stronger sense of psychological ownership and offer deeper insights into its effects on SAV acceptance.

It should be noted that the current framework is not designed to propose new theoretical frameworks or investigate causal relationships among factors. Future work could enhance the framework by integrating methods that indicate the directionality (positive or negative effects) of the relationships between predictors and outcome variables. Techniques such as examining decision trees generated during Random Forest construction or conducting partial dependence analysis---which illustrates the marginal effects of variables---would provide a deeper understanding of how each predictor influences the outcomes. Sample decision trees (Figure \ref{fig:tree_example}) and partial dependence plots (Figure \ref{fig:Partial_Dependence_A3}) based on our case study are presented in Appendix \ref{appendix:sample_tree}.

Lastly, there is an opportunity for subsequent research to broaden the scope by incorporating additional psychological and situational factors such as driving pleasure, parking options, pricing strategies, waiting times, and trip lengths. Including these variables may offer a more comprehensive view of the multifaceted nature of SAV acceptance and capture factors relevant to a broader audience.

\bigskip

\section*{CRediT authorship contribution statement}

\textbf{Lirui Guo:} Conceptualization, Data curation, Investigation, Formal analysis, Methodology, Validation, Software, Visualization, Writing – original draft. 
\textbf{Michael G. Burke:} Conceptualization, Methodology, Formal analysis, Supervision, Validation, Visualization, Writing – review \& editing. 
\textbf{Wynita M. Griggs:} Conceptualization, Project administration, Resources, Methodology, Supervision, Validation, Visualization, Writing – review \& editing.

\section*{Role of the funding source}

This work was partially supported by internal funding provided by Monash University, Australia.

\section*{Declaration of competing interest}

The authors declare that they have no known competing financial interests or personal relationships that could have appeared
to influence the work reported in this paper.

\section*{Declaration of generative AI and AI-assisted technologies in the writing process}

During the preparation of this work the authors used OpenAI ChatGPT in order to improve language and readability. After using this tool/service, the authors reviewed and edited the content as needed and take full responsibility for the content of the publication.

\appendix

\section*{Appendices}

\section{Survey Items}\label{Appendix_survey_item}

\subsection{Part I: Travel Modes}

1. Have you heard of Autonomous Vehicles / Automated Vehicles / Self-driving Cars before?
\begin{itemize}
    \item [$\bigcirc$] Yes
    \item [$\bigcirc$] No
\end{itemize}

\vspace{1em} 

2. Where did you hear about Autonomous Vehicles / Automated Vehicles / Self-driving Cars? (You can choose \textbf{multiple answers}.)
\begin{itemize}
    \item [$\square$] Movies
    \item [$\square$] News
    \item [$\square$] Science fiction
    \item [$\square$] Advertisement
    \item [$\square$] Other
\end{itemize}

\vspace{1em} 

3. How much do you know about Autonomous Vehicles?
\begin{itemize}
    \item [$\bigcirc$] Nothing
    \item [$\bigcirc$] A little
    \item [$\bigcirc$] A moderate amount
    \item [$\bigcirc$] A lot
    \item [$\bigcirc$] A great deal
\end{itemize}

\vspace{1em} 

4. Which of the following automated driver assistance systems have you used? (You can choose \textbf{multiple answers}.)
\begin{itemize}
    \item [$\square$] Adaptive cruise control (automatically adjusts the vehicle speed to maintain a safe distance between vehicles)
    \item [$\square$] Lane keeping assistance
    \item [$\square$] Automated parking system
    \item [$\square$] Autopilot
    \item [$\square$] None
\end{itemize}

\vspace{1em}

5. How do you currently \textbf{travel}? As a percentage, please give an estimated fraction of travel time spent using each option. (For example, 30\% private car, 60\% train, 10\% walking, 0\% shared mobility, 0\% others. The total should sum to 100\%.)

\vspace{0.5em}

\begin{tabularx}{\textwidth}{X>{\centering\arraybackslash}m{2cm}}
    \hline
    \textbf{Mode of Travel} & \textbf{Percentage (\%)} \\ \hline
    \%, Private car & \fbox{0} \\ 
    \%, Shared mobility (e.g., taxi, Uber, Didi) & \fbox{0} \\ 
    \%, Public transport (i.e., bus, tram, train) & \fbox{0} \\ 
    \%, Active transport (i.e., bicycle, scooter, walk) & \fbox{0} \\ 
    \%, Other & \fbox{0} \\ \hline
    \textbf{Total} & \fbox{0} \\ \hline
\end{tabularx}

\vspace{1em}

6. Do you own a car?
\begin{itemize}
    \item [$\bigcirc$] No
    \item [$\bigcirc$] Yes
\end{itemize}

\vspace{1em} 

7. How frequently do you drive a vehicle?
\begin{itemize}
    \item [$\bigcirc$] Never
    \item [$\bigcirc$] Less than four days per month
    \item [$\bigcirc$] One day a week
    \item [$\bigcirc$] 2--3 days a week
    \item [$\bigcirc$] 4--6 days a week
    \item [$\bigcirc$] Daily
\end{itemize}

\vspace{1em} 

8. Have you used shared mobility services before? (i.e., taxi, Uber, Didi)
\begin{itemize}
    \item [$\bigcirc$] No
    \item [$\bigcirc$] Yes
\end{itemize}

\vspace{1em} 

9. How do you rate your experience with these shared mobility services? (i.e., taxi, Uber, Didi)

\begin{tikzpicture}
\draw[thick] (-6, 0) -- (6, 0);
\foreach \x in {-5, -2.5, 0, 2.5, 5}
    \draw[thick] (\x, 0.25) -- (\x, -0.25);
\node at (-6, -0.5) {-100};
\node at (0, -0.5) {0};
\node at (6, -0.5) {100};
\node at (-6, 0.5) {Dissatisfied};
\node at (0, 0.5) {Neutral};
\node at (6, 0.5) {Satisfied};
\end{tikzpicture}

\vspace{1em} 

10. How many \textbf{hours} do you approximately spend on transportation every day? (Any transport mode)

\fbox{\rule{3cm}{0pt}\rule[0cm]{0pt}{0.3cm}} 

\bigskip

\subsection{Part II: Be Driven Into The Future}\label{Appendix_survey_part2}

Table \ref{tab:questionnaire} below provides a comprehensive summary of the questionnaire items and their respective sources, which were employed to examine the selected factors in our study. These factors include Perceived Risk (PR), Trust (T), Perceived Usefulness (PU), Perceived Ease of Use (PEOU), Attitude towards using (A), and Behavioral Intention to Use (BI). Each factor offers insight into the specific items and considerations that underpinned our investigation.

\begin{table}[H]
\centering
\caption{Questionnaire Items and Sources}
\label{tab:questionnaire}
\resizebox{\textwidth}{!}{%
\begin{adjustbox}{max width=\textwidth}
\begin{tabular}{llrlrr}
  \hline
  Questionnaire items & Modified from \\ 
  \hline
PR1: Do you think using SAVs would decrease or increase your risk of being involved in an accident? & \cite{lirui-047} \\
PR2: Do you think SAVs interacting with other road users would be safe or dangerous? & \cite{lirui-037}\\
PR3: How concerned are you about the SAV disclosing my personal information? & \cite{lirui-047} \\
PR4: How concerned are you about the SAV equipment and system failures? & \cite{lirui-043} \\
PR5: How concerned are you about the legal liability of users or owners of SAVs? &  \cite{lirui-001} \\
PR6: How worried are you about SAVs may not function properly? &  \cite{lirui-036} \\
PR7: How worried are you about failures or malfunctions in SAVs causing harmful accidents? &  \cite{lirui-047} \\
\textbf{PR8: Overall, do you think using SAVs would be safe or risky?} &  \cite{lirui-001} \\

T1: How much do you trust government authorities to regulate and supervise SAVs? & \cite{lirui-047} \\
T2: How much do you trust SAVs manufactured by traditional automakers? & \cite{lirui-047} \\
T3: How much do you trust SAVs manufactured by Information Technology companies? & \cite{lirui-001} \\
T4: How much do you trust Information Technology companies that produce SAV apps? & \cite{lirui-001} \\
T5: How much do you trust SAVs to keep your best interests in mind? & \cite{lirui-047} \\
T6: Do you think SAVs would be reliable or not reliable? & \cite{lirui-002} \\
\textbf{T7: Overall, do you trust or distrust SAVs?} & \cite{lirui-002} \\

PU1: How useful do you think using SAVs would be in meeting your transport needs? & \cite{lirui-019} \\
PU2: How useful do you think using SAVs would be in undertaking your daily routine? & \cite{lirui-037} \\
PU3: How useful do you think using SAVs would be when you're impaired? & \cite{lirui-019} \\
PU4: How useful do you think using SAVs would be in improving your productivity? & \cite{lirui-036} \\
PU5: How useful do you think using SAVs would be in improving your work-life balance? & \cite{lirui-002} \\
PU6: Do you think using SAVs could reduce or increase your travel cost? & \cite{lirui-047} \\
\textbf{PU7: Overall, do you think SAVs would be useful or not useful?} & \cite{lirui-036} \\

PEOU1: Do you think SAVs would be easy or difficult to operate? & \cite{lirui-036} \\
PEOU2: How long do you think it would take you to learn how to use a SAV? & \cite{lirui-036} \\
PEOU3: Compared to traditional travel modes, do you think using SAVs would be more convenient? & \cite{lirui-047} \\
\textbf{PEOU4: Overall, do you think using SAVs would be easy or difficult?} & \cite{lirui-036} \\

A1: Do you think using SAVs would be attractive or unattractive? &  \cite{Christos_thesis_2020} \\
A2: Do you think using SAVs would be relaxing or stressful? & \cite{lirui-037} \\
A3: How much do you agree with, I think SAVs would be a good addition to towns and cities? & \cite{lirui-019}\\
A4: How much do you agree with, I think using SAVs would be enjoyable? & \cite{Christos_thesis_2020} \\
A5: How much do you agree with, I think using SAVs would be beneficial to me? & \cite{lirui-019} \\
A6: How much do you agree with, I think the availability of SAVs could change my travel behavior? & \cite{lirui-047} \\
\textbf{A7: Overall, are you holding a positive or negative attitude towards using a SAV?} & \cite{lirui-019}\\

BI1: Do you agree or disagree with, I would likely consider SAVs as my choice of mobility in the future? & \cite{lirui-036} \\
BI2: Do you agree or disagree with, I would likely recommend my friends and colleagues to use SAVs? & \cite{lirui-037} \\
BI3: Do you agree or disagree with, I would likely use SAVs repeatedly in the future? & \cite{lirui-037} \\
\textbf{BI4: Overall, would you use an SAV if given the chance?} & \cite{lirui-036} \\
  \hline
  \multicolumn{6}{l}{*Note: PR = perceived risk; T = trust; PU = perceived usefulness; PEOU = perceived ease of use; A = attitude; BI = behavioural intention to use.} \\

\end{tabular}
\end{adjustbox}
}
\end{table}

\subsection{Part III: Demographics}

1. What is your gender?
\begin{itemize}
    \item [$\bigcirc$] Male
    \item [$\bigcirc$] Female
    \item [$\bigcirc$] Non-binary
    \item [$\bigcirc$] Prefer not to say
    \item [$\bigcirc$] Prefer to self-describe: \fbox{\rule{3cm}{0pt}\rule[0cm]{0pt}{0.3cm}} 
\end{itemize}

\vspace{1em} 

2. What is your age?
\begin{itemize}
    \item [$\bigcirc$] 18 -- 34
    \item [$\bigcirc$] 35 -- 44
    \item [$\bigcirc$] 45 -- 54
    \item [$\bigcirc$] 55 -- 64
    \item [$\bigcirc$] $\geq 65$
    \item [$\bigcirc$] Prefer not to say
\end{itemize}

\vspace{1em} 

3. What type of area do you live in?
\begin{itemize}
    \item [$\bigcirc$] Inner metropolitan
    \item [$\bigcirc$] Outer metropolitan
    \item [$\bigcirc$] Regional
    \item [$\bigcirc$] Country/rural
    \item [$\bigcirc$] Prefer not to say
\end{itemize}

\vspace{1em} 

4. What is your annual gross household income?
\begin{itemize}
    \item [$\bigcirc$] Less than \$30,000
    \item [$\bigcirc$] \$30,000 -- \$74,999
    \item [$\bigcirc$] \$75,000 -- \$149,999
    \item [$\bigcirc$] \$150,000 or more
    \item [$\bigcirc$] Prefer not to say
\end{itemize}

\vspace{1em} 

5. Are you responsible for transporting anyone with special care needs (e.g., elderly or disabled)?
\begin{itemize}
    \item [$\bigcirc$] No
    \item [$\bigcirc$] Yes
    \item [$\bigcirc$] Prefer not to say
\end{itemize}

\vspace{1em} 

6. Do you have any permanent injuries or illnesses that affect your driving?
\begin{itemize}
    \item [$\bigcirc$] No
    \item [$\bigcirc$] Yes
    \item [$\bigcirc$] Prefer not to say
\end{itemize}

\vspace{1em} 

7. What is your highest education level (completed)?
\begin{itemize}
    \item [$\bigcirc$] None (no studies)
    \item [$\bigcirc$] Primary studies
    \item [$\bigcirc$] Secondary studies / high school
    \item [$\bigcirc$] TAFE
    \item [$\bigcirc$] University studies
    \item [$\bigcirc$] Prefer not to say
\end{itemize}

\section{Survey Story Design}\label{Appendix_survey_story}

The two distinct narratives crafted to illustrate the potential applications of Shared Autonomous Vehicles (SAVs) are provided below. Version 1 serves as the standard narrative, while Version 2 integrates specific applications of three `routes' of psychological ownership, control, self-investment, and intimate knowledge. This integration aimed to trigger a sense of psychological ownership in participants as they engaged with the stories. In these narratives, the `routes' to psychological ownership are highlighted in red, with the corresponding applications emphasized in italics.

\bigskip

\begin{tcolorbox}[boxrule=0.4pt, colframe=black]
\textbf{Version 1 (standard):}

``Oliver wants to visit his friend living 15km away from his home. He requests a ride using an app on his smartphone. An autonomous vehicle arrives at his door within minutes. Oliver enters his friend's address using the Car Infotainment System (i.e., the car user interface screen), and the vehicle transports him directly to his destination. During the trip, Oliver continues watching the movie he had started viewing at home. Upon arrival, he doesn’t need to park the vehicle, which continues on to a nearby location to pick up another rider.''

``Amelia has pre-set a pick-up time at 8:30 am in front of her house, and another at 17:10 pm out the front of her company, for every workday. It is now 8 am in the morning, and Amelia has just woken up. She finishes her breakfast and walks out of her house. An autonomous vehicle is waiting at her door. During the trip to work, she checks her emails and plans her day. Upon arrival, Amelia walks to her office, and the autonomous vehicle drives away. At 17:10 pm, when she finishes work, an autonomous vehicle is waiting on the roadside to drive her home.''
\end{tcolorbox}

\bigskip

\begin{tcolorbox}[boxrule=0.4pt, colframe=black]

\textbf{Version 2 (with psychological ownership):}

``Oliver wants to visit his friend living 15km away from his home. He requests a ride using an app on his smartphone. An autonomous vehicle arrives at his door within minutes. \textit{Oliver puts his bag inside the car boot and gets into the car} (Control: spatial control). \textit{He adjusts the seat to a comfortable position and puts the window down like he did when he used traditional cars with human drivers} (Control: configuration control). He enters his friend's address using the Car Infotainment System (i.e., the car user interface screen), and the vehicle transports him directly to his destination. During the trip, Oliver continues watching the movie he had started viewing at home. Upon arrival, \textit{he takes his empty drink bottles with him to keep the vehicle clean} (Remove the evidence of other users to build possession) and doesn’t need to worry about parking the vehicle. His ``SAV app" pops up a notification to say this is \textit{his 400th trip using SAVs, with a total travelled distance of 5679 kilometers so far} (Intimate knowledge). He \textit{wins an award ``Level 2 SAV Silver User''} (Self-investment: emblems signal information about the user's identity) and \textit{he decides to share it to his social media} (Social Media).''

``Amelia is a big fan of shared autonomous vehicle services. In her ``SAV app'' on her phone, \textit{she creates her own virtual persona (i.e., a cat named ‘Mickey’)}  (Self-investment: creation) which will also be displayed as a virtual assistant on the Car Infotainment System (i.e., the car user interface screen) when she rides in an SAV. She additionally pre-sets a pick-up time of 8:30 am in front of her house, and another at 17:10 pm out the front of her company, for every workday. When she first used the shared autonomous vehicle service some months ago, \textit{she had adjusted the seat, light, and temperature settings to her favorite positions using her phone and saved the settings in her ``SAV app''} (Control: configuration control; Self-investment). It is now 8 am in the morning, and Amelia has just woken up. Amelia finishes her breakfast and walks out of her house. An autonomous vehicle is waiting at her door. When she gets into the vehicle, the seat, light, and temperature have already been adjusted to her favorite position and levels. \textit{‘Mickey' appears on the car's dashboard screen and says: ``Good morning, Amelia. Do you want to go to work now following your usual route?''} (Intimate knowledge: enabling)  Amelia says ``Yes'' and the vehicle drives her to work. During the trip, Amelia checks her emails and plans her day. Upon arrival, ‘Mickey’ says, ``Have a nice day Amelia. I will pick you up at 5:10 pm. See you soon!'' Amelia walks to her office and the autonomous vehicle drives away. At 17:10 pm, when she finishes work, an autonomous vehicle is waiting on the roadside with `Mickey' on the car's dashboard screen to drive her home.''

\end{tcolorbox}

\bigskip

\section{Sample Tree and Partial Dependence Analysis}\label{appendix:sample_tree}

In our case study, we used the Random Forest method to train predictive models and compute the importance of each predictor variable, summarized in Figure \ref{fig:cir_plot}. While variable importance scores indicate how much each predictor contributes to the model's predictions, they do not reveal the direction of the effect---whether it is positive or negative. To gain deeper insights into the relationships between predictors and the target variable, we can examine individual decision trees generated during the Random Forest construction and perform partial dependence analysis.

\subsection{Sample Decision Tree}

\begin{figure}[H]
\captionsetup{font=small}
    \centering
    \includegraphics[width=\linewidth]{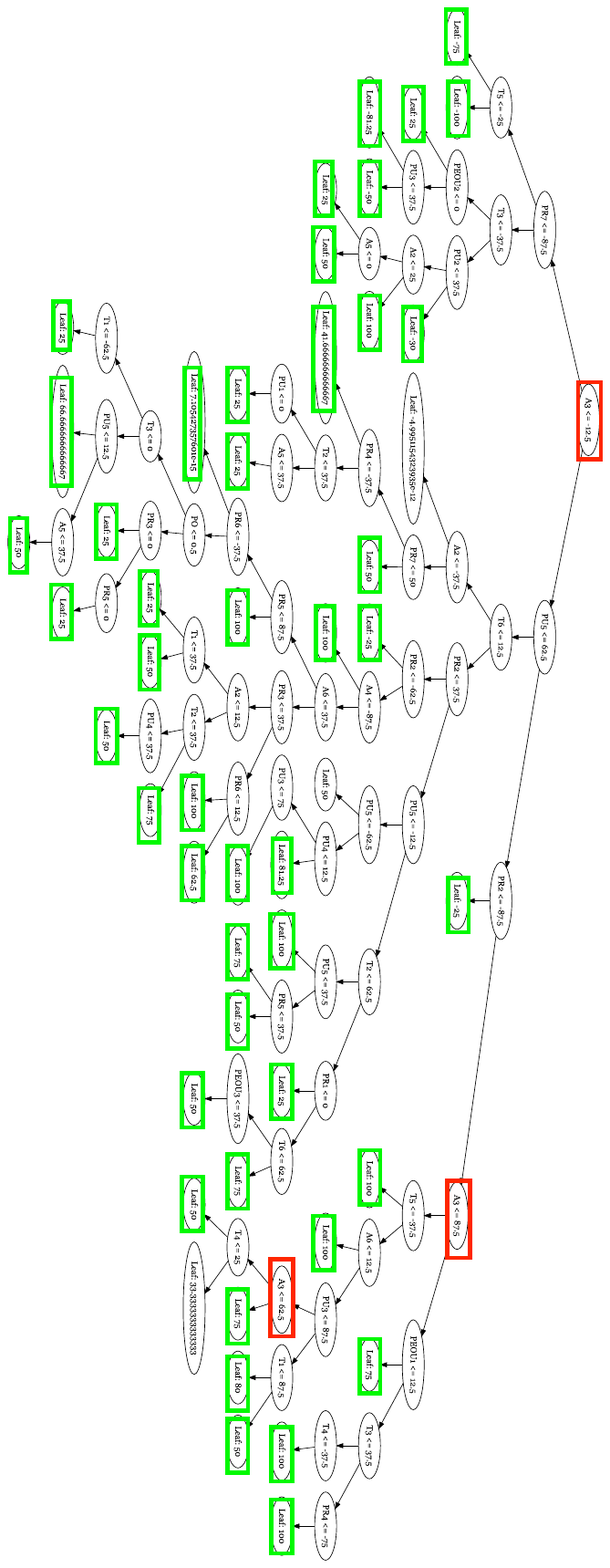} 
    \caption{A decision tree from the trained Random Forest model in predicting BI.}
    \label{fig:tree_example}
\end{figure}

Figure \ref{fig:tree_example} displays a sample decision tree extracted from the Random Forest model used to predict Behavioral Intention (BI) to use SAVs.

\begin{itemize}
    \item \textbf{Nodes}: Represent predictor variables (e.g., \textbf{A3}, \textbf{PR7}, \textbf{T5}). Each node includes a splitting rule based on the value of the predictor.

    \item \textbf{Branches}: Illustrate the decision paths based on these splitting rules, leading to different outcomes.

    \item \textbf{Leaves}: The terminal nodes (highlighted in green) show the predicted BI outcome at the end of each decision path.
\end{itemize}

To interpret this plot:

\begin{enumerate}
    \item First Split: The tree starts with A3 (``SAVs would be a good addition to towns and cities"). If A3 \(\leq\) -12.5, we follow the left branch.

    \item Second Split: Next, PR7 (``Malfunctions causing harm") is considered. If PR7 \(\leq\) -87.5, we continue left.

    \item Third Split: Then, T5 (``SAVs keeping user interests in mind") is evaluated. If T5 \(\leq\) -25, we reach a leaf node.

    \item Outcome: At this leaf, the predicted BI is -75, indicating a strong negative behavioral intention to use SAVs.
\end{enumerate}

This example illustrates that A3 is a critical decision point, appearing at the top of the tree and influencing subsequent splits. While Random Forest aggregates many such trees to enhance predictive accuracy, examining a single tree provides an interpretable example of key decision points and how specific predictors interact to influence the outcome.

\subsection{Partial Dependence Analysis}

We can also perform a partial dependence analysis to examine the marginal effect of a predictor variable on the outcome variable, averaging out the effects of other predictors. The `partialPlot' function from the `randomForest' package \citep{Package_randomForest} can be used to compute the partial dependence plots. The function being plotted is defined as Equation \ref{eq:partial_dependence}, where $x$ is the predictor for which partial dependence is sought (i.e., A3), and $x_{iC}$ is the other predictors in the data (e.g., A1, A2, PR1, PR2, etc.). 

\begin{equation}
    \tilde{f}(x) = \frac{1}{n} \sum_{i=1}^{n} f(x, x_{iC})
    \label{eq:partial_dependence}
\end{equation}

\begin{figure}
    \centering
    \includegraphics[width=1\linewidth]{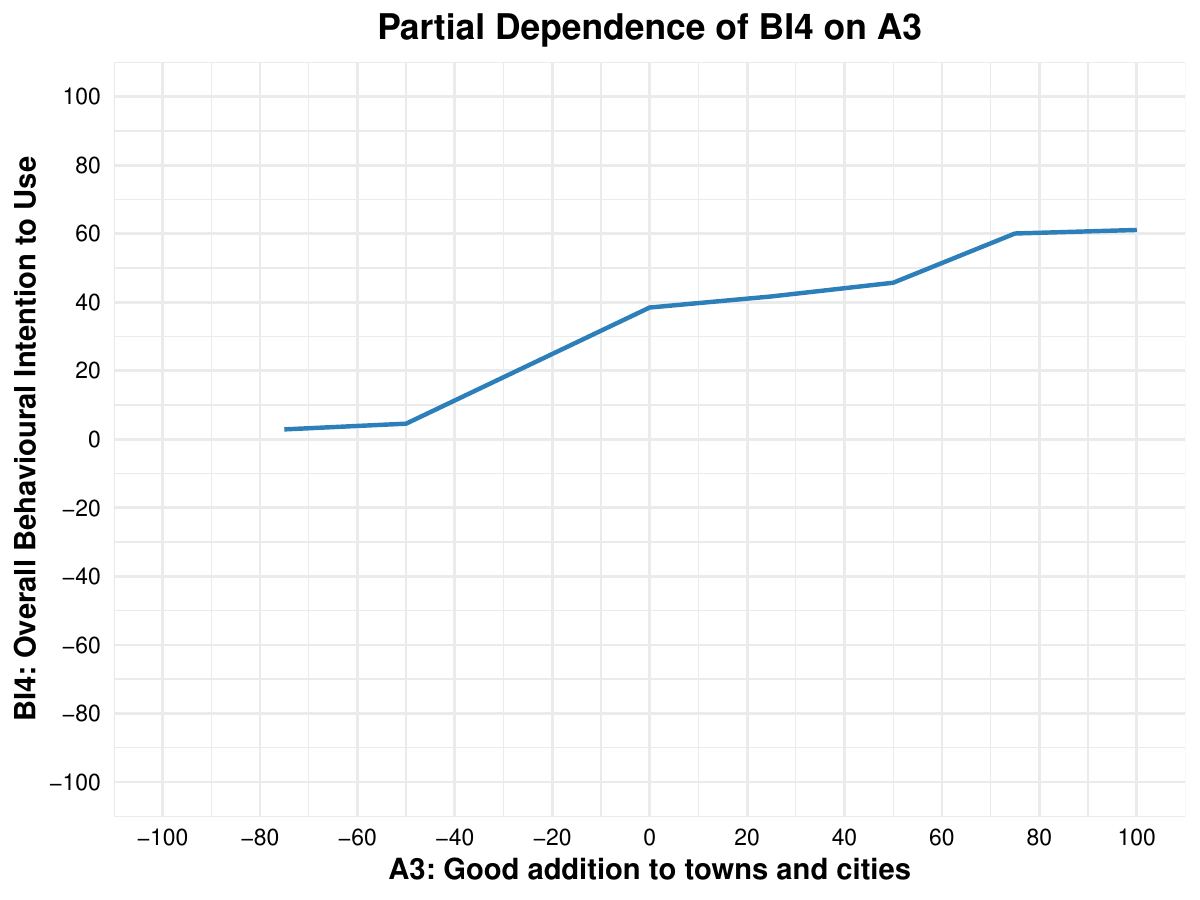}
    \caption{\textbf{Partial Dependence Plot of Behavioral Intention (BI4) on A3}: The graph shows the marginal effect of the perception of A3 (a good addition to towns and cities) on the overall behavioral intention to use (BI4).}
    \label{fig:Partial_Dependence_A3}
\end{figure}

The same example of A3 is used for demonstration. We generate a plot (Figure \ref{fig:Partial_Dependence_A3}) that shows the relationship between A3 and the predicted BI. The plot displays how changes in A3 (from its minimum to maximum observed values) affect the predicted BI, holding all other variables constant. A positive slope in the plot indicates that higher values of A3 are associated with higher predicted BI, suggesting a positive relationship. This analysis helps us understand whether a predictor has a positive or negative influence on the outcome and the nature of that relationship (linear, non-linear).

\section{Relative Importance}

This section presents the relative importance results from all the Random Forest models, providing both an overall view and detailed insights. 

Table \ref{table:factor_impor} summarizes the overall relative importance of each factor (construct) in predicting the target variables. This table highlights which constructs, such as Attitude (A), Perceived Risk (PR), or Trust (T), are most influential in predicting outcomes like Behavioral Intention (BI).

\begin{table}[H]
\centering
\caption{Overall Relative Importance of Each Factor in Predicting the Target Variables.}\label{table:factor_impor}
\begin{tabular}{llr}
  \hline
Predictor & Target Variable & Relative Importance \\ 
  \hline
Attitude & Behavioural Intention to Use & 61.94 \\ 
  Attitude & Perceived Usefulness & 43.81 \\ 
  Attitude & Perceived Ease of Use & 30.08 \\ 
  Attitude & Perceived Risk & 34.59 \\ 
  Attitude & Trust & 57.45 \\ 
  Perceived Ease of Use & Behavioural Intention to Use & 7.75 \\ 
  Perceived Ease of Use & Attitude & 18.05 \\ 
  Perceived Ease of Use & Perceived Usefulness & 31.24 \\ 
  Perceived Ease of Use & Perceived Risk & 12.30 \\ 
  Perceived Ease of Use & Trust & 9.75 \\ 
  Psychological Ownership & Behavioural Intention to Use & 0.32 \\ 
  Psychological Ownership & Attitude & 0.31 \\ 
  Psychological Ownership & Perceived Usefulness & 0.59 \\ 
  Psychological Ownership & Perceived Ease of Use & 1.01 \\ 
  Psychological Ownership & Perceived Risk & 0.96 \\ 
  Psychological Ownership & Trust & 0.59 \\ 
  Perceived Risk & Behavioural Intention to Use & 11.85 \\ 
  Perceived Risk & Attitude & 20.74 \\ 
  Perceived Risk & Perceived Usefulness & 11.44 \\ 
  Perceived Risk & Perceived Ease of Use & 24.51 \\ 
  Perceived Risk & Trust & 22.61 \\ 
  Perceived Usefulness & Behavioural Intention to Use & 9.54 \\ 
  Perceived Usefulness & Attitude & 25.94 \\ 
  Perceived Usefulness & Perceived Ease of Use & 22.98 \\ 
  Perceived Usefulness & Perceived Risk & 19.27 \\ 
  Perceived Usefulness & Trust & 9.60 \\ 
  Trust & Behavioural Intention to Use & 8.60 \\ 
  Trust & Attitude & 34.96 \\ 
  Trust & Perceived Usefulness & 12.92 \\ 
  Trust & Perceived Ease of Use & 21.42 \\ 
  Trust & Perceived Risk & 32.87 \\ 
   \hline
\end{tabular}
\end{table}

\bigskip

Table \ref{table:item_impor} provides a detailed breakdown of the relative importance of each individual factor item (specific survey questions) in predicting the target variables. 

\small
\begin{longtable}{llr}
\caption{Overall Relative Importance of Each Factor-items In Predicting the Target Variables.}\label{table:item_impor} \\
  \hline
Predictor & Target Variable & Relative Importance \\ 
  \hline
A1: Attractiveness & BI4: Overall behavioural intention to use & 6.89 \\ 
  A1: Attractiveness & PR8: Overall perceived risk & 5.33 \\ 
  A1: Attractiveness & T7: Overall trust & 4.09 \\ 
  A1: Attractiveness & PU7: Overall perceived usefulness & 3.68 \\ 
  A1: Attractiveness & PEOU4: Overall perceived ease of use & 3.23 \\ 
  A2: Stress level & T7: Overall trust & 19.04 \\ 
  A2: Stress level & PR8: Overall perceived risk & 7.86 \\ 
  A2: Stress level & PEOU4: Overall perceived ease of use & 7.52 \\ 
  A2: Stress level & BI4: Overall behavioural intention to use & 2.28 \\ 
  A2: Stress level & PU7: Overall perceived usefulness & 1.75 \\ 
  A3: Good addition to towns and cities & BI4: Overall behavioural intention to use & 27.44 \\ 
  A3: Good addition to towns and cities & T7: Overall trust & 15.22 \\ 
  A3: Good addition to towns and cities & PU7: Overall perceived usefulness & 12.84 \\ 
  A3: Good addition to towns and cities & PR8: Overall perceived risk & 8.35 \\ 
  A3: Good addition to towns and cities & PEOU4: Overall perceived ease of use & 5.84 \\ 
  A4: Enjoyment & PU7: Overall perceived usefulness & 9.71 \\ 
  A4: Enjoyment & BI4: Overall behavioural intention to use & 6.07 \\ 
  A4: Enjoyment & PEOU4: Overall perceived ease of use & 5.99 \\ 
  A4: Enjoyment & T7: Overall trust & 4.98 \\ 
  A4: Enjoyment & PR8: Overall perceived risk & 4.55 \\ 
  A5: Benefit & BI4: Overall behavioural intention to use & 15.41 \\ 
  A5: Benefit & T7: Overall trust & 12.33 \\ 
  A5: Benefit & PU7: Overall perceived usefulness & 11.09 \\ 
  A5: Benefit & PR8: Overall perceived risk & 5.62 \\ 
  A5: Benefit & PEOU4: Overall perceived ease of use & 3.88 \\ 
  A6: Changed travel behaviours & PU7: Overall perceived usefulness & 4.74 \\ 
  A6: Changed travel behaviours & BI4: Overall behavioural intention to use & 3.84 \\ 
  A6: Changed travel behaviours & PEOU4: Overall perceived ease of use & 3.61 \\ 
  A6: Changed travel behaviours & PR8: Overall perceived risk & 2.90 \\ 
  A6: Changed travel behaviours & T7: Overall trust & 1.78 \\ 
  PEOU1: Operation & PR8: Overall perceived risk & 5.50 \\ 
  PEOU1: Operation & T7: Overall trust & 2.93 \\ 
  PEOU1: Operation & PU7: Overall perceived usefulness & 2.85 \\ 
  PEOU1: Operation & A7: Overall attitude & 2.08 \\ 
  PEOU1: Operation & BI4: Overall behavioural intention to use & 1.55 \\ 
  PEOU2: Learning time & PR8: Overall perceived risk & 3.50 \\ 
  PEOU2: Learning time & T7: Overall trust & 3.08 \\ 
  PEOU2: Learning time & BI4: Overall behavioural intention to use & 2.95 \\ 
  PEOU2: Learning time & PU7: Overall perceived usefulness & 2.01 \\ 
  PEOU2: Learning time & A7: Overall attitude & 1.53 \\ 
  PEOU3: Convenience & PU7: Overall perceived usefulness & 26.38 \\ 
  PEOU3: Convenience & A7: Overall attitude & 14.43 \\ 
  PEOU3: Convenience & T7: Overall trust & 3.74 \\ 
  PEOU3: Convenience & PR8: Overall perceived risk & 3.30 \\ 
  PEOU3: Convenience & BI4: Overall behavioural intention to use & 3.25 \\ 
  PO: Psychological ownership & PEOU4: Overall perceived ease of use & 1.01 \\ 
  PO: Psychological ownership & PR8: Overall perceived risk & 0.96 \\ 
  PO: Psychological ownership & T7: Overall trust & 0.59 \\ 
  PO: Psychological ownership & PU7: Overall perceived usefulness & 0.59 \\ 
  PO: Psychological ownership & BI4: Overall behavioural intention to use & 0.32 \\ 
  PO: Psychological ownership & A7: Overall attitude & 0.31 \\ 
  PR1: Accident risk & A7: Overall attitude & 5.72 \\ 
  PR1: Accident risk & T7: Overall trust & 4.83 \\ 
  PR1: Accident risk & PEOU4: Overall perceived ease of use & 4.38 \\ 
  PR1: Accident risk & BI4: Overall behavioural intention to use & 3.12 \\ 
  PR1: Accident risk & PU7: Overall perceived usefulness & 1.97 \\ 
  PR2: Interactions with other road users & A7: Overall attitude & 7.78 \\ 
  PR2: Interactions with other road users & T7: Overall trust & 6.49 \\ 
  PR2: Interactions with other road users & PEOU4: Overall perceived ease of use & 4.23 \\ 
  PR2: Interactions with other road users & BI4: Overall behavioural intention to use & 2.03 \\ 
  PR2: Interactions with other road users & PU7: Overall perceived usefulness & 1.59 \\ 
  PR3: Disclosure of personal information & PEOU4: Overall perceived ease of use & 3.05 \\ 
  PR3: Disclosure of personal information & T7: Overall trust & 3.05 \\ 
  PR3: Disclosure of personal information & PU7: Overall perceived usefulness & 1.95 \\ 
  PR3: Disclosure of personal information & A7: Overall attitude & 1.82 \\ 
  PR3: Disclosure of personal information & BI4: Overall behavioural intention to use & 1.40 \\ 
  PR4: SAV equipment and system failures & PEOU4: Overall perceived ease of use & 3.55 \\ 
  PR4: SAV equipment and system failures & PU7: Overall perceived usefulness & 2.14 \\ 
  PR4: SAV equipment and system failures & T7: Overall trust & 1.64 \\ 
  PR4: SAV equipment and system failures & BI4: Overall behavioural intention to use & 1.48 \\ 
  PR4: SAV equipment and system failures & A7: Overall attitude & 1.35 \\ 
  PR5: Legal liability & PEOU4: Overall perceived ease of use & 3.20 \\ 
  PR5: Legal liability & T7: Overall trust & 1.67 \\ 
  PR5: Legal liability & A7: Overall attitude & 1.50 \\ 
  PR5: Legal liability & BI4: Overall behavioural intention to use & 1.06 \\ 
  PR5: Legal liability & PU7: Overall perceived usefulness & 1.05 \\ 
  PR6: Improper SAV functioning & PEOU4: Overall perceived ease of use & 2.96 \\ 
  PR6: Improper SAV functioning & T7: Overall trust & 2.72 \\ 
  PR6: Improper SAV functioning & PU7: Overall perceived usefulness & 1.49 \\ 
  PR6: Improper SAV functioning & BI4: Overall behavioural intention to use & 1.48 \\ 
  PR6: Improper SAV functioning & A7: Overall attitude & 1.37 \\ 
  PR7: Malfunctions causing harm & PEOU4: Overall perceived ease of use & 3.15 \\ 
  PR7: Malfunctions causing harm & T7: Overall trust & 2.22 \\ 
  PR7: Malfunctions causing harm & BI4: Overall behavioural intention to use & 1.27 \\ 
  PR7: Malfunctions causing harm & PU7: Overall perceived usefulness & 1.24 \\ 
  PR7: Malfunctions causing harm & A7: Overall attitude & 1.21 \\ 
  PU1: Transport needs & A7: Overall attitude & 9.54 \\ 
  PU1: Transport needs & PEOU4: Overall perceived ease of use & 4.46 \\ 
  PU1: Transport needs & PR8: Overall perceived risk & 4.01 \\ 
  PU1: Transport needs & T7: Overall trust & 2.26 \\ 
  PU1: Transport needs & BI4: Overall behavioural intention to use & 1.20 \\ 
  PU2: Daily routine & A7: Overall attitude & 3.94 \\ 
  PU2: Daily routine & PEOU4: Overall perceived ease of use & 3.62 \\ 
  PU2: Daily routine & PR8: Overall perceived risk & 3.61 \\ 
  PU2: Daily routine & T7: Overall trust & 1.85 \\ 
  PU2: Daily routine & BI4: Overall behavioural intention to use & 1.39 \\ 
  PU3: Impairment & PEOU4: Overall perceived ease of use & 4.27 \\ 
  PU3: Impairment & PR8: Overall perceived risk & 2.58 \\ 
  PU3: Impairment & BI4: Overall behavioural intention to use & 2.03 \\ 
  PU3: Impairment & A7: Overall attitude & 1.31 \\ 
  PU3: Impairment & T7: Overall trust & 1.02 \\ 
  PU4: Productivity & PEOU4: Overall perceived ease of use & 4.00 \\ 
  PU4: Productivity & PR8: Overall perceived risk & 3.00 \\ 
  PU4: Productivity & A7: Overall attitude & 2.88 \\ 
  PU4: Productivity & BI4: Overall behavioural intention to use & 1.96 \\ 
  PU4: Productivity & T7: Overall trust & 1.34 \\ 
  PU5: Work-life balance & A7: Overall attitude & 4.27 \\ 
  PU5: Work-life balance & PEOU4: Overall perceived ease of use & 3.06 \\ 
  PU5: Work-life balance & PR8: Overall perceived risk & 2.90 \\ 
  PU5: Work-life balance & BI4: Overall behavioural intention to use & 1.46 \\ 
  PU5: Work-life balance & T7: Overall trust & 1.27 \\ 
  PU6: Travel cost & A7: Overall attitude & 4.00 \\ 
  PU6: Travel cost & PEOU4: Overall perceived ease of use & 3.57 \\ 
  PU6: Travel cost & PR8: Overall perceived risk & 3.18 \\ 
  PU6: Travel cost & T7: Overall trust & 1.87 \\ 
  PU6: Travel cost & BI4: Overall behavioural intention to use & 1.49 \\ 
  T1: Government regulation & A7: Overall attitude & 3.92 \\ 
  T1: Government regulation & PR8: Overall perceived risk & 3.26 \\ 
  T1: Government regulation & PEOU4: Overall perceived ease of use & 3.24 \\ 
  T1: Government regulation & PU7: Overall perceived usefulness & 1.32 \\ 
  T1: Government regulation & BI4: Overall behavioural intention to use & 1.12 \\ 
  T2: Traditional automakers manufacturing SAVs & PR8: Overall perceived risk & 5.34 \\ 
  T2: Traditional automakers manufacturing SAVs & PEOU4: Overall perceived ease of use & 2.90 \\ 
  T2: Traditional automakers manufacturing SAVs & A7: Overall attitude & 2.83 \\ 
  T2: Traditional automakers manufacturing SAVs & PU7: Overall perceived usefulness & 1.55 \\ 
  T2: Traditional automakers manufacturing SAVs & BI4: Overall behavioural intention to use & 1.53 \\ 
  T3: IT companies manufacturing SAVs & PR8: Overall perceived risk & 5.02 \\ 
  T3: IT companies manufacturing SAVs & PEOU4: Overall perceived ease of use & 3.27 \\ 
  T3: IT companies manufacturing SAVs & A7: Overall attitude & 2.77 \\ 
  T3: IT companies manufacturing SAVs & PU7: Overall perceived usefulness & 1.48 \\ 
  T3: IT companies manufacturing SAVs & BI4: Overall behavioural intention to use & 1.10 \\ 
  T4: IT companies developing SAV apps & PU7: Overall perceived usefulness & 5.07 \\ 
  T4: IT companies developing SAV apps & A7: Overall attitude & 4.65 \\ 
  T4: IT companies developing SAV apps & PR8: Overall perceived risk & 4.21 \\ 
  T4: IT companies developing SAV apps & PEOU4: Overall perceived ease of use & 2.76 \\ 
  T4: IT companies developing SAV apps & BI4: Overall behavioural intention to use & 1.25 \\ 
  T5: SAVs keeping user interests in mind & PR8: Overall perceived risk & 4.67 \\ 
  T5: SAVs keeping user interests in mind & A7: Overall attitude & 3.66 \\ 
  T5: SAVs keeping user interests in mind & PEOU4: Overall perceived ease of use & 3.61 \\ 
  T5: SAVs keeping user interests in mind & PU7: Overall perceived usefulness & 1.22 \\ 
  T5: SAVs keeping user interests in mind & BI4: Overall behavioural intention to use & 1.22 \\ 
  T6: Reliability & A7: Overall attitude & 17.14 \\ 
  T6: Reliability & PR8: Overall perceived risk & 10.37 \\ 
  T6: Reliability & PEOU4: Overall perceived ease of use & 5.63 \\ 
  T6: Reliability & BI4: Overall behavioural intention to use & 2.39 \\ 
  T6: Reliability & PU7: Overall perceived usefulness & 2.28 \\ 
   \hline
\end{longtable}
\normalsize

\begin{figure}[H]
    \centering
    \includegraphics[width=1\textwidth]{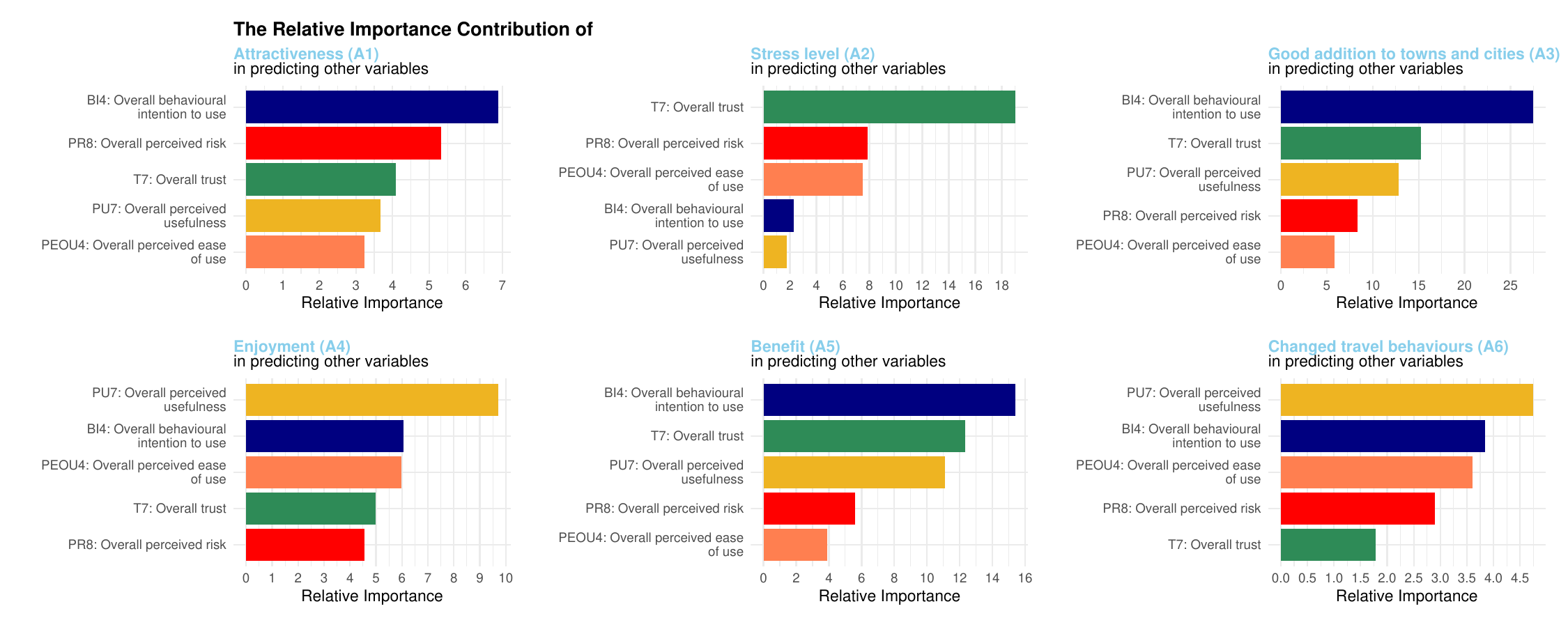}
    \caption{The relative importance contribution of Attitude items in predicting target variables, highlighting key predictors in behavioral intention, perceived usefulness, and other psychological factors.}
    \label{fig:attitude_importance}
\end{figure}

\begin{figure}[H]
    \centering
    \includegraphics[width=1\textwidth]{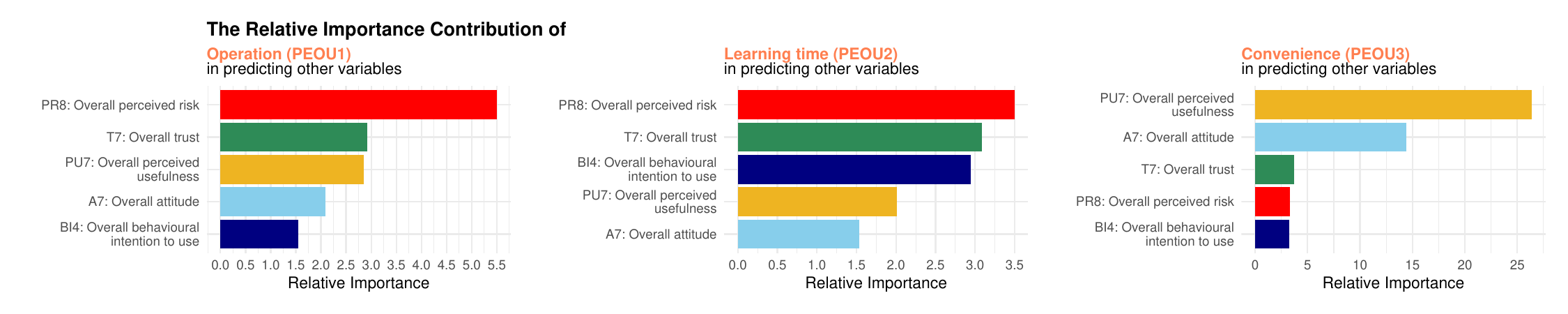}
    \caption{The relative importance contribution of Perceived Ease of Use (PEOU) items in predicting target variables, including behavioral intention and related psychological constructs.
    \label{fig:peou_importance}}
\end{figure}

\begin{figure}[H]
    \centering
    \includegraphics[width=1\textwidth]{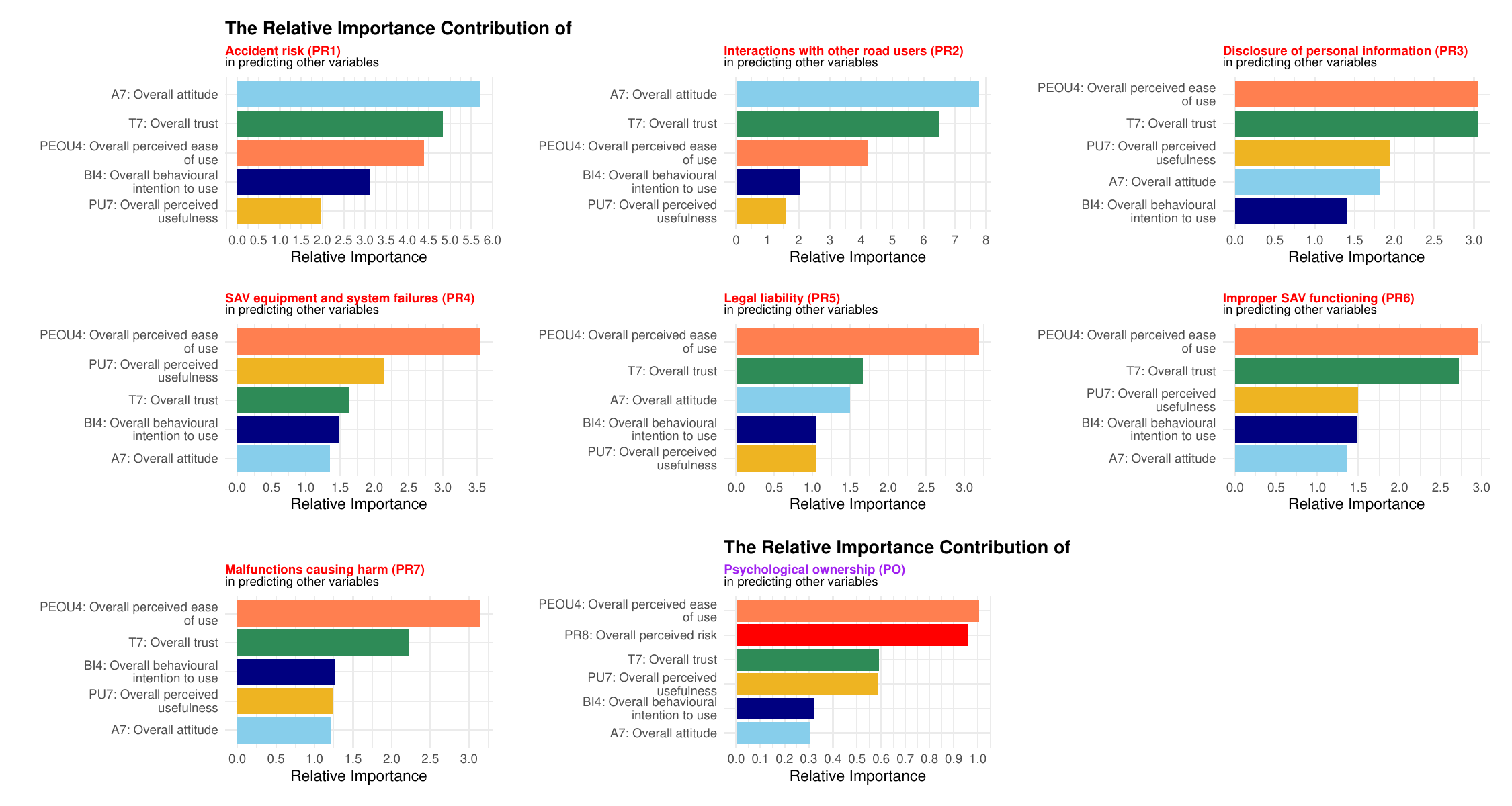}
    \caption{The relative importance contribution of Perceived Risk (PR) and Psychological Ownership (PO) items in predicting target variables, including behavioral intention and related constructs.}
    \label{fig:pr_po_importance}
\end{figure}

\begin{figure}[H]
    \centering
    \includegraphics[width=1\textwidth]{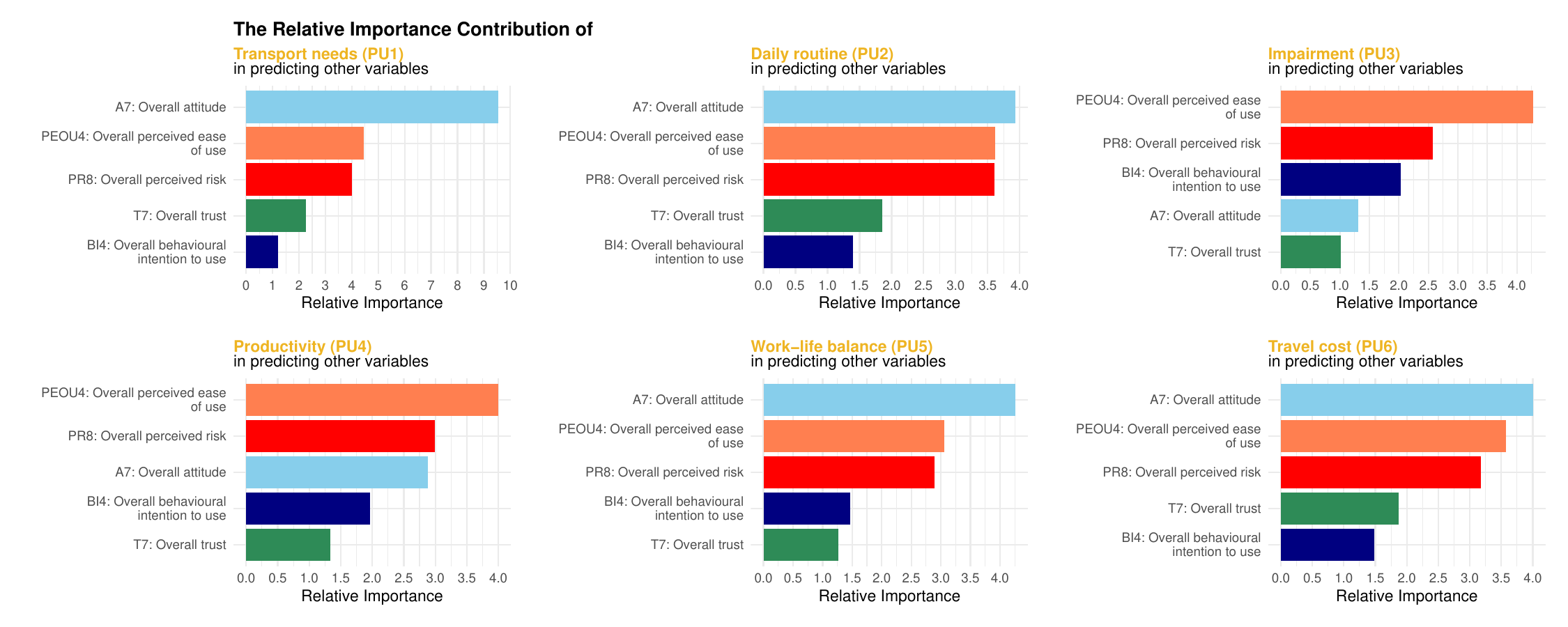}
    \caption{The relative importance contribution of Perceived Usefulness (PU) items in predicting target variables across key psychological factors such as attitude, behavioral intention, and trust.}
    \label{fig:pu_importance}
\end{figure}

\begin{figure}[H]
    \centering
    \includegraphics[width=1\textwidth]{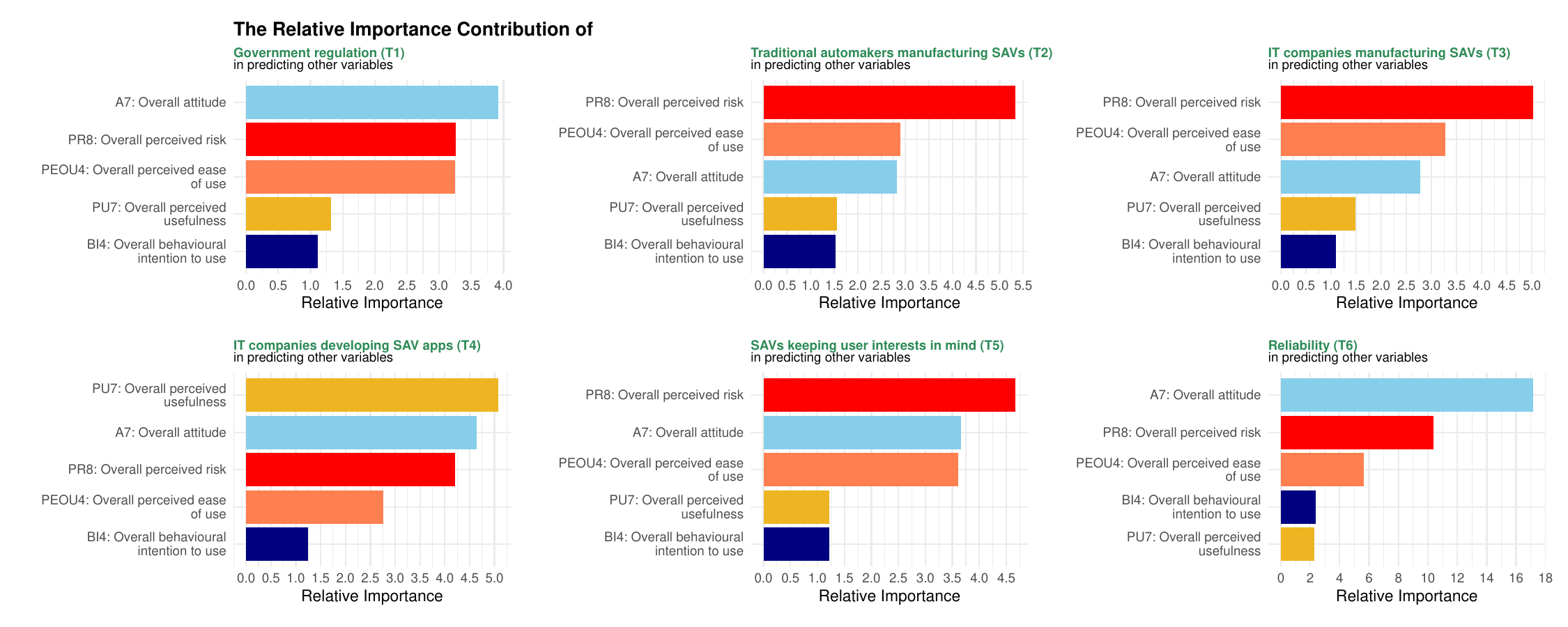}
    \caption{The relative importance contribution of Trust items in predicting target variables, focusing on behavioral intention, perceived risk, and related constructs.}
    \label{fig:trust_importance}
\end{figure}

To further illustrate the contributions of each predictor, Figure \ref{fig:imp_vip_all} displays the predictor importance plots, showing the relative importance of \textbf{all} predictors in predicting \textbf{each} target variable. This helps identify the most significant predictors for a specific outcome Figures \ref{fig:attitude_importance}--\ref{fig:trust_importance} present the relative importance contributions of \textbf{individual} predictors in predicting \textbf{all} target variables. Each figure focuses on one key predictor (e.g., A1) and shows its relative importance across all target variables (e.g., BI4, PR8, T7, PU7, PEOU7). This allows us to see how a single predictor influences multiple outcomes.

\newpage
{\footnotesize\bibliographystyle{elsarticle-harv}
\bibliography{reference}}

\begin{thebibliography}{74}
\expandafter\ifx\csname natexlab\endcsname\relax\def\natexlab#1{#1}\fi
\providecommand{\url}[1]{\texttt{#1}}
\providecommand{\href}[2]{#2}
\providecommand{\path}[1]{#1}
\providecommand{\DOIprefix}{doi:}
\providecommand{\ArXivprefix}{arXiv:}
\providecommand{\URLprefix}{URL: }
\providecommand{\Pubmedprefix}{pmid:}
\providecommand{\doi}[1]{\href{http://dx.doi.org/#1}{\path{#1}}}
\providecommand{\Pubmed}[1]{\href{pmid:#1}{\path{#1}}}
\providecommand{\bibinfo}[2]{#2}
\ifx\xfnm\relax \def\xfnm[#1]{\unskip,\space#1}\fi
\bibitem[{Abel(2022)}]{website-chord}
\bibinfo{author}{Abel, G.}, \bibinfo{year}{2022}.
\newblock \bibinfo{title}{Chord diagram for directional origin-destination data}.
\newblock \bibinfo{howpublished}{Available at: \url{https://guyabel.github.io/migest/reference/mig_chord.html}}.
\newblock \bibinfo{note}{Accessed: May 20, 2023}.
\bibitem[{Abel and Sander(2014)}]{lirui-093}
\bibinfo{author}{Abel, G.J.}, \bibinfo{author}{Sander, N.}, \bibinfo{year}{2014}.
\newblock \bibinfo{title}{Quantifying global international migration flows}.
\newblock \bibinfo{journal}{Science} \bibinfo{volume}{343}, \bibinfo{pages}{1520--1522}.
\newblock \URLprefix \url{https://www.science.org/doi/abs/10.1126/science.1248676}, \DOIprefix\doi{10.1126/science.1248676}, \href{http://arxiv.org/abs/https://www.science.org/doi/pdf/10.1126/science.1248676}{{\tt arXiv:https://www.science.org/doi/pdf/10.1126/science.1248676}}.
\bibitem[{Aghelpour et~al.(2021)Aghelpour, Singh and Varshavian}]{lirui-095}
\bibinfo{author}{Aghelpour, P.}, \bibinfo{author}{Singh, V.}, \bibinfo{author}{Varshavian, V.}, \bibinfo{year}{2021}.
\newblock \bibinfo{title}{Time series prediction of seasonal precipitation in iran, using data-driven models: a comparison under different climatic conditions}.
\newblock \bibinfo{journal}{Arabian Journal of Geosciences} \bibinfo{volume}{14}.
\newblock \DOIprefix\doi{10.1007/s12517-021-06910-0}.
\bibitem[{Ahmed et~al.(2022)Ahmed, Iqbal, Karyotis, Palade and Amin}]{lirui-094}
\bibinfo{author}{Ahmed, M.L.}, \bibinfo{author}{Iqbal, R.}, \bibinfo{author}{Karyotis, C.}, \bibinfo{author}{Palade, V.}, \bibinfo{author}{Amin, S.A.}, \bibinfo{year}{2022}.
\newblock \bibinfo{title}{Predicting the public adoption of connected and autonomous vehicles}.
\newblock \bibinfo{journal}{IEEE Transactions on Intelligent Transportation Systems} \bibinfo{volume}{23}, \bibinfo{pages}{1680--1688}.
\newblock \DOIprefix\doi{10.1109/TITS.2021.3109846}.
\bibitem[{Ajzen(1991)}]{lirui-084}
\bibinfo{author}{Ajzen, I.}, \bibinfo{year}{1991}.
\newblock \bibinfo{title}{The theory of planned behavior}.
\newblock \bibinfo{journal}{Organizational Behavior and Human Decision Processes} \bibinfo{volume}{50}, \bibinfo{pages}{179--211}.
\newblock \URLprefix \url{https://www.sciencedirect.com/science/article/pii/074959789190020T}, \DOIprefix\doi{https://doi.org/10.1016/0749-5978(91)90020-T}. \bibinfo{note}{theories of Cognitive Self-Regulation}.
\bibitem[{Alwabel and Zeng(2021)}]{Alwabel_data-driven_modeling_2021}
\bibinfo{author}{Alwabel, A.S.A.}, \bibinfo{author}{Zeng, X.J.}, \bibinfo{year}{2021}.
\newblock \bibinfo{title}{Data-driven modeling of technology acceptance: A machine learning perspective}.
\newblock \bibinfo{journal}{Expert Systems with Applications} \bibinfo{volume}{185}, \bibinfo{pages}{115584}.
\newblock \DOIprefix\doi{10.1016/j.eswa.2021.115584}.
\bibitem[{Barbour et~al.(2019)Barbour, Menon, Zhang and Mannering}]{lirui-082}
\bibinfo{author}{Barbour, N.}, \bibinfo{author}{Menon, N.}, \bibinfo{author}{Zhang, Y.}, \bibinfo{author}{Mannering, F.}, \bibinfo{year}{2019}.
\newblock \bibinfo{title}{Shared automated vehicles: A statistical analysis of consumer use likelihoods and concerns}.
\newblock \bibinfo{journal}{Transport Policy} \bibinfo{volume}{80}, \bibinfo{pages}{86--93}.
\newblock \URLprefix \url{https://www.sciencedirect.com/science/article/pii/S0967070X18306917}, \DOIprefix\doi{https://doi.org/10.1016/j.tranpol.2019.05.013}.
\bibitem[{Baxter et~al.(2015)Baxter, Aurisicchio and Childs}]{lirui-040}
\bibinfo{author}{Baxter, W.L.}, \bibinfo{author}{Aurisicchio, M.}, \bibinfo{author}{Childs, P.R.}, \bibinfo{year}{2015}.
\newblock \bibinfo{title}{A psychological ownership approach to designing object attachment}.
\newblock \bibinfo{journal}{Journal of Engineering Design} \bibinfo{volume}{26}, \bibinfo{pages}{140--156}.
\newblock \URLprefix \url{https://doi.org/10.1080/09544828.2015.1030371}, \DOIprefix\doi{10.1080/09544828.2015.1030371}, \href{http://arxiv.org/abs/https://doi.org/10.1080/09544828.2015.1030371}{{\tt arXiv:https://doi.org/10.1080/09544828.2015.1030371}}.
\bibitem[{Becker and Axhausen(2017)}]{lirui-078}
\bibinfo{author}{Becker, F.}, \bibinfo{author}{Axhausen, K.W.}, \bibinfo{year}{2017}.
\newblock \bibinfo{title}{Literature review on surveys investigating the acceptance of automated vehicles}.
\newblock \bibinfo{journal}{Transportation} \bibinfo{volume}{44}, \bibinfo{pages}{1293–1306}.
\newblock \DOIprefix\doi{https://doi.org/10.1007/s11116-017-9808-9}.
\bibitem[{Biau(2012)}]{biau_2012_analysis}
\bibinfo{author}{Biau, G.}, \bibinfo{year}{2012}.
\newblock \bibinfo{title}{Analysis of a random forests model}.
\newblock \bibinfo{journal}{The Journal of Machine Learning Research} \bibinfo{volume}{13}, \bibinfo{pages}{1063--1095}.
\bibitem[{Breiman(2001a)}]{breiman_random_2001}
\bibinfo{author}{Breiman, L.}, \bibinfo{year}{2001}a.
\newblock \bibinfo{title}{Random forests}.
\newblock \bibinfo{journal}{Machine Learning} \bibinfo{volume}{45}, \bibinfo{pages}{5--32}.
\newblock \URLprefix \url{https://doi.org/10.1023/A:1010933404324}, \DOIprefix\doi{10.1023/A:1010933404324}.
\bibitem[{Breiman(2001b)}]{breimanStatisticalModelingTwo2001}
\bibinfo{author}{Breiman, L.}, \bibinfo{year}{2001}b.
\newblock \bibinfo{title}{Statistical {{Modeling}}: {{The Two Cultures}} (with comments and a rejoinder by the author)}.
\newblock \bibinfo{journal}{Statistical Science} \bibinfo{volume}{16}, \bibinfo{pages}{199--231}.
\newblock \URLprefix \url{https://projecteuclid.org/journals/statistical-science/volume-16/issue-3/Statistical-Modeling--The-Two-Cultures-with-comments-and-a/10.1214/ss/1009213726.full}, \DOIprefix\doi{10.1214/ss/1009213726}.
\bibitem[{Breiman et~al.(2022)Breiman, Cutler, Liaw and Wiener}]{Package_randomForest}
\bibinfo{author}{Breiman, L.}, \bibinfo{author}{Cutler, A.}, \bibinfo{author}{Liaw, A.}, \bibinfo{author}{Wiener, M.}, \bibinfo{year}{2022}.
\newblock \bibinfo{title}{Breiman and cutlers random forests for classification and regression}.
\newblock \bibinfo{howpublished}{Comprehensive R Archive Network}.
\bibitem[{Choi and Ji(2015)}]{lirui-004}
\bibinfo{author}{Choi, J.K.}, \bibinfo{author}{Ji, Y.G.}, \bibinfo{year}{2015}.
\newblock \bibinfo{title}{Investigating the importance of trust on adopting an autonomous vehicle}.
\newblock \bibinfo{journal}{International Journal of Human–Computer Interaction} \bibinfo{volume}{31}, \bibinfo{pages}{692--702}.
\newblock \URLprefix \url{https://doi.org/10.1080/10447318.2015.1070549}, \DOIprefix\doi{10.1080/10447318.2015.1070549}, \href{http://arxiv.org/abs/https://doi.org/10.1080/10447318.2015.1070549}{{\tt arXiv:https://doi.org/10.1080/10447318.2015.1070549}}.
\bibitem[{Cunningham et~al.(2019a)Cunningham, Regan, Horberry, Weeratunga and Dixit}]{lirui-017}
\bibinfo{author}{Cunningham, M.L.}, \bibinfo{author}{Regan, M.A.}, \bibinfo{author}{Horberry, T.}, \bibinfo{author}{Weeratunga, K.}, \bibinfo{author}{Dixit, V.}, \bibinfo{year}{2019}a.
\newblock \bibinfo{title}{Public opinion about automated vehicles in australia: Results from a large-scale national survey}.
\newblock \bibinfo{journal}{Transportation Research Part A: Policy and Practice} \bibinfo{volume}{129}, \bibinfo{pages}{1--18}.
\newblock \URLprefix \url{https://www.sciencedirect.com/science/article/pii/S0965856419302745}, \DOIprefix\doi{https://doi.org/10.1016/j.tra.2019.08.002}.
\bibitem[{Cunningham et~al.(2019b)Cunningham, Regan, Horberry, Weeratunga and Dixit}]{lirui-074}
\bibinfo{author}{Cunningham, M.L.}, \bibinfo{author}{Regan, M.A.}, \bibinfo{author}{Horberry, T.}, \bibinfo{author}{Weeratunga, K.}, \bibinfo{author}{Dixit, V.}, \bibinfo{year}{2019}b.
\newblock \bibinfo{title}{Public opinion about automated vehicles in australia: Results from a large-scale national survey}.
\newblock \bibinfo{journal}{Transportation Research Part A: Policy and Practice} \bibinfo{volume}{129}, \bibinfo{pages}{1--18}.
\newblock \URLprefix \url{https://www.sciencedirect.com/science/article/pii/S0965856419302745}, \DOIprefix\doi{https://doi.org/10.1016/j.tra.2019.08.002}.
\bibitem[{Dai et~al.(2021)Dai, Li, Liu and Lin}]{lirui-037}
\bibinfo{author}{Dai, J.}, \bibinfo{author}{Li, R.}, \bibinfo{author}{Liu, Z.}, \bibinfo{author}{Lin, S.}, \bibinfo{year}{2021}.
\newblock \bibinfo{title}{Impacts of the introduction of autonomous taxi on travel behaviors of the experienced user: Evidence from a one-year paid taxi service in guangzhou, china}.
\newblock \bibinfo{journal}{Transportation Research Part C: Emerging Technologies} \bibinfo{volume}{130}, \bibinfo{pages}{103311}.
\newblock \URLprefix \url{https://www.sciencedirect.com/science/article/pii/S0968090X21003193}, \DOIprefix\doi{https://doi.org/10.1016/j.trc.2021.103311}.
\bibitem[{Davis and Davis(1989)}]{lirui-053}
\bibinfo{author}{Davis, F.}, \bibinfo{author}{Davis, F.}, \bibinfo{year}{1989}.
\newblock \bibinfo{title}{Perceived usefulness, perceived ease of use, and user acceptance of information technology}.
\newblock \bibinfo{journal}{MIS Quarterly} \bibinfo{volume}{13}, \bibinfo{pages}{319--}.
\newblock \DOIprefix\doi{10.2307/249008}.
\bibitem[{Davis and Venkatesh(1996)}]{Davis_A-critical-assessment_1996}
\bibinfo{author}{Davis, F.D.}, \bibinfo{author}{Venkatesh, V.}, \bibinfo{year}{1996}.
\newblock \bibinfo{title}{A critical assessment of potential measurement biases in the technology acceptance model: three experiments}.
\newblock \bibinfo{journal}{International Journal of Human-Computer Studies} \bibinfo{volume}{45}, \bibinfo{pages}{19--45}.
\newblock \DOIprefix\doi{https://doi.org/10.1006/ijhc.1996.0040}.
\bibitem[{Dichabeng et~al.(2021)Dichabeng, Merat and Markkula}]{Dichabeng_Factors_2021}
\bibinfo{author}{Dichabeng, P.}, \bibinfo{author}{Merat, N.}, \bibinfo{author}{Markkula, G.}, \bibinfo{year}{2021}.
\newblock \bibinfo{title}{Factors that influence the acceptance of future shared automated vehicles – a focus group study with united kingdom drivers}.
\newblock \bibinfo{journal}{Transportation Research Part F: Traffic Psychology and Behaviour} \bibinfo{volume}{82}, \bibinfo{pages}{121--140}.
\newblock \URLprefix \url{https://www.sciencedirect.com/science/article/pii/S1369847821001856}, \DOIprefix\doi{https://doi.org/10.1016/j.trf.2021.08.009}.
\bibitem[{Ellis(2022)}]{website-mtry}
\bibinfo{author}{Ellis, C.}, \bibinfo{year}{2022}.
\newblock \bibinfo{title}{Mtry in random forests}.
\newblock \URLprefix \url{https://crunchingthedata.com/mtry-in-random-forests/#:~:text=The%20mtry%20parameter%20does%20exactly,any%20given%20point%20in%20time.}
\bibitem[{Ernst and Reinelt(2017)}]{lirui-033}
\bibinfo{author}{Ernst, C.P.H.}, \bibinfo{author}{Reinelt, P.}, \bibinfo{year}{2017}.
\newblock \bibinfo{title}{Autonomous car acceptance: Safety vs. personal driving enjoyment}, in: \bibinfo{booktitle}{Americas Conference on Information Systems}, pp. \bibinfo{pages}{1--8}.
\bibitem[{Finnegan et~al.(2019)Finnegan, Sao and Huchko}]{Finnegan_using_2019}
\bibinfo{author}{Finnegan, A.}, \bibinfo{author}{Sao, S.S.}, \bibinfo{author}{Huchko, M.J.}, \bibinfo{year}{2019}.
\newblock \bibinfo{title}{Using a chord diagram to visualize dynamics in contraceptive use: Bringing data into practice}.
\newblock \bibinfo{journal}{Global Health: Science and Practice} \bibinfo{volume}{7}, \bibinfo{pages}{598--605}.
\newblock \URLprefix \url{https://www.ghspjournal.org/content/7/4/598}, \DOIprefix\doi{10.9745/GHSP-D-19-00205}, \href{http://arxiv.org/abs/https://www.ghspjournal.org/content/7/4/598.full.pdf}{{\tt arXiv:https://www.ghspjournal.org/content/7/4/598.full.pdf}}.
\bibitem[{Gkartzonikas(2020)}]{Christos_thesis_2020}
\bibinfo{author}{Gkartzonikas, C.}, \bibinfo{year}{2020}.
\newblock \bibinfo{title}{A Stated Preference Study for Assessing Public Acceptance Towards Autonomous Vehicles}.
\newblock Ph.D. thesis. Purdue University Graduate School.
\newblock \DOIprefix\doi{https://doi.org/10.25394/PGS.12210293.v1}.
\bibitem[{Greenwell and Boehmke(2023)}]{greenwell_package_2023}
\bibinfo{author}{Greenwell, B.M.}, \bibinfo{author}{Boehmke, B.C.}, \bibinfo{year}{2023}.
\newblock \bibinfo{title}{Variable {{Importance Plots}}}.
\bibitem[{Hair et~al.(2010)Hair, Black, Babin and Anderson}]{mda_2010}
\bibinfo{author}{Hair, J.}, \bibinfo{author}{Black, W.}, \bibinfo{author}{Babin, B.}, \bibinfo{author}{Anderson, R.}, \bibinfo{year}{2010}.
\newblock \bibinfo{title}{Multivariate Data Analysis: A Global Perspective}.
\bibitem[{Hair et~al.(2021)Hair, Hult, Ringle, Sarstedt, Danks and Ray}]{PLS-SEM_R_book}
\bibinfo{author}{Hair, J.}, \bibinfo{author}{Hult, G.T.M.}, \bibinfo{author}{Ringle, C.}, \bibinfo{author}{Sarstedt, M.}, \bibinfo{author}{Danks, N.}, \bibinfo{author}{Ray, S.}, \bibinfo{year}{2021}.
\newblock \bibinfo{title}{Partial Least Squares Structural Equation Modeling (PLS-SEM) Using R: A workbook}.
\newblock \DOIprefix\doi{10.1007/978-3-030-80519-7}.
\bibitem[{Iarossi(2006)}]{lirui-042}
\bibinfo{author}{Iarossi, G.}, \bibinfo{year}{2006}.
\newblock \bibinfo{title}{Power of Survey Design : A User's Guide for Managing Surveys, Interpreting Results, and Influencing Respondents}.
\newblock \bibinfo{publisher}{World Bank Publications}.
\newblock \DOIprefix\doi{10.1596/978-0-8213-6392-8}.
\bibitem[{Jami et~al.(2020)Jami, Kouchaki and Gino}]{lirui-064}
\bibinfo{author}{Jami, A.}, \bibinfo{author}{Kouchaki, M.}, \bibinfo{author}{Gino, F.}, \bibinfo{year}{2020}.
\newblock \bibinfo{title}{{I Own, So I Help Out: How Psychological Ownership Increases Prosocial Behavior}}.
\newblock \bibinfo{journal}{Journal of Consumer Research} \bibinfo{volume}{47}, \bibinfo{pages}{698--715}.
\newblock \URLprefix \url{https://doi.org/10.1093/jcr/ucaa040}, \DOIprefix\doi{10.1093/jcr/ucaa040}, \href{http://arxiv.org/abs/https://academic.oup.com/jcr/article-pdf/47/5/698/36105780/ucaa040.pdf}{{\tt arXiv:https://academic.oup.com/jcr/article-pdf/47/5/698/36105780/ucaa040.pdf}}.
\bibitem[{Janatabadi and Ermagun(2022)}]{lirui-083}
\bibinfo{author}{Janatabadi, F.}, \bibinfo{author}{Ermagun, A.}, \bibinfo{year}{2022}.
\newblock \bibinfo{title}{Empirical evidence of bias in public acceptance of autonomous vehicles}.
\newblock \bibinfo{journal}{Transportation Research Part F: Traffic Psychology and Behaviour} \bibinfo{volume}{84}, \bibinfo{pages}{330--347}.
\newblock \URLprefix \url{https://www.sciencedirect.com/science/article/pii/S1369847821002837}, \DOIprefix\doi{https://doi.org/10.1016/j.trf.2021.12.005}.
\bibitem[{Javid and Al-Hashimi(2020)}]{lirui-031}
\bibinfo{author}{Javid, M.A.}, \bibinfo{author}{Al-Hashimi, A.R.}, \bibinfo{year}{2020}.
\newblock \bibinfo{title}{Significance of attitudes, passion and cultural factors in driver’s speeding behavior in oman: application of theory of planned behavior}.
\newblock \bibinfo{journal}{International Journal of Injury Control and Safety Promotion} \bibinfo{volume}{27}, \bibinfo{pages}{172--180}.
\newblock \URLprefix \url{https://doi.org/10.1080/17457300.2019.1695632}, \DOIprefix\doi{10.1080/17457300.2019.1695632}, \href{http://arxiv.org/abs/https://doi.org/10.1080/17457300.2019.1695632}{{\tt arXiv:https://doi.org/10.1080/17457300.2019.1695632}}. \bibinfo{note}{pMID: 31790324}.
\bibitem[{Kenesei et~al.(2022)Kenesei, Ásványi, Kökény, Jászberényi, Miskolczi, Gyulavári and Syahrivar}]{lirui-047}
\bibinfo{author}{Kenesei, Z.}, \bibinfo{author}{Ásványi, K.}, \bibinfo{author}{Kökény, L.}, \bibinfo{author}{Jászberényi, M.}, \bibinfo{author}{Miskolczi, M.}, \bibinfo{author}{Gyulavári, T.}, \bibinfo{author}{Syahrivar, J.}, \bibinfo{year}{2022}.
\newblock \bibinfo{title}{Trust and perceived risk: How different manifestations affect the adoption of autonomous vehicles}.
\newblock \bibinfo{journal}{Transportation Research Part A: Policy and Practice} \bibinfo{volume}{164}, \bibinfo{pages}{379--393}.
\newblock \URLprefix \url{https://www.sciencedirect.com/science/article/pii/S0965856422002269}, \DOIprefix\doi{https://doi.org/10.1016/j.tra.2022.08.022}.
\bibitem[{Kirk and Swain(2018)}]{POchapter5}
\bibinfo{author}{Kirk, C.P.}, \bibinfo{author}{Swain, S.D.}, \bibinfo{year}{2018}.
\newblock \bibinfo{title}{Consumer Psychological Ownership of Digital Technology}. \bibinfo{publisher}{Springer International Publishing}, \bibinfo{address}{Cham}. chapter~\bibinfo{chapter}{5}.
\newblock pp. \bibinfo{pages}{69--90}.
\newblock \URLprefix \url{https://doi.org/10.1007/978-3-319-77158-8_5}, \DOIprefix\doi{10.1007/978-3-319-77158-8_5}.
\bibitem[{Kline and St(2022)}]{Principles_and_Practice_of_SEM_2022}
\bibinfo{author}{Kline, R.}, \bibinfo{author}{St, C.}, \bibinfo{year}{2022}.
\newblock \bibinfo{title}{Principles and Practice of Structural Equation Modeling}.
\bibitem[{Krueger et~al.(2016)Krueger, Rashidi and Rose}]{lirui-025}
\bibinfo{author}{Krueger, R.}, \bibinfo{author}{Rashidi, T.H.}, \bibinfo{author}{Rose, J.M.}, \bibinfo{year}{2016}.
\newblock \bibinfo{title}{Preferences for shared autonomous vehicles}.
\newblock \bibinfo{journal}{Transportation Research Part C: Emerging Technologies} \bibinfo{volume}{69}, \bibinfo{pages}{343--355}.
\newblock \URLprefix \url{https://www.sciencedirect.com/science/article/pii/S0968090X16300870}, \DOIprefix\doi{https://doi.org/10.1016/j.trc.2016.06.015}.
\bibitem[{Kuhn et~al.(2023)Kuhn, Wing, Weston, Williams, Keefer, Engelhardt, Cooper, Mayer, Kenkel, Team, Benesty, Lescarbeau, Ziem, Scrucca, Tang, Candan and Hunt}]{R_package_caret}
\bibinfo{author}{Kuhn, M.}, \bibinfo{author}{Wing, J.}, \bibinfo{author}{Weston, S.}, \bibinfo{author}{Williams, A.}, \bibinfo{author}{Keefer, C.}, \bibinfo{author}{Engelhardt, A.}, \bibinfo{author}{Cooper, T.}, \bibinfo{author}{Mayer, Z.}, \bibinfo{author}{Kenkel, B.}, \bibinfo{author}{Team, R.C.}, \bibinfo{author}{Benesty, M.}, \bibinfo{author}{Lescarbeau, R.}, \bibinfo{author}{Ziem, A.}, \bibinfo{author}{Scrucca, L.}, \bibinfo{author}{Tang, Y.}, \bibinfo{author}{Candan, C.}, \bibinfo{author}{Hunt, T.}, \bibinfo{year}{2023}.
\newblock \bibinfo{title}{Caret: {{Classification}} and regression training}.
\bibitem[{Kyriakidis et~al.(2015)Kyriakidis, Happee and {de Winter}}]{lirui-034}
\bibinfo{author}{Kyriakidis, M.}, \bibinfo{author}{Happee, R.}, \bibinfo{author}{{de Winter}, J.}, \bibinfo{year}{2015}.
\newblock \bibinfo{title}{Public opinion on automated driving: Results of an international questionnaire among 5000 respondents}.
\newblock \bibinfo{journal}{Transportation Research Part F: Traffic Psychology and Behaviour} \bibinfo{volume}{32}, \bibinfo{pages}{127--140}.
\newblock \URLprefix \url{https://www.sciencedirect.com/science/article/pii/S1369847815000777}, \DOIprefix\doi{https://doi.org/10.1016/j.trf.2015.04.014}.
\bibitem[{Lee et~al.(2019)Lee, Lee, Park, Lee and Ha}]{lirui-036}
\bibinfo{author}{Lee, J.}, \bibinfo{author}{Lee, D.}, \bibinfo{author}{Park, Y.}, \bibinfo{author}{Lee, S.}, \bibinfo{author}{Ha, T.}, \bibinfo{year}{2019}.
\newblock \bibinfo{title}{Autonomous vehicles can be shared, but a feeling of ownership is important: Examination of the influential factors for intention to use autonomous vehicles}.
\newblock \bibinfo{journal}{Transportation Research Part C: Emerging Technologies} \bibinfo{volume}{107}, \bibinfo{pages}{411--422}.
\newblock \URLprefix \url{https://www.sciencedirect.com/science/article/pii/S0968090X19301895}, \DOIprefix\doi{https://doi.org/10.1016/j.trc.2019.08.020}.
\bibitem[{Liljamo et~al.(2021)Liljamo, Liimatainen, Pöllänen and Viri}]{lirui-015}
\bibinfo{author}{Liljamo, T.}, \bibinfo{author}{Liimatainen, H.}, \bibinfo{author}{Pöllänen, M.}, \bibinfo{author}{Viri, R.}, \bibinfo{year}{2021}.
\newblock \bibinfo{title}{The effects of mobility as a service and autonomous vehicles on people’s willingness to own a car in the future}.
\newblock \bibinfo{journal}{Sustainability} \bibinfo{volume}{13}.
\newblock \URLprefix \url{https://www.mdpi.com/2071-1050/13/4/1962}, \DOIprefix\doi{10.3390/su13041962}.
\bibitem[{Liu et~al.(2019a)Liu, Guo, Ren, Wang and Xu}]{lirui-008}
\bibinfo{author}{Liu, P.}, \bibinfo{author}{Guo, Q.}, \bibinfo{author}{Ren, F.}, \bibinfo{author}{Wang, L.}, \bibinfo{author}{Xu, Z.}, \bibinfo{year}{2019}a.
\newblock \bibinfo{title}{Willingness to pay for self-driving vehicles: Influences of demographic and psychological factors}.
\newblock \bibinfo{journal}{Transportation Research Part C: Emerging Technologies} \bibinfo{volume}{100}, \bibinfo{pages}{306--317}.
\newblock \URLprefix \url{https://www.sciencedirect.com/science/article/pii/S0968090X18311306}, \DOIprefix\doi{https://doi.org/10.1016/j.trc.2019.01.022}.
\bibitem[{Liu et~al.(2019b)Liu, Yang and Xu}]{lirui-001}
\bibinfo{author}{Liu, P.}, \bibinfo{author}{Yang, R.}, \bibinfo{author}{Xu, Z.}, \bibinfo{year}{2019}b.
\newblock \bibinfo{title}{Public acceptance of fully automated driving: Effects of social trust and risk/benefit perceptions}.
\newblock \bibinfo{journal}{Risk Analysis} \bibinfo{volume}{39}, \bibinfo{pages}{326--341}.
\newblock \URLprefix \url{https://onlinelibrary.wiley.com/doi/abs/10.1111/risa.13143}, \DOIprefix\doi{https://doi.org/10.1111/risa.13143}, \href{http://arxiv.org/abs/https://onlinelibrary.wiley.com/doi/pdf/10.1111/risa.13143}{{\tt arXiv:https://onlinelibrary.wiley.com/doi/pdf/10.1111/risa.13143}}.
\bibitem[{Man et~al.(2020)Man, Xiong, Chang and Chan}]{lirui-071}
\bibinfo{author}{Man, S.S.}, \bibinfo{author}{Xiong, W.}, \bibinfo{author}{Chang, F.}, \bibinfo{author}{Chan, A.H.S.}, \bibinfo{year}{2020}.
\newblock \bibinfo{title}{Critical factors influencing acceptance of automated vehicles by hong kong drivers}.
\newblock \bibinfo{journal}{IEEE Access} \bibinfo{volume}{8}, \bibinfo{pages}{109845--109856}.
\newblock \DOIprefix\doi{10.1109/ACCESS.2020.3001929}.
\bibitem[{Menon et~al.(2019)Menon, Barbour, Zhang, Pinjari and Mannering}]{lirui-081}
\bibinfo{author}{Menon, N.}, \bibinfo{author}{Barbour, N.}, \bibinfo{author}{Zhang, Y.}, \bibinfo{author}{Pinjari, A.R.}, \bibinfo{author}{Mannering, F.}, \bibinfo{year}{2019}.
\newblock \bibinfo{title}{Shared autonomous vehicles and their potential impacts on household vehicle ownership: An exploratory empirical assessment}.
\newblock \bibinfo{journal}{International Journal of Sustainable Transportation} \bibinfo{volume}{13}, \bibinfo{pages}{111--122}.
\newblock \URLprefix \url{https://doi.org/10.1080/15568318.2018.1443178}, \DOIprefix\doi{10.1080/15568318.2018.1443178}, \href{http://arxiv.org/abs/https://doi.org/10.1080/15568318.2018.1443178}{{\tt arXiv:https://doi.org/10.1080/15568318.2018.1443178}}.
\bibitem[{Merfeld et~al.(2019)Merfeld, Wilhelms, Henkel and Kreutzer}]{lirui-020}
\bibinfo{author}{Merfeld, K.}, \bibinfo{author}{Wilhelms, M.P.}, \bibinfo{author}{Henkel, S.}, \bibinfo{author}{Kreutzer, K.}, \bibinfo{year}{2019}.
\newblock \bibinfo{title}{Carsharing with shared autonomous vehicles: Uncovering drivers, barriers and future developments – a four-stage delphi study}.
\newblock \bibinfo{journal}{Technological Forecasting and Social Change} \bibinfo{volume}{144}, \bibinfo{pages}{66--81}.
\newblock \URLprefix \url{https://www.sciencedirect.com/science/article/pii/S0040162517316293}, \DOIprefix\doi{https://doi.org/10.1016/j.techfore.2019.03.012}.
\bibitem[{Milakis et~al.(2017)Milakis, van Arem and van Wee}]{lirui-024}
\bibinfo{author}{Milakis, D.}, \bibinfo{author}{van Arem, B.}, \bibinfo{author}{van Wee, B.}, \bibinfo{year}{2017}.
\newblock \bibinfo{title}{Policy and society related implications of automated driving: A review of literature and directions for future research}.
\newblock \bibinfo{journal}{Journal of Intelligent Transportation Systems} \bibinfo{volume}{21}, \bibinfo{pages}{324--348}.
\newblock \URLprefix \url{https://doi.org/10.1080/15472450.2017.1291351}, \DOIprefix\doi{10.1080/15472450.2017.1291351}, \href{http://arxiv.org/abs/https://doi.org/10.1080/15472450.2017.1291351}{{\tt arXiv:https://doi.org/10.1080/15472450.2017.1291351}}.
\bibitem[{Mohammadzadeh(2021)}]{lirui-014}
\bibinfo{author}{Mohammadzadeh, M.}, \bibinfo{year}{2021}.
\newblock \bibinfo{title}{Sharing or owning autonomous vehicles? comprehending the role of ideology in the adoption of autonomous vehicles in the society of automobility}.
\newblock \bibinfo{journal}{Transportation Research Interdisciplinary Perspectives} \bibinfo{volume}{9}, \bibinfo{pages}{100294}.
\newblock \URLprefix \url{https://www.sciencedirect.com/science/article/pii/S2590198220302050}, \DOIprefix\doi{https://doi.org/10.1016/j.trip.2020.100294}.
\bibitem[{Montoro et~al.(2019)Montoro, Useche, Alonso, Lijarcio, Bosó-Seguí and Martí-Belda}]{lirui-030}
\bibinfo{author}{Montoro, L.}, \bibinfo{author}{Useche, S.A.}, \bibinfo{author}{Alonso, F.}, \bibinfo{author}{Lijarcio, I.}, \bibinfo{author}{Bosó-Seguí, P.}, \bibinfo{author}{Martí-Belda, A.}, \bibinfo{year}{2019}.
\newblock \bibinfo{title}{Perceived safety and attributed value as predictors of the intention to use autonomous vehicles: A national study with spanish drivers}.
\newblock \bibinfo{journal}{Safety Science} \bibinfo{volume}{120}, \bibinfo{pages}{865--876}.
\newblock \URLprefix \url{https://www.sciencedirect.com/science/article/pii/S0925753519301894}, \DOIprefix\doi{https://doi.org/10.1016/j.ssci.2019.07.041}.
\bibitem[{Narayanan et~al.(2020)Narayanan, Chaniotakis and Antoniou}]{lirui-038}
\bibinfo{author}{Narayanan, S.}, \bibinfo{author}{Chaniotakis, E.}, \bibinfo{author}{Antoniou, C.}, \bibinfo{year}{2020}.
\newblock \bibinfo{title}{Shared autonomous vehicle services: A comprehensive review}.
\newblock \bibinfo{journal}{Transportation Research Part C: Emerging Technologies} \bibinfo{volume}{111}, \bibinfo{pages}{255--293}.
\newblock \URLprefix \url{https://www.sciencedirect.com/science/article/pii/S0968090X19303493}, \DOIprefix\doi{https://doi.org/10.1016/j.trc.2019.12.008}.
\bibitem[{O'Malley and Neelon(2014)}]{OMALLEY2014131}
\bibinfo{author}{O'Malley, A.}, \bibinfo{author}{Neelon, B.}, \bibinfo{year}{2014}.
\newblock \bibinfo{title}{Latent factor and latent class models to accommodate heterogeneity, using structural equation}.
\newblock \bibinfo{journal}{Encyclopedia of Health Economics} , \bibinfo{pages}{131--140}\URLprefix \url{https://www.sciencedirect.com/science/article/pii/B9780123756787007124}, \DOIprefix\doi{https://doi.org/10.1016/B978-0-12-375678-7.00712-4}.
\bibitem[{Orsot-Dessi et~al.(2023)Orsot-Dessi, Ashta and Mor}]{lirui-092}
\bibinfo{author}{Orsot-Dessi, P.}, \bibinfo{author}{Ashta, A.}, \bibinfo{author}{Mor, S.}, \bibinfo{year}{2023}.
\newblock \bibinfo{title}{The determinants of the intention to use autonomous vehicles}.
\newblock \bibinfo{journal}{African Journal of Science, Technology, Innovation and Development} \bibinfo{volume}{0}, \bibinfo{pages}{1--11}.
\newblock \URLprefix \url{https://doi.org/10.1080/20421338.2023.2174754}, \DOIprefix\doi{10.1080/20421338.2023.2174754}.
\bibitem[{Peck et~al.(2021)Peck, Kirk, Luangrath and Shu}]{lirui-066}
\bibinfo{author}{Peck, J.}, \bibinfo{author}{Kirk, C.P.}, \bibinfo{author}{Luangrath, A.W.}, \bibinfo{author}{Shu, S.B.}, \bibinfo{year}{2021}.
\newblock \bibinfo{title}{Caring for the commons: Using psychological ownership to enhance stewardship behavior for public goods}.
\newblock \bibinfo{journal}{Journal of Marketing} \bibinfo{volume}{85}, \bibinfo{pages}{33--49}.
\newblock \URLprefix \url{https://doi.org/10.1177/0022242920952084}, \DOIprefix\doi{10.1177/0022242920952084}, \href{http://arxiv.org/abs/https://doi.org/10.1177/0022242920952084}{{\tt arXiv:https://doi.org/10.1177/0022242920952084}}.
\bibitem[{Peck and Shu(2018)}]{lirui-039}
\bibinfo{author}{Peck, J.}, \bibinfo{author}{Shu, S.B.}, \bibinfo{year}{2018}.
\newblock \bibinfo{title}{Psychological Ownership and Consumer Behavior}.
\newblock \bibinfo{publisher}{Springer Cham}.
\newblock \DOIprefix\doi{https://doi.org/10.1007/978-3-319-77158-8}.
\bibitem[{Pettigrew et~al.(2019)Pettigrew, Dana and Norman}]{lirui-021}
\bibinfo{author}{Pettigrew, S.}, \bibinfo{author}{Dana, L.M.}, \bibinfo{author}{Norman, R.}, \bibinfo{year}{2019}.
\newblock \bibinfo{title}{Clusters of potential autonomous vehicles users according to propensity to use individual versus shared vehicles}.
\newblock \bibinfo{journal}{Transport Policy} \bibinfo{volume}{76}, \bibinfo{pages}{13--20}.
\newblock \URLprefix \url{https://www.sciencedirect.com/science/article/pii/S0967070X18306395}, \DOIprefix\doi{https://doi.org/10.1016/j.tranpol.2019.01.010}.
\bibitem[{Pierce et~al.(2003)Pierce, Kostova and Dirks}]{lirui-061}
\bibinfo{author}{Pierce, J.L.}, \bibinfo{author}{Kostova, T.}, \bibinfo{author}{Dirks, K.T.}, \bibinfo{year}{2003}.
\newblock \bibinfo{title}{The state of psychological ownership: Integrating and extending a century of research}.
\newblock \bibinfo{journal}{Review of General Psychology} \bibinfo{volume}{7}, \bibinfo{pages}{84--107}.
\newblock \URLprefix \url{https://doi.org/10.1037/1089-2680.7.1.84}, \DOIprefix\doi{10.1037/1089-2680.7.1.84}, \href{http://arxiv.org/abs/https://doi.org/10.1037/1089-2680.7.1.84}{{\tt arXiv:https://doi.org/10.1037/1089-2680.7.1.84}}.
\bibitem[{Pierce and Peck(2018)}]{POchapter1}
\bibinfo{author}{Pierce, J.L.}, \bibinfo{author}{Peck, J.}, \bibinfo{year}{2018}.
\newblock \bibinfo{title}{The History of Psychological Ownership and Its Emergence in Consumer Psychology}. \bibinfo{publisher}{Springer International Publishing}, \bibinfo{address}{Cham}. chapter~\bibinfo{chapter}{1}.
\newblock pp. \bibinfo{pages}{1--18}.
\newblock \URLprefix \url{https://doi.org/10.1007/978-3-319-77158-8_1}, \DOIprefix\doi{10.1007/978-3-319-77158-8_1}.
\bibitem[{Rosa et~al.(2021)Rosa, Sharma, Tully, Giannella and Rino}]{lirui-062}
\bibinfo{author}{Rosa, W.D.L.}, \bibinfo{author}{Sharma, E.}, \bibinfo{author}{Tully, S.M.}, \bibinfo{author}{Giannella, E.}, \bibinfo{author}{Rino, G.}, \bibinfo{year}{2021}.
\newblock \bibinfo{title}{Psychological ownership interventions increase interest in claiming government benefits}.
\newblock \bibinfo{journal}{Proceedings of the National Academy of Sciences} \bibinfo{volume}{118}, \bibinfo{pages}{e2106357118}.
\newblock \URLprefix \url{https://www.pnas.org/doi/abs/10.1073/pnas.2106357118}, \DOIprefix\doi{10.1073/pnas.2106357118}, \href{http://arxiv.org/abs/https://www.pnas.org/doi/pdf/10.1073/pnas.2106357118}{{\tt arXiv:https://www.pnas.org/doi/pdf/10.1073/pnas.2106357118}}.
\bibitem[{Rousseau et~al.(1998)Rousseau, Sitkin, Burt and Camerer}]{lirui-097}
\bibinfo{author}{Rousseau, D.}, \bibinfo{author}{Sitkin, S.}, \bibinfo{author}{Burt, R.}, \bibinfo{author}{Camerer, C.}, \bibinfo{year}{1998}.
\newblock \bibinfo{title}{Not so different after all: A cross-discipline view of trust}.
\newblock \bibinfo{journal}{Academy of Management Review} \bibinfo{volume}{23}.
\newblock \DOIprefix\doi{10.5465/AMR.1998.926617}.
\bibitem[{Rudin(2019)}]{rudin_stop_2019}
\bibinfo{author}{Rudin, C.}, \bibinfo{year}{2019}.
\newblock \bibinfo{title}{Stop explaining black box machine learning models for high stakes decisions and use interpretable models instead}.
\newblock \bibinfo{journal}{Nature Machine Intelligence} \bibinfo{volume}{1}, \bibinfo{pages}{206--215}.
\newblock \URLprefix \url{https://www.nature.com/articles/s42256-019-0048-x}, \DOIprefix\doi{10.1038/s42256-019-0048-x}. \bibinfo{note}{number: 5 Publisher: Nature Publishing Group}.
\bibitem[{SAE International, 2021()}]{SAEwebsite}
SAE International, 2021, \bibinfo{year}{2021}.
\newblock \bibinfo{title}{Sae levels of driving automation™ refined for clarity and international audience}.
\newblock \URLprefix \url{https://www.sae.org/blog/sae-j3016-update}.
\bibitem[{Sener et~al.(2019)Sener, Zmud and Williams}]{lirui-010}
\bibinfo{author}{Sener, I.N.}, \bibinfo{author}{Zmud, J.}, \bibinfo{author}{Williams, T.}, \bibinfo{year}{2019}.
\newblock \bibinfo{title}{Measures of baseline intent to use automated vehicles: A case study of texas cities}.
\newblock \bibinfo{journal}{Transportation Research Part F: Traffic Psychology and Behaviour} \bibinfo{volume}{62}, \bibinfo{pages}{66--77}.
\newblock \URLprefix \url{https://www.sciencedirect.com/science/article/pii/S1369847818303413}, \DOIprefix\doi{https://doi.org/10.1016/j.trf.2018.12.014}.
\bibitem[{Tian et~al.(2021)Tian, Feng, Timmermans and Yao}]{lirui-055}
\bibinfo{author}{Tian, Z.}, \bibinfo{author}{Feng, T.}, \bibinfo{author}{Timmermans, H.J.}, \bibinfo{author}{Yao, B.}, \bibinfo{year}{2021}.
\newblock \bibinfo{title}{Using autonomous vehicles or shared cars? results of a stated choice experiment}.
\newblock \bibinfo{journal}{Transportation Research Part C: Emerging Technologies} \bibinfo{volume}{128}, \bibinfo{pages}{103117}.
\newblock \URLprefix \url{https://www.sciencedirect.com/science/article/pii/S0968090X21001364}, \DOIprefix\doi{https://doi.org/10.1016/j.trc.2021.103117}.
\bibitem[{Tomarken and Waller(2005)}]{tomarken_structural_2005}
\bibinfo{author}{Tomarken, A.J.}, \bibinfo{author}{Waller, N.G.}, \bibinfo{year}{2005}.
\newblock \bibinfo{title}{Structural equation modeling: Strengths, limitations, and misconceptions}.
\newblock \bibinfo{journal}{Annual Review of Clinical Psychology} \bibinfo{volume}{1}, \bibinfo{pages}{31--65}.
\newblock \DOIprefix\doi{10.1146/annurev.clinpsy.1.102803.144239}.
\bibitem[{Venkatesh and Bala(2008)}]{venkateshTechnologyAcceptanceModel2008}
\bibinfo{author}{Venkatesh, V.}, \bibinfo{author}{Bala, H.}, \bibinfo{year}{2008}.
\newblock \bibinfo{title}{Technology acceptance model 3 and a research agenda on interventions}.
\newblock \bibinfo{journal}{Decision Sciences} \bibinfo{volume}{39}, \bibinfo{pages}{273--315}.
\newblock \URLprefix \url{https://onlinelibrary.wiley.com/doi/abs/10.1111/j.1540-5915.2008.00192.x}, \DOIprefix\doi{10.1111/j.1540-5915.2008.00192.x}.
\bibitem[{Venkatesh and Davis(1996)}]{lirui-098}
\bibinfo{author}{Venkatesh, V.}, \bibinfo{author}{Davis, F.D.}, \bibinfo{year}{1996}.
\newblock \bibinfo{title}{A model of the antecedents of perceived ease of use: Development and test}.
\newblock \bibinfo{journal}{Decision Sciences} \bibinfo{volume}{27}, \bibinfo{pages}{451 – 481}.
\newblock \URLprefix \url{https://www.scopus.com/inward/record.uri?eid=2-s2.0-0040008172&doi=10.1111%2fj.1540-5915.1996.tb00860.x&partnerID=40&md5=64adb7c289779f8b875913870e9a6f8b}, \DOIprefix\doi{10.1111/j.1540-5915.1996.tb00860.x}. \bibinfo{note}{cited by: 2167}.
\bibitem[{Venkatesh et~al.(2003)Venkatesh, Morris, Davis and Davis}]{lirui-085}
\bibinfo{author}{Venkatesh, V.}, \bibinfo{author}{Morris, M.G.}, \bibinfo{author}{Davis, G.B.}, \bibinfo{author}{Davis, F.D.}, \bibinfo{year}{2003}.
\newblock \bibinfo{title}{User acceptance of information technology: Toward a unified view}.
\newblock \bibinfo{journal}{MIS Quarterly} \bibinfo{volume}{27}, \bibinfo{pages}{425--478}.
\newblock \URLprefix \url{http://www.jstor.org/stable/30036540}.
\bibitem[{Wang et~al.(2023)Wang, Zhang and Lin}]{lirui-091}
\bibinfo{author}{Wang, F.}, \bibinfo{author}{Zhang, Z.}, \bibinfo{author}{Lin, S.}, \bibinfo{year}{2023}.
\newblock \bibinfo{title}{Purchase intention of autonomous vehicles and industrial policies: Evidence from a national survey in china}.
\newblock \bibinfo{journal}{Transportation Research Part A: Policy and Practice} \bibinfo{volume}{173}, \bibinfo{pages}{103719}.
\newblock \URLprefix \url{https://www.sciencedirect.com/science/article/pii/S0965856423001398}, \DOIprefix\doi{https://doi.org/10.1016/j.tra.2023.103719}.
\bibitem[{Xu et~al.(2018)Xu, Zhang, Min, Wang, Zhao and Liu}]{lirui-005}
\bibinfo{author}{Xu, Z.}, \bibinfo{author}{Zhang, K.}, \bibinfo{author}{Min, H.}, \bibinfo{author}{Wang, Z.}, \bibinfo{author}{Zhao, X.}, \bibinfo{author}{Liu, P.}, \bibinfo{year}{2018}.
\newblock \bibinfo{title}{What drives people to accept automated vehicles? findings from a field experiment}.
\newblock \bibinfo{journal}{Transportation Research Part C: Emerging Technologies} \bibinfo{volume}{95}, \bibinfo{pages}{320--334}.
\newblock \URLprefix \url{https://www.sciencedirect.com/science/article/pii/S0968090X18302316}, \DOIprefix\doi{https://doi.org/10.1016/j.trc.2018.07.024}.
\bibitem[{Ye et~al.(2022)Ye, Sui, Wang, Yan and Chen}]{lirui-043}
\bibinfo{author}{Ye, X.}, \bibinfo{author}{Sui, X.}, \bibinfo{author}{Wang, T.}, \bibinfo{author}{Yan, X.}, \bibinfo{author}{Chen, J.}, \bibinfo{year}{2022}.
\newblock \bibinfo{title}{Research on parking choice behavior of shared autonomous vehicle services by measuring users’ intention of usage}.
\newblock \bibinfo{journal}{Transportation Research Part F: Traffic Psychology and Behaviour} \bibinfo{volume}{88}, \bibinfo{pages}{81--98}.
\newblock \URLprefix \url{https://www.sciencedirect.com/science/article/pii/S1369847822000973}, \DOIprefix\doi{https://doi.org/10.1016/j.trf.2022.05.012}.
\bibitem[{Yuen et~al.(2020)Yuen, Wong, Ma and Wang}]{yuenDeterminantsPublicAcceptance2020}
\bibinfo{author}{Yuen, K.F.}, \bibinfo{author}{Wong, Y.D.}, \bibinfo{author}{Ma, F.}, \bibinfo{author}{Wang, X.}, \bibinfo{year}{2020}.
\newblock \bibinfo{title}{The determinants of public acceptance of autonomous vehicles: {{An}} innovation diffusion perspective}.
\newblock \bibinfo{journal}{Journal of Cleaner Production} \bibinfo{volume}{270}, \bibinfo{pages}{121904}.
\newblock \DOIprefix\doi{10.1016/j.jclepro.2020.121904}.
\bibitem[{Zhang et~al.(2019)Zhang, Tao, Qu, Zhang, Lin and Zhang}]{lirui-019}
\bibinfo{author}{Zhang, T.}, \bibinfo{author}{Tao, D.}, \bibinfo{author}{Qu, X.}, \bibinfo{author}{Zhang, X.}, \bibinfo{author}{Lin, R.}, \bibinfo{author}{Zhang, W.}, \bibinfo{year}{2019}.
\newblock \bibinfo{title}{The roles of initial trust and perceived risk in public's acceptance of automated vehicles}.
\newblock \bibinfo{journal}{Transportation Research Part C: Emerging Technologies} \bibinfo{volume}{98}, \bibinfo{pages}{207--220}.
\newblock \URLprefix \url{https://www.sciencedirect.com/science/article/pii/S0968090X18308398}, \DOIprefix\doi{https://doi.org/10.1016/j.trc.2018.11.018}.
\bibitem[{Zhang et~al.(2021)Zhang, Zeng, Zhang, Tao, Li and Qu}]{lirui-002}
\bibinfo{author}{Zhang, T.}, \bibinfo{author}{Zeng, W.}, \bibinfo{author}{Zhang, Y.}, \bibinfo{author}{Tao, D.}, \bibinfo{author}{Li, G.}, \bibinfo{author}{Qu, X.}, \bibinfo{year}{2021}.
\newblock \bibinfo{title}{What drives people to use automated vehicles? a meta-analytic review}.
\newblock \bibinfo{journal}{Accident Analysis \& Prevention} \bibinfo{volume}{159}, \bibinfo{pages}{106270}.
\newblock \URLprefix \url{https://www.sciencedirect.com/science/article/pii/S0001457521003018}, \DOIprefix\doi{https://doi.org/10.1016/j.aap.2021.106270}.
\bibitem[{Zhao et~al.(2016)Zhao, Parvinzamir, Wei, Liu, Deng, Dong, Third, Lukosevicius, Marozas, Kaldoudi and Clapworthy}]{Zhao_visual_2016}
\bibinfo{author}{Zhao, Y.}, \bibinfo{author}{Parvinzamir, F.}, \bibinfo{author}{Wei, S.}, \bibinfo{author}{Liu, E.}, \bibinfo{author}{Deng, Z.}, \bibinfo{author}{Dong, F.}, \bibinfo{author}{Third, A.}, \bibinfo{author}{Lukosevicius, A.}, \bibinfo{author}{Marozas, V.}, \bibinfo{author}{Kaldoudi, E.}, \bibinfo{author}{Clapworthy, G.}, \bibinfo{year}{2016}.
\newblock \bibinfo{title}{Visual analytics for health monitoring and risk management in carre}, in: \bibinfo{booktitle}{International Conference on Technologies for E-Learning and Digital Entertainment}, pp. \bibinfo{pages}{380--391}.
\newblock \DOIprefix\doi{10.1007/978-3-319-40259-8_33}.
\bibitem[{Zmud et~al.(2016)Zmud, Sener and Wagner}]{lirui-026}
\bibinfo{author}{Zmud, J.}, \bibinfo{author}{Sener, I.N.}, \bibinfo{author}{Wagner, J.}, \bibinfo{year}{2016}.
\newblock \bibinfo{title}{Self-driving vehicles: Determinants of adoption and conditions of usage}.
\newblock \bibinfo{journal}{Transportation Research Record} \bibinfo{volume}{2565}, \bibinfo{pages}{57--64}.
\newblock \URLprefix \url{https://doi.org/10.3141/2565-07}, \DOIprefix\doi{10.3141/2565-07}, \href{http://arxiv.org/abs/https://doi.org/10.3141/2565-07}{{\tt arXiv:https://doi.org/10.3141/2565-07}}.
\bibitem[{Zong et~al.(2019)Zong, Zhang and Jiang}]{lirui-080}
\bibinfo{author}{Zong, W.}, \bibinfo{author}{Zhang, J.}, \bibinfo{author}{Jiang, Y.}, \bibinfo{year}{2019}.
\newblock \bibinfo{title}{Long-term changes in japanese young people’s car ownership and usage from an expenditure perspective}.
\newblock \bibinfo{journal}{Transportation Research Part D: Transport and Environment} \bibinfo{volume}{75}, \bibinfo{pages}{23--41}.
\newblock \URLprefix \url{https://www.sciencedirect.com/science/article/pii/S1361920919303281}, \DOIprefix\doi{https://doi.org/10.1016/j.trd.2019.08.013}.

\end{thebibliography}

\end{document}